\pdfoutput=0
\documentclass[11pt]{article}
\usepackage{axodraw2,pix}
\usepackage{epsfig}
\usepackage{amsfonts}
\usepackage{amsmath,amssymb}
\usepackage{bbm,bm}
\usepackage{cite}
\allowdisplaybreaks[1]

\usepackage[
	colorlinks=true,
	linkcolor=black,
	citecolor=black,
	filecolor=black,
	urlcolor=black,
        breaklinks=true
        ]{hyperref}

\renewcommand{\rmi}[1]{{\mbox{\scriptsize #1}}}
\newcommand{\rmii}[1]{{\mbox{\tiny\rm{#1}}}}
\newcommand{\clog}{\Big(c+\ln\Big(\frac{3T}{\Lamd}\Big)\Big)}
\newcommand{\bmu}{\bar{\mu}}
\newcommand{\Lamd}{\bmu_{\rmii{3d}}}
\newcommand{\LamD}{\bmu}

\newcommand{\nf}{n_{\rm f}}

\newcommand{\Nc}{N_{\rm c}}

\newcommand{\Tc}{T_{\rm c}}
\newcommand{\Tn}{T_{\rm n}}

\newcommand{\Yl}{Y_{\ell}}
\newcommand{\Yq}{Y_{q}}
\newcommand{\Ye}{Y_{e}}
\newcommand{\Yu}{Y_{u}}
\newcommand{\Yd}{Y_{d}}
\newcommand{\Yf}{Y_{\rmi{f}}}
\newcommand{\Ys}{Y_{\phi}}
\newcommand{\gY}{g_\rmii{$Y$}}
\newcommand{\go}{g_1}
\newcommand{\gt}{g_2}
\newcommand{\gs}{g_\rmi{s}}
\newcommand{\mDi}[1]{m_{\rmii{D}#1}}
\newcommand{\mA}{m_\rmii{$A$}}
\newcommand{\mG}{m_\rmii{$G$}}

\newcommand{\MZ}{M_\rmii{$Z$}}
\newcommand{\MW}{M_\rmii{$W$}}

\newcommand{\Mpl}{M_\rmii{Pl}}
\newcommand{\vrel}{v_\rmi{rel}}

\newcommand{\re}{\mathop{\mbox{Re}}}

\newcommand{\T}{\rmii{$T$}}

\newcommand{\msl}[1]{\,\slash\!\!\!{#1}\,}
\newcommand\MSbar{$\overline{\rm MS}$}

\DeclareMathOperator{\arccot}{arccot}

\renewcommand{\vec}[1]{{\bf #1}}
\renewcommand{\nn}{\nonumber \\}
\newcommand{\gammaE}{{\gamma_\rmii{E}}}

\newcommand{\nF}{n_\rmii{F}}
\newcommand{\nB}{n_\rmii{B}}

\newcommand{\sumint}[1]{{\hbox{$\sum$}\!\!\!\!\!\!\!\int\,}_{\!\!\!\!\raise-0.9ex\hbox{$\scriptstyle{#1}$}}}
\newcommand{\Tint}[1]{{\hbox{$\sum$}\!\!\!\!\!\!\!\int\,}_{\!\!\!\!\raise-0.9ex\hbox{$\scriptstyle{#1}$}}}
\newcommand{\Tinti}[1]{{{\Sigma}\!\!\!\!\raise0.3ex\hbox{$\int$}_\rmii{${#1}$}}}
\newcommand{\Tintip}[1]{{{\Sigma'}\!\!\!\!\!\raise0.3ex\hbox{$\int$}_\rmii{${#1}$}}}

\def\scfc{0.7}  % picture scale factor
\def\phgt{21}   % all picture height 30 * \scfc
\def\pwc{21}    % c   picture width  30 * \scfc
\def\pwcb{31.5} % cb  picture width  45 * \scfc
\def\pwcc{42} % cb  picture width  60 * \scfc
\newcommand{\PIC}[4]{\;\parbox[c]{#2 pt}{\begin{picture}(#2,#3)(0,0)
\SetWidth{1.0}\SetScale{#4} #1 \end{picture}}\;}
\renewcommand{\pic}[1]{\PIC{#1}{\pwc}{\phgt}{\scfc}}
\renewcommand{\picb}[1]{\PIC{#1}{\pwcb}{\phgt}{\scfc}}
\renewcommand{\picc}[1]{\PIC{#1}{\pwcc}{\phgt}{\scfc}}

\makeatletter \@addtoreset{equation}{section} \makeatother
\renewcommand{\theequation}{\arabic{section}.\arabic{equation}}
\makeatletter
\renewcommand\section{\@startsection{section}{1}{\z@}%
  {-5.5ex \@plus -1ex \@minus -.2ex}% bfr-skip
  {2.3ex \@plus.2ex}%
  {\normalfont\large\bfseries}}
\renewcommand\subsection{\@startsection{subsection}{2}{\z@}%
  {-3.25ex\@plus -1ex \@minus -.2ex}%
  {1.5ex \@plus .2ex}%
  {\normalfont\normalsize\bfseries}}
\renewcommand\thesection{\@arabic\c@section}
\renewcommand\thesubsection{\thesection.\@arabic\c@subsection}
\renewcommand{\@seccntformat}[1]{%
  \csname the#1\endcsname.\hspace{1.0em}}
\makeatother

\begin{document}

\flushbottom

\begin{titlepage}

\begin{flushright}
%Notes SB,PS,TT
HIP-2022-19/TH\\
NORDITA 2022-050
\end{flushright}
\begin{centering}

\vfill

{\Large{\bf%
Strong electroweak phase transition\\
in $t$-channel simplified dark matter models
}}

\vspace{0.8cm}

\renewcommand{\thefootnote}{\fnsymbol{footnote}}
Simone Biondini$^{\rm a,}$%
\footnotemark[1],
Philipp Schicho$^{\rm b,}$%
\footnotemark[2],
Tuomas V.~I.~Tenkanen$^{\rm c,d,e,}$%
\footnotemark[3]

\vspace{0.8cm}

$^\rmi{a}$%
{\em
Department of Physics, University of Basel,\\
Klingelbergstr.~82,
CH-4056 Basel,
Switzerland\\}
\vspace{0.3cm}

$^{\rmi{b}}$%
{\em
Department of Physics and Helsinki Institute of Physics,\\
P.O.~Box 64,
FI-00014 University of Helsinki,
Finland\\}
\vspace{0.3cm}

$^\rmi{c}$%
{\em
Nordita,
KTH Royal Institute of Technology and Stockholm University,\\
Roslagstullsbacken 23,
SE-106 91 Stockholm,
Sweden\\}
\vspace{0.3cm}

$^\rmi{d}$%
{\em
Tsung-Dao Lee Institute \& School of Physics and Astronomy,
Shanghai Jiao Tong University,
Shanghai 200240, China\\}
\vspace{0.3cm}

$^\rmi{e}$%
{\em
Shanghai Key Laboratory for Particle Physics and Cosmology,
Key Laboratory for Particle Astrophysics and Cosmology (MOE),
Shanghai Jiao Tong University,\\
Shanghai 200240, China\\}

\vspace*{0.8cm}

\mbox{\bf Abstract}
 
\end{centering}

\vspace*{0.3cm}

\noindent
Beyond the Standard Model physics is required to explain both
dark matter and
the baryon asymmetry of the universe,
the latter possibly generated during a strong first-order electroweak phase transition.
While many proposed models tackle these problems
independently,
it is interesting to inquire whether the same model can explain both.
In this context,
we link state-of-the-art perturbative assessments of
the phase transition thermodynamics with
the extraction of the dark matter energy density.
These techniques are applied to
a next-to-minimal dark matter model containing
an inert
Majorana fermion that is coupled to
Standard Model leptons via
a scalar mediator,
where the mediator interacts directly with the Higgs boson.
For dark matter masses
$
180~{\rm GeV} < M_\chi \lesssim
300~{\rm GeV}$,
we discern regions of the model parameter space that reproduce
the observed dark matter energy density and
allow for a first-order phase transition, while evading the most stringent collider constraints.

\vfill
\end{titlepage}

\tableofcontents
\renewcommand{\thefootnote}{\fnsymbol{footnote}}
\footnotetext[1]{simone.biondini@unibas.ch}
\footnotetext[2]{philipp.schicho@helsinki.fi}
\footnotetext[3]{tuomas.tenkanen@su.se}
\clearpage

\renewcommand{\thefootnote}{\arabic{footnote}}
\setcounter{footnote}{0}

%%%%%%%%%%%%%%%%%%%%%%%%%%%%% SECTION %%%%%%%%%%%%%%%%%%%%%%%%%%%%%%%%%%%%
%
\section{Introduction}

Beyond the Standard Model (BSM) physics is invoked to explain at least
two compelling observations:
the matter-antimatter, or
baryon, asymmetry of the universe and
dark matter (DM).
In both cases new degrees of freedom are introduced and assumed to interact with
the SM particles to have
a model testable by collider probes.

New particles that couple to the SM Higgs boson can affect
the electroweak phase transition (EWPT) thermodynamics.
Such new particles can even render the character of the transition from
smooth crossover to
first-order.
A strong transition opens up the possibility for
a successful baryogenesis mechanism once additional CP phases are included.
Generating the matter-antimatter asymmetry
during the EWPT is known as
electroweak baryogenesis~(EWBG)~\cite{Kuzmin:1985mm,Shaposhnikov:1987tw}.
It is particularly appealing as it is perhaps
the only proposed mechanism of baryogenesis
directly testable at energies of
present-day collider experiments~\cite{Morrissey:2012db,Ramsey-Musolf:2019lsf}.
Moreover, a strong first-order phase transition can trigger
gravitational wave (GW) production, that can well be in reach of
forthcoming space-based interferometers~\cite{%
  LISA:2017pwj,Kawamura:2006up,Guo:2018npi,Harry:2006fi,Caprini:2019egz}.
During recent years,
studies of cosmological phase transitions have increased the hope for
their concrete probing by gravitational wave astronomy.
In complementarity with collider experiments,
such probes could scope
the underlying theories of elementary particle physics.
In turn, stable massive particles are required
to explain the dark matter component of our universe.
While evidence for DM is merely based on its gravitational effects,
nothing prevents DM
from interacting feebly with the visible sector.

It is compelling
to ask whether a single BSM model can accommodate {\em both}
dark matter and
a strong EWPT.
In addition to following a minimalist approach,
it is important
to investigate all the imprints that additional degrees of freedom and
their interactions may have left during the cosmological history.
In this context, rather extensive investigations have been carried out
e.g.~for scalar extensions of the SM~\cite{%
  Barger:2008jx,Barger:2010yn,Espinosa:2011ax,Ahriche:2012ei,Chowdhury:2011ga,
  Borah:2012pu,Gonderinger:2012rd,Gil:2012ya,
  Cline:2012hg,Cline:2017qpe,
  Cho:2021itv,Alanne:2014bra,Basler:2020nrq,Jiang:2015cwa,Chiang:2017nmu,Chen:2019ebq,Ghorbani:2018yfr,Ghorbani:2019itr,Ertas:2021xeh},
super-symmetry~\cite{Kumar:2011np,Kozaczuk:2011vr,Carena:2011jy}, composite Higgs models~\cite{Espinosa:2011eu,Chala:2016ykx} and
simplified dark matter models~\cite{Ghorbani:2017jls,Liu:2021mhn}.

One aim of this article
is to link perturbative state-of-art assessments of
the thermodynamics of the EWPT with
the extraction of the dark matter energy density.
Two main aspects, that we improve upon, are the following.
First, we compute the phase transition thermodynamics utilising
dimensionally reduced effective field theories at high temperature.
This allows for including all necessary thermal corrections required for
the leading renormalisation group (RG) improvement~\cite{Gould:2021oba}.
Most literature determines
the regions of the model parameters compatible with
a first-order phase transition (FOPT) using
one-loop computations based on
a daisy-resummed thermal effective potential~\cite{Quiros:1999jp}.
Such computations do not admit RG improvement and are consequently plagued by
a potentially large renormalisation scale dependence.
This large dependence is
an intrinsic, theoretical uncertainty and
reflects that
missing higher loop order corrections are large~\cite{Croon:2020cgk}.
As a second aspect, we improve
the accurate extraction of the dark matter energy density by assessing
the relevance of Sommerfeld enhancement~\cite{Sommerfeld:1931,Sakharov:1991pia,Hisano:2004ds} and
bound-state effects~\cite{vonHarling:2014kha} on
the annihilation processes that drive the freeze-out mechanism.%
\footnote{
  This work focuses on the freeze-out production mechanism.
  Complementary production mechanisms are extensively discussed in
  sec.~\ref{sec:dark_matter}.
}

To implement our program and in contrast to earlier literature,
we consider a {\em simplified} dark matter model that features
a SM singlet Majorana fermion coupled to
SM leptons via a scalar mediator.
The latter has SM quantum numbers as dictated by gauge invariance,
and
it couples also to the Higgs boson.
The portal interaction between the Higgs and the mediator
can affect the EWPT thermodynamics.
The model belongs to a broader next-to-minimal family of models that offer
a rich phenomenology and diverse production mechanisms in the early universe.
They are often referred to as
``$t$-channel mediator models''~\cite{An:2013xka,Kopp:2014tsa,Garny:2015wea,Arina:2020udz},
where the mediator is indeed the degree of freedom that couples
the visible sector in a gauge-invariant and renormalisable way to
the actual DM particle,
that is sterile under the SM gauge groups;
see ref.~\cite{Garny:2015wea} for a review.
Simplified models are specifically conceived to involve only
a few new particles and interactions, and many of them can be understood as
a limit of a more general new-physics scenario.
For example the model we consider has ties with supersymmetry
(cf.\ sec.~\ref{sec:model}).
The main advantages of a simplified-model approach are
(i) to carry out the relevant phenomenology with a handful of parameters,
(ii) scrutinise the DM production mechanisms in the early universe and
(iii) recast experimental constraints on the model parameters.

Within the simplified model considered, we study
the thermodynamics of the EWPT where the Higgs boson interacts with
a complex ${\rm U(1)}_\rmii{Y}$ charged scalar, that in turn couples via
a Yukawa interaction to
a Majorana fermion and
a right-handed SM lepton.
This model gives rise to a mechanism for
a two-step phase transition.
The presence of a light, dynamical new scalar
generates a barrier to the leading-order (LO) Higgs potential and
strengthens the final transition to the electroweak minimum.
The dark matter particle enters the dynamics of the phase transition as
a further, despite milder, perturbation.

This article is organised follows.
Section~\ref{sec:model}
introduces our setup and
describes the salient features of the model Lagrangian by
making contact with the framework of simplified dark matter models.
Section~\ref{sec:ewpt}
summarises the procedures to address the EWPT thermodynamics
and
sec.~\ref{sec:dark_matter} discusses the extraction of the dark matter energy density.
Section~\ref{sec:interplay} comprises the results of our numerical analysis, with a focus on the overlap between the EWPT and DM.
Finally, we discuss our findings together with an outlook in sec.~\ref{sec:outlook}.
While the main body of the article
includes the ingredients and results for
a self-contained discussion on EWPT and DM,
technical details to aid the analyses are collected in
appendix~\ref{sec:model:match} and \ref{sec:app:DM}.

%%%%%%%%%%%%%%%%%%%%%%%%%%%%% SECTION %%%%%%%%%%%%%%%%%%%%%%%%%%%%%%%%%%%%
%
\section{Model}
\label{sec:model}

We focus on a model that augments the Standard Model by
a gauge singlet Majorana fermion~$(\chi)$ and
a complex scalar field~$(\eta)$, which is a singlet under
${\rm SU(2)}_\rmii{L}$ and
${\rm SU(3)}$
but charged under ${\rm U(1)}_\rmii{Y}$ with
hypercharge $Y_\eta$.
The scalar then mediates the interaction between
the dark fermion and the SM degrees of freedom, more precisely charged right-handed leptons.
We impose a $Z_2$
discrete symmetry under which $\chi$ and $\eta$ are odd while
the Standard Model particles are even, to guarantee
the stability of the dark matter particle~\cite{Garny:2015wea,Arina:2020udz}.
The Majorana fermion is assumed to be lighter than the accompanying scalar state $\eta$,
hence the decay process $\chi \to \eta \, e$ involving right-handed leptons, $e$,
is kinematically not allowed despite the $Z_2$~symmetry is respected.
An additional Yukawa interaction between the Standard Model Higgs doublet,
lepton doublets and the singlet Majorana fermion is forbidden by the $Z_2$~symmetry.%
\footnote{
  The operator reads
  $\mathcal{L}^{\rmi{portal}'}_{\rmi{Yukawa }}= \tilde{y} \, \bar{\chi} \tilde{\phi}^\dagger P_{\rmii{L}} \ell + \rm{h.c.}$,
  where the Higgs doublet appears in the following form
  $\tilde{\phi} = i \sigma^2 \phi^*$, and
  $\ell$ is a SU(2) lepton doublet.
  This very interaction, whenever the mass of the dark fermion is larger than
  the Higgs and lepton mass, would trigger the decay process for the dark matter particle.
}
The fermion $\chi$ is then stable and is the actual DM particle.
Since the complex scalar $\eta$ is charged under
the ${\rm U(1)}_\rmii{Y}$ gauge group and
interacts with photons,
it does not qualify for a DM candidate.

The choice of the SM gauge charges of the scalar $\eta$ is two-fold.
We avoid QCD interactions since new coloured scalars are
bounded
to be heavier than 1~TeV~\cite{CMS:2017abv,CMS:2017mbm,ATLAS:2017eoo,ATLAS:2021twp}
by collider searches.
This lower mass bound renders such states nonviable for a sizeable impact on the EWPT.
Instead, the scalar mediator can be charged under
${\rm SU(2)}_\rmii{L}$.
Here,
we limit ourselves to a SU(2)-singlet state to simplify the framework
of linking the thermodynamics of the phase transition and the dark matter relic density.

The corresponding Lagrangian in
Minkowskian spacetime
takes the form
\begin{align} 
\label{eq:lag:4d}
\mathcal{L}_{\rmi{4d}} &=
    \mathcal{L}_{\rmii{SM}}
  + \mathcal{L}_{\eta}
  + \mathcal{L}_{\chi}
  - \mathcal{L}^{\rmi{portal}}_{}
  \;,
\end{align}
with
the dark Majorana fermion $\mathcal{L}_{\chi}$ and
the complex scalar $\mathcal{L}_{\eta}$ sectors
\begin{align} 
\mathcal{L}_{\chi} &=
    \frac{1}{2} \bar{\chi} (i \msl{\partial} - \mu_{\chi} ) \chi
    \;,\\
\mathcal{L}_{\eta} &=
    (D_\mu \eta)^\dagger (D_\mu \eta)
  - \mu^2_\eta \eta^\dagger\eta
  - \lambda_{2} (\eta^\dagger\eta)^2
    \;,
\end{align}
where
the covariant derivative is
$
D_\mu \eta =
(\partial_\mu - i\go\frac{Y_\eta}{2} B_\mu)\eta
$,
$\go$ is the ${\rm U(1)}_\rmii{Y}$ gauge coupling,
and
$\lambda_{1}$ is reserved for the SM Higgs doublet self-coupling.
The two sectors above interact through
both scalar and Yukawa (fermion) portal couplings
$\mathcal{L}^{\rmi{portal}}_{ } =
  \mathcal{L}^{\rmi{portal}}_{\rmi{scalar}}
+ \mathcal{L}^{\rmi{portal}}_{\rmi{Yukawa}}$
\begin{align} 
\label{eq:L:portal:scalar}
\mathcal{L}^{\rmi{portal}}_{\rmi{scalar}} &=
  \lambda_{3} (\eta^\dagger\eta) (\phi^\dagger \phi)
  \;,\\[2mm]
\label{eq:L:portal:yukawa}
\mathcal{L}^{\rmi{portal}}_{\rmi{Yukawa}} &=
    y\,\eta \, \bar{\chi} P_{\rmii{R}}\, e 
  + \text{h.c.} 
  \;,
\end{align}
where 
$P_{\rmii{R/L}} = (\mathbbm{1} \pm \gamma_{5})/2$ are chiral projectors and
$e$ is a right-handed SM lepton i.e.\
electron,
muon, or
tau.
As commonly adopted in the literature, we assume the dark matter particle only couples to one generation
of fermions, which can be ensured by introducing
a family global quantum numbers carried by $\eta$.%
\footnote{
  Lifting this requirement comes at the cost of an excess in flavour-violating effects.
  The appendix of~\cite{Kopp:2014tsa} details the conditions that must be fulfilled to satisfy constraints from flavour physics for coupling to leptons.
}
This Yukawa coupling term requires the following relation amongst hypercharges
$Y_\eta= - \Ye$,
where $\Ye$ is the hypercharge of the SM (right-handed) lepton.
The free parameters of the theory are
the couplings
$\lambda_2$,
$\lambda_3$,
$|y|^2$, as well as
the mass scales
$\mu_\chi^{ }$ and
$\mu_\eta^{2}$, often written in terms of the mass splitting
$\Delta M = M_\eta - M_\chi$,
where
$M_\eta$ and
$M_\chi$ are the physical pole masses.
The model Lagrangian for the
$
{\rm SU(2)}_\rmii{L}\times
{\rm U(1)}_\rmii{Y}$ charged scalar can be found in
ref.~\cite{Garny:2015wea}.

The simplified model has ties with
the minimal supersymmetric Standard Model (MSSM).
Supersymmetry postulates the existence of partners of
the Standard Model degrees of freedom, referred to as {\em sparticles},
with a spin that
differs by one half unit from each corresponding
SM particle.
The conservation of the $R$-parity guarantees the stability of
the lightest supersymmetric particle (LSP).
In the simplified model,
the $Z_2$~symmetry plays the same role of the $R$-parity.
Moreover, if the LSP is electromagnetically neutral and weakly interacting then it is
a natural dark matter candidate.
In most cases, the LSP is assumed to be the lightest neutralino,
which is one of the mass eigenstates formed from the linear combination of
the super-partners of
the neutral Higgs bosons and
electroweak gauge bosons.
There are four neutralinos usually indicated by
$\tilde{\chi}^0_j,$ where
$j=1,\dots,4$ with increasing mass.
The lightest neutralino
is then a Majorana fermion, like
the fermion $\chi$ of the model~\eqref{eq:lag:4d}.
Moreover, the lightest neutralino interacts with sleptons or squarks,
that are heavier states (next-to-LSP) and resemble the scalar mediators of the $t$-channel simplified models,
depending on the SM gauge group charges
(in our case $\eta$ is a slepton-like particle).
A major difference with MSSM parameters lies in
the portal coupling $\lambda_3$, which we take to be $\mathcal{O}(1)$ rather than negligible, and
the freedom in the Yukawa coupling $y$.
In the MSSM one has $y = \sqrt{2} \go \approx 0.48$
for right handed leptons at the electroweak scale (see e.g.~\cite{Garny:2015wea}).

Supersymmetric particles have been intensively searched for at the LHC, and
stringent bounds have been put on the mass of the QCD coloured states such as
squarks and gluinos~\cite{%
  CMS:2017abv,CMS:2017mbm,ATLAS:2017eoo,CMS:2020fia,CMS:2020pyk,ATLAS:2021twp}.
This pushes the masses of the new states beyond $1$~TeV.
Conversely, for the colourless states, namely sleptons, the bounds are less severe and exclude masses
$M_{\rmi{slepton}} \lesssim 350 (430)$~GeV for
neutralinos lighter than
$140(180)$~GeV~\cite{ATLAS:2019gti,ATLAS:2019lff}, respectively for staus and smuons.
These collider searches can also be applied to
the simplified model in eq.~\eqref{eq:lag:4d}
since it features the same field content and type of interactions~\cite{Garny:2015wea}.
However, one does not have to stick to the MSSM parameters,
as the collider searches do not rely on the specific values of $y$ and $\lambda_3$.
This is further discussed in sec.~\ref{sec:dark_matter}.

Complementary experimental constraints on
the model
could be direct or indirect~\cite{Garny:2015wea}.
Direct detection is not a viable option because
the dark matter fermion does not couple to quarks, that are
the constituents of the nuclear targets at the direct detection facilities.
Even loop-induced interactions are ineffective, since they have to proceed via
the couplings between SM leptons and the Higgs boson, which are fairly small.
Indirect detection, that probe the dark matter annihilation occurring today, can 
potentially put constraints on the model parameters.
However, as shown in~\cite{Garny:2015wea},
the Fermi and HESS collaborations~\cite{Fermi-LAT:2013thd,HESS:2013rld,HESS:2018cbt} are
insensitive to the parameter space compatible with
the thermal freeze-out for this model.

%%%%%%%%%%%%%%%%%%%%%%%%%%%%% SUBSECTION %%%%%%%%%%%%%%%%%%%%%%%%%%%%%%%%%%%%
%
\section{Strong electroweak phase transition}
\label{sec:ewpt}

Since we will analyse the equilibrium thermodynamic properties of the EWPT,
we employ the imaginary-time formalism~\cite{Matsubara:1955ws}.
The starting point in this formalism is the Euclidean version of the Lagrangian~\eqref{eq:lag:4d}.
To by-pass infrared sensitive effects,
we employ effective field theory (EFT) techniques at high temperature.
Concretely, we utilise
the dimensionally reduced effective theory~\cite{Ginsparg:1980ef,Appelquist:1981vg}
for the fundamental theory given by the Lagrangian eq.~\eqref{eq:lag:4d}.

%%%%%%%%%%%%%%%%%%%%%%%%%%%%% SUBSECTION %%%%%%%%%%%%%%%%%%%%%%%%%%%%%%%%%%%%
%
\subsection{Dimensionally reduced model}

The high-temperature plasma exhibits a multi-scale hierarchy close to
the critical temperature ($\Tc$) of the phase transition.
In this context,
heavy, non-dynamical degrees of freedom can be integrated out.
The corresponding modes contain
the non-zero bosonic and
all fermionic Matsubara modes in
the imaginary time formalism, and
the additional Debye screened remnants of gauge fields.
A set of generic rules for such reductions in electroweak theories were established
in~\cite{Farakos:1994kx,Braaten:1995cm,Braaten:1995jr,Kajantie:1995dw}, and
recently automated in a package in~\cite{Ekstedt:2022bff}.

The resulting theory is an EFT for the original zero Matsubara modes that live in
three spatial dimensions.
Due to the distinct rest~frame of the heat bath,
Lorentz symmetry is not manifest and
additional interactions including temporal-scalars are introduced.
For the SM gauge group, these are the
electric ($B_{0}^{ }$),
isospin-electric ($A_{0}^{a}$), and
colour-electric ($C_{0}^{\alpha}$)
fields that get Debye screened at
$\mDi{i} \sim \mathcal{O}(g_{i}T)$
where
$g_{i} \in \{g_1,g_2,g_3\}$
is the gauge coupling for the respective gauge group.
These fields
are described by the Lagrangian
\begin{align}
\label{eq:lag:3d}
\mathcal{L}_{\rmii{temp}} &=
    \frac{1}{2} (\partial_r^{ } B_{0}^{ })^2
  + \frac{1}{2} \mDi{\rmii{1}}^{2} B_{0}^2
  + \frac{1}{2} (D_r^{ } A_{0}^{a})^2
  + \frac{1}{2} \mDi{\rmii{2}}^{2} A_{0}^{a} A_{0}^{a} 
  \nn &
  + \frac{1}{2} (D_r^{ } C_{0}^{\alpha})^2
  + \frac{1}{2} \mDi{\rmii{3}}^{2} C_{0}^{\alpha} C_{0}^{\alpha}
  + \frac{1}{2} h_{3}' B_{0}^2 \phi^\dagger \phi
  + \frac{1}{2} \rho_{3}' B_{0}^2 \eta^\dagger \eta
  + \dots
  \;,
\end{align}
where the ellipsis implicitly denotes
self-interaction terms and
portal couplings to Higgs and $\eta$.
The matching coefficients for such operators are either known
for the pure SM~\cite{Niemi:2021qvp,Schicho:2021gca} or
are subleading~\cite{Ekstedt:2022bff} --
especially when coupling to colour-electric fields.
The covariant derivatives act on the adjoint scalars as
$D_{r}^{ }A_{0}^{a} =
    \partial_{r}^{ } A_0^{a}
  + g_{2,3}^{ } \epsilon^a_{\phantom{a}bc}A^b_rA^c_0$,
$D_{r}^{ }C_{0}^{\alpha} =
    \partial_{r}^{ } C_{0}^\alpha
  + g_{3,3}^{ } f^\alpha_{\phantom{\alpha}\beta\rho}C^\beta_rC^\rho_0$,
$g_{i,3}$ are the $3$-dimensional effective couplings,
and
the indices $r\in\{1,\dots,d\}$.
The corresponding matching relations are collected in appendix~\ref{sec:model:match}.
At the scale of the phase transition
$\mathcal{O}(g_{2}^{2}T)$
modes of the scale
$\mathcal{O}(g_{2}^{ }T)$
can be integrated out since they are heavy~\cite{Kajantie:1995dw}.
Hence,
structurally the Lagrangian of the EFT remains the same but
in three dimensions and
is defined by
the scalar sector in eq.~\eqref{eq:lag:4d} and
the scalar interaction eq.~\eqref{eq:L:portal:scalar}.

The dark sector fermionic parameters
$\mu_\chi$ and
$|y|^2$ only affect the dimensional reduction matching relations.
In particular,
in the high-temperature expansion,
$\mu_\chi$ enters the next-to-leading order (NLO) dimensional reduction merely via
the mass correction~\eqref{eq:m:eta:2l} of
the one-loop fermionic diagram contributing to
the $\eta$ two-point function
\begin{equation}
  \TopSi(\Lsc2,1) \supset
  \TopoSB(\Lsc2,\Aqum,\Aque)
  \;.
\end{equation}
Here,
the dark fermion $\chi$ is displayed by a double-solid line,
the complex scalar $\eta$ by an arrowed double-dashed line,
and
the SM lepton $e$ by an arrowed line.
For large Majorana fermion masses
$\mu_{\chi} \gtrsim T$
the high-temperature-expansion in $\mu_{\chi}/T$ can converge badly,
and even be invalidated.
In this case,
the fermionic mass should be kept explicit~\cite{Laine:2019uua}
to cover a broader
$\mu_{\chi}/T$
range where fermionic mass effects become relevant.
The resulting sum-integrals that appear for multiple correlation functions
need to be evaluated numerically without high-temperature expansion.
The corresponding fermionic thermal integrals are listed
in appendix~\ref{sec:eft:3d};
in case of bosons cf.\
refs.~\cite{Laine:2000kv,Brauner:2016fla,Laine:2017hdk,Niemi:2018asa}.

Subsequently, we apply the high-temperature expansion, and formally count
$\mu_\chi \sim gT$.
The validity of this assumption is assessed carefully
in appendix~\ref{sec:eft:3d}.

%%%%%%%%%%%%%%%%%%%%%%%%%%%%% SUBSECTION %%%%%%%%%%%%%%%%%%%%%%%%%%%%%%%%%%%%
%
\subsection{Phase transition thermodynamics}
\label{sec:pttd}

The equilibrium thermodynamics of the system can be analysed in terms of
the effective potential that describes the free energy of the plasma~\cite{Farakos:1994kx}. Following~\cite{Schicho:2022wty}, we compute
the effective potential
at one-loop level
within the 3d EFT of the previous section.
Here, the EFT constructed by NLO dimensional reduction
includes the leading RG improvement stemming from
the hard scale.
This significantly improves~\cite{Croon:2020cgk,Gould:2021oba}
typical one-loop studies with
a daisy-resummed thermal effective potential~\cite{Arnold:1992rz,Quiros:1999jp}. 
Despite RG improvement via the 4d theory RG scale,
we do not have the full, leading RG improvement related to RG scale of
the final ultrasoft scale EFT.
The latter would require a two-loop computation of the effective potential
which can be obtained using external software
such as {\tt DRalgo}~\cite{Ekstedt:2022bff}.

\begin{figure}[t]
\centering
\includegraphics[width=0.31\textwidth]{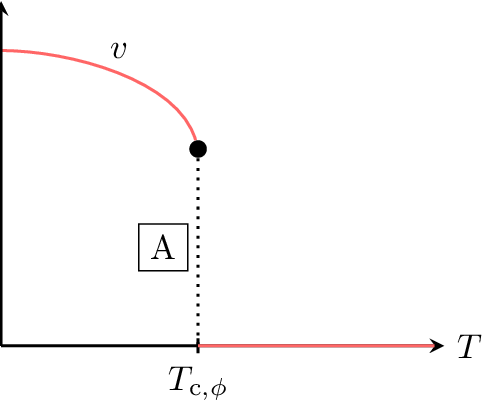}%
\hfill
\includegraphics[width=0.31\textwidth]{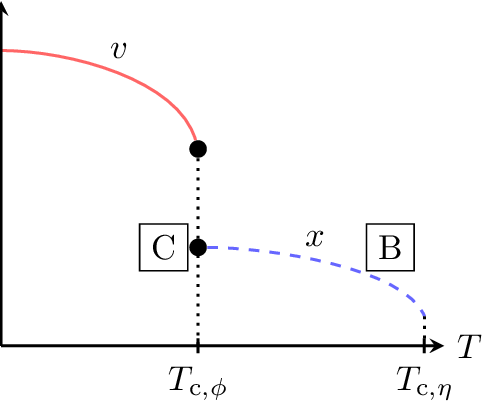}%
\hfill
\includegraphics[width=0.35\textwidth]{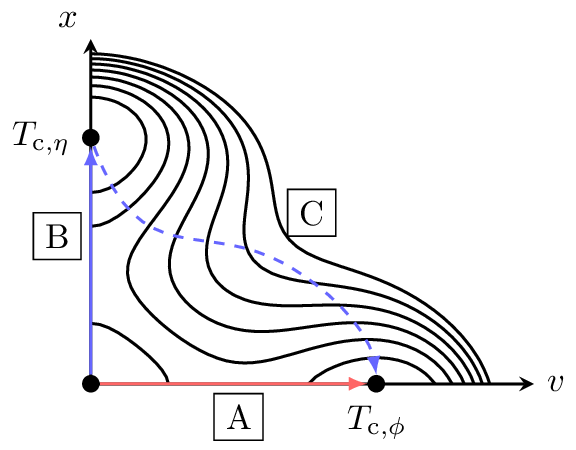}%
\caption{%
  Multi-step transition for
  the Higgs ($\phi$) with background field $v$ and
  complex singlet ($\eta$) scalar with background field $x$
  (cf.\ eq.~\eqref{eq:phi:eta:param}).
  The two left plots trace the $T$-dependence of the minima of $x$ and $v$.
  A single step (A) transitions from a symmetric
  to the Higgs phase at
  the critical temperature
  $T_{\rmi{c},\phi}$.
  During a two-step
  transition, one
  transitions first
  to the singlet direction (B) at the critical temperature $T_{\rmi{c},\eta}$
  before transitioning to the Higgs direction at $T_{\rmi{c},\phi}$ (C).
  The contour lines in the right plot schematically depict the effective potential.
}
\label{fig:multi_step}
\end{figure}

The effective potential is computed using the background field method,
within the 3d EFT~\cite{Farakos:1994kx}.
Scalar fields are parameterised as
\begin{align}
\label{eq:phi:eta:param}
\phi &=
\begin{pmatrix}
  G^+ \\ 
  \frac{1}{\sqrt{2}} (v + h + i z) 
\end{pmatrix}
  \;,\qquad
\eta = \frac{1}{\sqrt{2}} (x + s + i A)
  \;,
\end{align}
where
$v$ and $x$ are assumed to be real background fields and
we ignore more general possible vacua, corresponding to imaginary parts
(cf.\ eg.~\cite{Basler:2021kgq}).
At zero temperature,
the background field
$v$ is identified with the Higgs vacuum-expectation-value (VEV) and
a corresponding singlet VEV $x$ vanishes.
The thermodynamics can be extracted from the effective potential evaluated at its minima, and fig.~\ref{fig:multi_step}
schematically illustrates
the minima as a function of
the temperature $T$
for a multi-step transition.
We focus on locating regions of first-order phase transitions,
since we want to determine
regions of the model parameter space where both
dark matter and
a strong phase transition are simultaneously possible.

Determining the character of phase transition in perturbation theory
is not entirely reliable.
A classic example is the SM itself.
There a perturbative treatment
--
based on finding a discontinuity in the Higgs background field at the critical temperature
-- indicates a weakly first-order phase transition for
Higgs masses $\gtrsim 70$~GeV.
Non-perturbative studies~\cite{
  Kajantie:1996mn,Kajantie:1995kf,Kajantie:1996qd,Csikor:1998eu,Aoki:1999fi}
demonstrated that for such large Higgs masses, the SM has no thermal phase transition.
Instead a crossover takes place,
one where the system transitions smoothly from the symmetric to broken Higgs phase.

In the SM,
the potential barrier that separates
symmetric and Higgs phase
is generated radiatively by loop corrections and non-existent at tree-level.
The situation differs when analysing the complex singlet $\eta$ as
two-step transitions allow for a barrier already at tree-level due to
a non-vanishing background in singlet direction.
It is these two-step transitions%
\footnote{
  An analogous model of the SM augmented with a real triplet scalar
  demonstrated~\cite{Niemi:2020hto}, by non-perturbative lattice simulations,
  that such two-step transitions exist and are generally strong.
  In perturbation theory, the significant strength of
  two-step transitions can be understood to be due to a tree-level barrier.
}
that we investigate in this work.

We find the global minimum of the potential as a function of the temperature, and
determine the critical temperature from a condition that minima are discontinuous.
Regions of strong first-order phase transitions
can be identified by the condition
$v_{3,c}/\sqrt{T_{{\rm c},\phi}} \geq 1$, where
$v_{3} = v/\sqrt{T}$
relates the background fields of
the 3d EFT and
4d parent theory;
see eg.~\cite{Basler:2020nrq} for an application of this strategy.
Since this condition is gauge-dependent,
it lacks direct physical meaning --
we determine it in Landau gauge.
However, it can give an indicative estimate for
the order parameters of the phase transition.
A theoretically more robust analysis utilises
the discontinuity of scalar condensates~\cite{Farakos:1994xh} at
the critical temperature, that can be computed in
a gauge-invariant manner~\cite{Niemi:2020hto,Croon:2020cgk,Schicho:2022wty}.
Here,
we choose
a practical approach in terms of background fields \cite{Schicho:2022wty}, and
expect that locating FOPT regions is insensitive to this choice.%
\footnote{
  It was concluded in~\cite{Schicho:2022wty} that
  gauge-dependent results obtained in Landau gauge
  do not differ from fully gauge-invariant results
  within error bounds related to varying the RG scale.
  The latter quantifies
  missing higher order contributions.
}

\begin{figure}[t]
\centering
\includegraphics[width=0.7\textwidth]{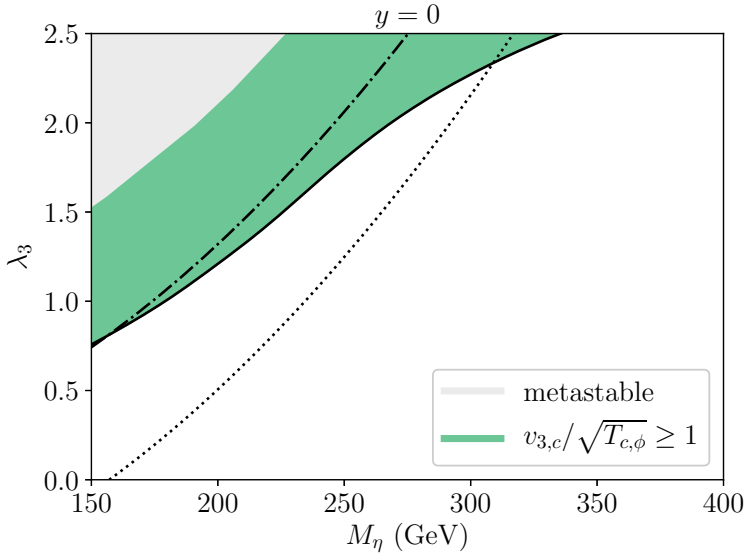}%
\caption{%
  Region of strong two-step phase transitions in the limit of
  decoupled dark matter ($y=0$) and
  fixed $\lambda_2=1.25$.
  The analysis uses a perturbative estimate via
  discontinuous background fields at
  the critical temperature $T_{{\rm c},\phi}$.
  We show
  $v_{3,{\rm c}}/\sqrt{T_{{\rm c},\phi}}=1$ as a solid black line and
  $v_{3,{\rm c}}/\sqrt{T_{{\rm c},\phi}}>1$ as a green band that increases towards
  the grey region.
  There the electroweak minimum of the zero-temperature tree-level potential is not global.
  The strongest transitions reside in a region of negative singlet mass parameter
  between the grey region and
  $\mu_\eta =0$ line (dash-dotted).
  The dotted contour
  $\mu_\eta = 0.5 \, \pi T$
  at fixed $T=100$~GeV
  is used to estimate the validity of the EFT construction, that assumes
  a light singlet.
  To the right of this line,
  such an assumption is compromised and
  $\mu_\eta^{ } \sim \pi T$.
}
\label{fig:EWPT-vanilla}
\end{figure}

Our result for a strong FOPT region is shown in fig.~\ref{fig:EWPT-vanilla}
in the case of a decoupled dark sector ($y=0$).
In this plot, we vary
the singlet mass and portal coupling for fixed
$\lambda_{2}=1.25$.
In the grey region (top left),
the minimum in Higgs direction is not global at zero temperature, and
the Higgs phase is metastable.
The green band depicts a FOPT region and
lies in a two-step transition regime, where above
$T_{{\rm c},\phi}$
the singlet has a non-zero background field ($x_3$) at the global minimum.
The dash-dotted line depicts a vanishing singlet mass parameter
$\mu_\eta = 0$, and
the FOPT region lies in the vicinity of that line.
The dotted line
($\mu_\eta = 0.5\,\pi T$,
$T=100$~GeV)
qualitatively confines the region of validity of the EFT construction.
To the right,
the singlet mass parameter becomes hard $\mu_\eta \sim \pi T$,
non-dynamical in the 3d EFT, and
should be integrated out along with the non-zero Matsubara modes.
From this observation, we cannot trust our result for
the upper right FOPT region,
and limit our study to
$\lambda_3 < 2.5$ and
$M_\eta < 400$~GeV.
For this qualitative guidance, we fixed
$T=100$~GeV which is a conservative lower bound for
$T_{{\rm c},\eta}$ that
we find in the FOPT region.
By varying $\lambda_2$,
the FOPT region moves moderately in
the $(M_\eta, \lambda_3)$-plane, but the general trend remains.
For fixed
$M_\eta$ and
$\lambda_3$, reducing
$\lambda_2$ weakens the transition.

\begin{figure}[t]
\centering
\includegraphics[width=0.7\textwidth]{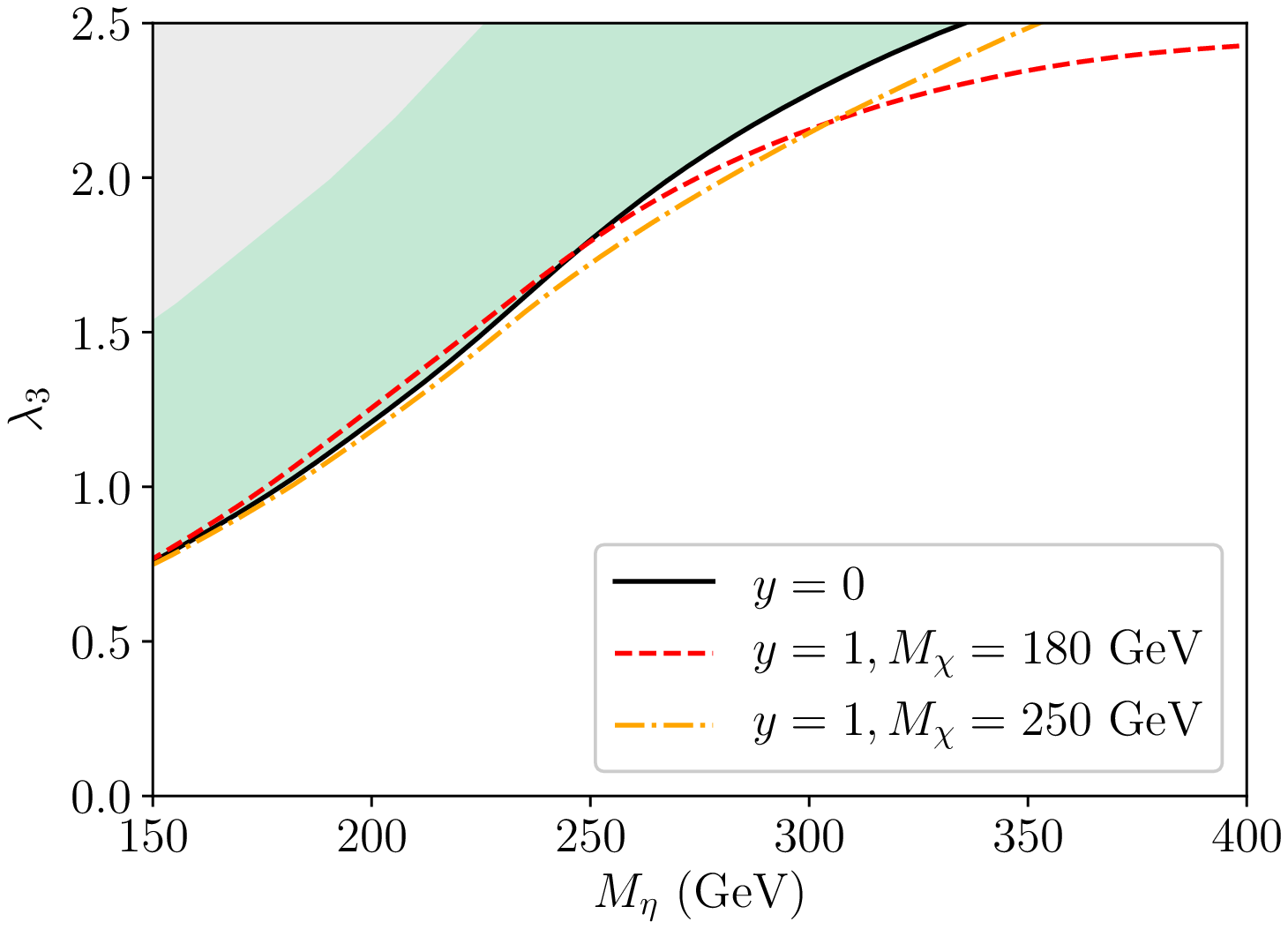}%
\caption{%
  Contours 
  $v_{3,{\rm c}}/\sqrt{T_{{\rm c},\phi}} = 1$ for
  non-vanishing $y$ and
  two different hypotheses for
  $M_\chi=180$~GeV (dashed) and
  $M_\chi=250$~GeV (dash-dotted) at
  fixed $\lambda_2=1.25$.
  The region above the shown contour lines,
  $v_{3,{\rm c}}/\sqrt{T_{{\rm c},\phi}} > 1$,
  corresponds to a strong phase transition.
  Only the band corresponding to $y=0$ is shaded.
}
\label{fig:EWPT-y-Mchi}
\end{figure}

While the parameters of the scalar potential
($\mu^2_\eta$,
$\lambda_3$,
$\lambda_2$)
dominate the phase structure,
the thermodynamic properties are also influenced by
the presence of a dark sector when $y \neq 0$.
This effect is illustrated
in fig.~\ref{fig:EWPT-y-Mchi} for
$y=1$ and two choices of
$M_\chi$.
The FOPT region stretches from
the dashed (dash-dotted) line to
the grey region, respectively
(only the $y=0$ FOPT band is shaded).
A non-zero $y$ mildly affects the region of a strong transition which is
qualitatively the same as at $y=0$ (solid line).
This is because the dark sector modifies
the dimensional reduction matching relations merely at NLO
(cf.\ eqs.~\eqref{eq:m:phi:2l} and \eqref{eq:m:eta:2l}), and
the transition is driven by the portal sector parameters.
However,
the different $y$ and $M_\chi$ contours behave non-trivially as a function of
$M_\eta$ and $\lambda_3$ and e.g.\
cross the solid black ($y=0$) line.
Here,
we refrain from further exploring the dependence of
the FOPT region on
$(y,M_\chi)$, but will inspect their role
in sec.~\ref{sec:interplay}, when
discussing the overlap with parameter regions that produce
the observed dark matter abundance.

Studying the nucleation of the new stable vacuum of the theory
is the next step in the analysis.
Here, we do not compute the bubble nucleation rate but address
some of the complications that can arise.
The full FOPT regions
in figs.~%
\ref{fig:EWPT-vanilla} and
\ref{fig:EWPT-y-Mchi} are
not viable for a cosmological electroweak phase transition, as
nucleation from the metastable singlet phase does not necessarily complete.
Since we do not compute the bubble nucleation rate, we are unable to
discriminate these unviable regions from those where nucleation completes.
These unviable regions can be expected to lie near the metastability region
at zero temperature.
Our purely perturbative analysis is unable to find one-step transitions, since non-perturbative studies are required to discriminate
first-order and crossover transitions.
However, in analogy to~\cite{
  Brauner:2016fla,Gould:2019qek,Niemi:2018asa,Andersen:2017ika,Gorda:2018hvi},
we could find these regions by integrating out
the singlet in regions of parameter space where it is hard or soft near
the critical temperature.
For the resulting SM-like 3d EFT,
the phase diagram is known non-perturbatively~\cite{Kajantie:1995kf}.
However, we do not perform such analyses here.
Due to its relevance for dark matter production, we determined
the typical critical temperature of the phase transition to be
$110~{\rm GeV} \lesssim T_{\rmi{c},\phi} \lesssim 155~{\rm GeV}$
inside the green band
in figs.~%
\ref{fig:EWPT-vanilla} and
\ref{fig:EWPT-y-Mchi}.

Another subtlety in this model is that
the nucleation temperature $\Tn$ could be significantly below
the critical temperature $\Tc$.
Such a potentially large degree of supercooling~\cite{Cline:1999wi}
could source
bubble walls that are extremely relativistic and
affect a successful EWBG~\cite{Bodeker:2017cim}.
Nucleation could also be affected by
the formation of topological defects.
One example are cosmic strings~\cite{Kajantie:1998zn}, which are generated due to the $\eta$-broken phase of
the Abelian ${\rm U(1)}_\rmii{Y}$ symmetry.
Investigating the resulting nucleation dynamics for this model is left for future work on
the subject.

%%%%%%%%%%%%%%%%%%%%%%%%%%%%% SUBSECTION %%%%%%%%%%%%%%%%%%%%%%%%%%%%%%%%%%%%
%
\section{Dark matter energy density}
\label{sec:dark_matter}

The dark matter model introduced in eq.~\eqref{eq:lag:4d}
can account for the observed energy density through different production mechanisms in the early universe.
The two main paradigms are
{\em freeze-out}~\cite{Gondolo:1990dk,Griest:1990kh} and
{\em freeze-in}~\cite{McDonald:2001vt,Hall:2009bx}.%
\footnote{
  Depending on the coupling strength of the actual DM particle,
  the setting can be more complicated
  via an intermediate regime between freeze-out and freeze-in, namely
  conversion-driven freeze-out~\cite{Garny:2017rxs}.
  This production mechanism is studied for the simplified model with
  a coloured~\cite{Garny:2017rxs} and
  a colourless mediator~\cite{Junius:2019dci}.
  Typical values of the Yukawa couplings are $y \approx 10^{-6}$.
}
The strategy
to distinguish between the two options is to inspect
the coupling of the actual DM particle, namely the Yukawa coupling $y$ in the portal interaction eq.~\eqref{eq:L:portal:yukawa}.
In this section,
we use the physical particle masses and indicate them with
$M_i$ for each species.

For $y$-values of the order of the electroweak SM gauge couplings,
the dark fermion $\chi$ and
anti-fermion $\bar{\chi}$ are
in thermal equilibrium in the early universe plasma.
DM particles follow an equilibrium abundance when the
temperature is larger than their mass,
that is also maintained when dark states enter a
non-relativistic regime.
Dark matter is mainly depleted by pair annihilations, which are very
efficient up until
$T/M_\chi \approx 1/25$.
Around this temperature, dark  particles decouple
and their comoving abundance is frozen ever since~\cite{Gondolo:1990dk,Griest:1990kh}.

For values
$y \lesssim \mathcal{O}(10^{-7})$~\cite{Enqvist:1992va,Hall:2009bx,Enqvist:2014zqa,Tenkanen:2016idg,Bernal:2017kxu},
dark matter particles
never reached thermal equilibrium due to their tiny coupling with the surrounding thermal bath.
This is
in contrast with the central assumption of the freeze-out mechanism.
The dark matter is feebly interacting and the production proceeds via freeze-in.
In this case, dark matter particles are generated through various processes, that comprise
decays of a heavier accompanying state in the dark sector, and
$2 \to 2$ scatterings that may involve SM particles.
Dark matter particles only appear in the final
state of the relevant reactions, and their abundance increases over the thermal
history up until the production rate is efficient.
Depending on the model parameters, in particular the mass splitting between
the Majorana fermion and the heavier scalar,
the super-WIMP mechanism contributes to a later-stage dark matter production through
the decays of the frozen-out
$\eta$ population~\cite{Garny:2018ali,Biondini:2020ric}.

This work focuses on the freeze-out scenario,
since we are also interested in shedding light on the effect of
a {\em sizeable} Yukawa coupling $y$ on the EWPT.
Yukawa coupling sizes of
$y \lesssim \mathcal{O}(10^{-7})$ required for freeze-in will
render Majorana fermions numerically irrelevant for
computing the phase transition thermodynamics.
A complementary production through freeze-in
is postponed for future work.
We describe the main differences with the freeze-out case and
corresponding challenges in deriving the DM energy density in
sec.~\ref{sec:freeze_in}.

%%%%%%%%%%%%%%%%%%%%%%%%%%%%% SUBSECTION %%%%%%%%%%%%%%%%%%%%%%%%%%%%%%%%%%%%
%
\subsection{Freeze-out}
\label{sec:freeze_out}

Three classes of processes are relevant for
pair annihilations in the present model.
One of them are pair annihilations of dark fermions.
In addition,
the presence of additional states that connect
the dark sector with the SM can significantly alter
the annihilation cross section~\cite{%
  Edsjo:1997bg,Griest:1990kh,Ellis:2014ipa,Garny:2015wea}.
Here, the role of the co-annihilating partner is played by the $\eta$ scalar.
Co-annihilations, i.e.\ processes with
$\eta$
and
$\chi$ as initial states,
and scalar-pair annihilations are typically relevant for
a relative mass splitting of
$\Delta M /M_{\chi} \sim 0.2$,
which is mildly model-dependent when assessed more precisely.
In general, small mass splittings correspond to a population for
$\eta$ as abundant as that of the dark fermion.
Therefore, the dynamics of the $\eta$ particles is as important as that of $\chi$.
Due to fast conversions as driven by the Yukawa coupling $y$,
the two populations are in thermal contact~\cite{Garny:2017rxs,Biondini:2019int}.
Conversely, for large mass splittings,
the equilibrium abundance of the scalar is suppressed by
$e^{-\Delta M /T}$ with respect to the lighter species,
without impacting the annihilation pattern.

Pair annihilations of the dark fermion give a $p$-wave leading contribution;
cf.\ eq.~\eqref{chi_chi_ann}.
This is a result of
the chiral suppression of
velocity-independent Majorana fermion pair annihilations
typical for
this model~\cite{Garny:2015wea,DeSimone:2016fbz}.%
\footnote{
  Chiral suppression occurs due to angular momentum conservation for
  Majorana fermions annihilating into two lighter fermions --
  here two SM right-handed leptons.
  Due to the small lepton masses, the cross section is suppressed by
  $(M_e/M_\chi)^2$, such that for
  $M_\chi \geq 100$~GeV it is rendered
  fairly smaller than the $p$-wave annihilation at typical freeze-out velocities 
  $\vrel^2 \approx 1/25$.
}
Hence, co-annihilation processes as well as pair-annihilations of singlet scalars,
which feature velocity-independent annihilation channels,
dictate the evolution of the dark matter energy density abundance for small mass splittings~\cite{Garny:2015wea,Biondini:2019int}.
This is even more pronounced when considering
for EWPT in sec.~\ref{sec:ewpt},
a non-vanishing Higgs portal coupling ($\lambda_3 \sim \mathcal{O}(1)$),
that opens up additional channels for
$\eta \eta^\dagger$ annihilations.
We also include the Higgs-portal coupling contribution to
the scalar mediator annihilations,
that are often neglected for the model at hand
(an exception is ref.\cite{Bollig:2021psb}).
We include them in the numerical extraction of the energy density, and list
the expression of the corresponding cross sections in the appendix~\ref{sec:app:DM}.

We assume that
$M_\chi \gtrsim 100$~GeV
when computing cross sections, and exploring the parameter space.
This is mainly motivated by the excluded regions reported
in ref.~\cite{Garny:2015wea} from
the LEP-SUSY working group~\cite{L3:2003fyi,ALEPH:2003acj,ALEPH:2001oot} and
further improvements performed at the LHC~\cite{ATLAS:2019gti,ATLAS:2019lff}.
Since in this model $M_\eta > M_\chi$, the in-vacuum masses of
the leptons are always negligible with respect to the dark matter particle and the scalar mediator.%
\footnote{
  We verified that thermal masses for the leptons of
  $\mathcal{O}(g T)$
  are negligible and smaller than 1~GeV at the freeze-out temperatures.
}
To affect the electroweak crossover,
the mass of the additional scalar cannot be too large
(cf.\ sec.~\ref{sec:ewpt})
and in this section
we consider
$M_\eta \lesssim 1$~TeV.
Once again, since
$M_\chi < M_\eta$, and the freeze-out occurs for
$M_\chi / T \sim 25$, the typical temperatures are below the electroweak transition
$T_{\rmi{c},\phi}$ as estimated in sec.~\ref{sec:ewpt}.
Therefore, we work in the broken electroweak theory where
the Higgs mechanism is active and responsible for
the mass generation of the SM fermions and gauge bosons.

Whenever the conversion processes between
the actual DM and the heavier co-annihilating partner are efficient during freeze-out, which is the case for typical $y$ values adopted here,
the effect of accompanying states can be captured by
a single Boltzmann equation~\cite{Gondolo:1990dk,Griest:1990kh}%
\footnote{
  Within $t$-channel models,
  inefficient conversion rates between the DM state and
  the co-annihilating species are addressed in~\cite{Garny:2017rxs}.
  Typical values for the loss of thermal equilibrium for
  the $\chi \leftrightarrow \eta$ conversion are $y \sim 10^{-6}$.
}
\begin{equation}
\label{BE_gen}
\frac{dn}{dt} + 3 Hn =
    -\langle \sigma_{{\rmi{eff}}} \,\vrel \rangle
    (n^2-n^2_{{\rmi{eq}}})
  \;,
\end{equation}
where the left side is
the covariant time derivative in an expanding background,
$H$ is the Hubble rate of the expanding universe,
$\vrel$ is the relative velocity of the annihilating pair, and
$n$ denotes the overall number density of both states
$\chi$ and
$\eta$.
Then, the total equilibrium number density, which accounts for both the particle species
$\chi$ and
$\eta$, is
\begin{equation}
\label{n_eq}
  n_{{\rm{eq}}} = \int_{\vec{p}} e^{-E_{p,\chi}/T}
  \Bigl[ g_{\chi} + g_{\eta} e^{-\Delta M_{\T}/T} \Bigr]
  \;,\quad
  E_{p,\chi} =
      M_\chi
    + \frac{\vec{p}^2}{2 M_\chi}
  \;,
\end{equation}
where
the internal degrees of freedom are
$g_\chi=2$ for the fermion (2 spin polarisations, Majorana fermion) and
$g_\eta=2$ for the scalar (particle and antiparticle, complex scalar singlet).
The mass difference $\Delta M_{\T}$ gets the vacuum contribution,
$\Delta M = M_\eta - M_\chi$, and
a thermal correction in the non-relativistic limit
(see also refs.~\cite{Biondini:2017ufr,Biondini:2018pwp,Biondini:2020ric})
\begin{align}
 \Delta M_{\T} = 
    \Delta M 
  &+ \frac{\lambda_3}{M_\eta} \int_{\vec{p}}\frac{ \nB(E_{p,\phi})}{E_{p,\phi}} 
  + \frac{\go^2 Y_\eta^2}{4 M_\eta} \int_{\vec{p}} \left(
      \frac{\tilde{c}^2 \, \nB(E_{p,\gamma})}{E_{p,\gamma}} 
    + \frac{\tilde{s}^2 \, \nB(E_{p,\rmii{$Z$}})}{E_{p,\rmii{$Z$}}}
    \right)
  \nn &
  - \frac{\go^2 Y_\eta^2}{32 \pi} \left(
        \tilde{c}^2 M_{\T,\gamma}
      + \tilde{s}^2 M_{\T,\rmii{$Z$}}
      - s^2 \MZ
    \right) 
    \;,
\end{align}
where
$E_{p,i} = \sqrt{\vec{p}^2 + M_{\T,i}^2}$.
The thermal contributions come from the gauge bosons and Higgs tadpoles, whereas the latter arises from the contribution
of screened soft gauge bosons at the scale $gT$
(also known as Salpeter correction~\cite{salpeter}).
The relevant thermal masses $M_{\T,i}$ and weak mixing angles are listed
in eqs.~\eqref{eq:mphi:T},
\eqref{eq:mZgamma:T},
\eqref{Weinberg_T0} and
\eqref{Weinberg_T}.
The integral measure is defined as
$\int_{\vec{p}} \equiv \int{\rm d}^3 \vec{p}/(2\pi)^3$, and
the effective thermally averaged annihilation cross section reads~\cite{Griest:1990kh}
 \begin{equation}
\langle \sigma_{{\rm{eff}}} \vrel \rangle =
   \sum_{i,j} \frac{n^{\rmi{eq}}_{i}\,n^{\rmi{eq}}_{j}}{(\sum_k n_k^{\rmi{eq}})^2}
   \langle \sigma_{ij} \vrel \rangle
   \;.
\label{co_cross}
\end{equation}
Here $\langle \sigma_{{\rm{eff}}} \vrel \rangle$
includes all combinations for the annihilating pairs, namely
$\chi \chi$,
$\chi \eta$,
$\eta \eta^\dagger$,
$\eta \eta$ and their conjugates when relevant.
Conventionally, it is calculated by thermally averaging the in-vacuum cross sections over
the centre-of-mass energies in the thermal environment~\cite{Gondolo:1990dk,Griest:1990kh}.
The in-vacuum annihilation cross sections in the broken phase of
the electroweak symmetry can be found in the appendix~\ref{sec:app:DM}.
The diagrams for the Majorana fermion (co-)annihilations, and
a representative process for pair-annihilation are collected
in fig.~\ref{fig:xx_and_xeta}.
\begin{figure}[t]
\centering
\begin{align}
  \mathcal{M}_{\chi\chi\to e\bar{e}} &=
      \VtxvT(\Lqu,\Lqum,\Lqum,\Luq,\Lscii)
    + \VtxvU(\Lqu,\Lqum,\Lqum,\Luq,\Lscii)
  \;, \nn[2mm]
  \mathcal{M}_{\chi\eta^\dagger \to e Z(\gamma)} &=
    \VtxvT(\Lgli,\Lscii,\Lqum,\Luq,\Lscii)
  + \VtxvS(\Lgli,\Lscii,\Lqum,\Luq,\Lqu)
\;, \nn[2mm]
   \mathcal{M}_{\eta\eta^\dagger \to \gamma\gamma} &=
  \Vtxvo(\Lgli,\Lscii,\Lcsii,\Lgli)
+ \VtxvT(\Lgli,\Lscii,\Lcsii,\Lgli,\Lscii)
+ \VtxvU(\Lgli,\Lscii,\Lcsii,\Lgli,\Lscii)
\; . \nonumber
\end{align}
\caption{%
  Diagrams for the dark matter pair annihilation and co-annihilation with
  the singlet scalar.
  The dark fermion $\chi$ is displayed by a double-solid line,
  the complex scalar $\eta$ by an arrowed double-dashed line,
  the SM lepton $e$ by an arrowed solid line, and
  photons and $Z$-bosons by wiggly lines.
}
\label{fig:xx_and_xeta}
\end{figure}
The impact of the processes
$\chi \eta$ and
$\eta \eta^\dagger$ is controlled by the functional dependence of
the equilibrium number densities in eq.~\eqref{co_cross}, that gives
$
  \langle \sigma_{{\rmi{eff}}} \vrel \rangle \approx
  \langle \sigma_{\chi \chi} \vrel \rangle
+ \langle \sigma_{\chi \eta} \vrel \rangle e^{-\Delta M_\T /T}
+ \langle \sigma_{\eta \eta^\dagger} \vrel \rangle e^{-2\Delta M_\T /T}$.
This manifests that a large ration of mass splittings over temperature
(at freeze-out and later stages) suppresses
the importance of the co-annihilations.

The Boltzmann equation~\eqref{BE_gen} is then as usual recast in terms of
the yield parameter $Y = n/s$, where
$s=2 \pi^2 h_{\rmi{eff}} \, T^3/45$ is the entropy density, and
the time evolution is traded for the variable $z = M_\chi/T$.
As for the temperature-dependent relativistic degrees of freedom
$h_{\rmi{eff}}$ entering the entropy density, we use the SM values from
ref.~\cite{Laine:2015kra}.%
\footnote{
  In the freeze-out case this is well justified since
  the states $\chi$ and $\eta$ are heavy and non-relativistic particles for
  the relevant temperature window.
}
The relativistic degrees of freedom for the energy density $g_{\rmi{eff}}$, that  enter
the Hubble rate $H=\sqrt{8\pi e/3}\,\Mpl$, where
$e=\pi^2 T^4  g_{\rmi{eff}}/30$, are also taken from~\cite{Laine:2015kra}. 

%%%%%%%%%%%%%%%%%%%%%%%%%%%%% SUBSUBSECTION %%%%%%%%%%%%%%%%%%%%%%%%%%%%%%%%%%%%
%
\subsubsection{$\eta \eta^\dagger$ annihilations}

An additional discussion is in order for the $\eta \eta^\dagger$ annihilations.
The scalar particle $\eta$ interacts with
the ${\rm U(1)}_\rmii{Y}$ gauge boson $B_\mu$ and
the Higgs boson.
The former interaction can also be understood in terms of the mass-diagonal fields i.e.\
the $Z$-boson and photon.
Since the pair-annihilations happen in a non-relativistic regime,
the scalar (anti-)particles
are heavy and slowly moving in the thermal plasma.
In this setting, repeated soft exchanges of the force carriers (vector or scalar particles)
can significantly alter the annihilation cross section of the incoming $\eta$ pair.
Two effects play an important role:
the Sommerfeld enhancement~\cite{Hisano:2004ds,Feng:2010zp,Iengo:2009ni} and
bound-state formation~\cite{Detmold:2014qqa,vonHarling:2014kha}.
Their main phenomenological consequence is that the annihilations are boosted for small velocities. Moreover,  whenever bound states are formed and not effectively dissociated
or melted away in the thermal plasma,
they provide an additional process for the depletion
of DM particles in the early universe.

The annihilation rate for the accompanying scalar contributes to
the overall cross section in eq.~\eqref{co_cross}
in the co-annihilation regime.
In turn,
the overall cross section
enters the extraction of the dark matter energy density.
\begin{table}
\centering
\begin{tabular}{c c c c l l }
\hline
\hline
  boson
  & \phantom{xx} & $\mathcal{V}(r)$  &
  \phantom{xx}
  & \multicolumn{1}{c}{coupling}
  \\
  \hline
  \\
  photon
  & \phantom{xx} & 
  $ \mathcal{V}_\gamma(r) = -\frac{\alpha_{\gamma}}{r} \, e^{-r M_{\T,\gamma} }$
  & \phantom{xx}
  & $\alpha_\gamma = \alpha_{\rmi{em}} \frac{Y_\eta^2}{4} \left( \frac{\tilde{c}_w}{c_w} \right)^2$ 
  \\[2mm]
  \hline
  & & & & & \\
  $Z$-boson
  & \phantom{xx}
  & $\mathcal{V}_{\rmii{$Z$}}(r) = -\frac{\alpha_Z }{r} e^{-r M_{\T,\rmii{$Z$}} }$
  & \phantom{xx}
  & $\alpha_{\rmii{$Z$}} = \alpha_{\rmi{em}} \frac{Y_\eta^2}{4} \left( \frac{\tilde{c}_w}{c_w} \right)^2 \tan^2(\tilde{\theta}_w) $
  \\[2mm]
  \hline
  & & & & & \\
  $H$-boson
  & \phantom{xx}
  & $\mathcal{V}_\phi (r) = -\frac{\alpha_\phi}{r} e^{-r M_{\T,\phi} }$
  & \phantom{xx}
  & $\alpha_\phi = \frac{\lambda_3^2 v_\T^2}{16\pi M_\eta}$ 
  \\[2mm]
  \hline
  \hline
\end{tabular}
\caption{%
  Attractive potentials and fine structure constants for
  the ${\rm U(1)}_\rmii{Y}$-charged scalar $\eta$ from
  the different force mediators.
  We abbreviate the Weinberg angle 
  at $T=0$ with
  $c \equiv \cos (\theta_w)$ in eq.~\eqref{Weinberg_T0} and
  at finite-temperature with
  $\tilde{c} \equiv \cos (\tilde{\theta}_w)$ in eq.~\eqref{Weinberg_T};
  $v_\T$ is the temperature-dependent Higgs VEV in eq.~\eqref{eq:mphi:T}.
  Here,
  $e=\go\cos(\theta_w)$, with
  $\alpha_{\rmi{em}}=e^2/(4 \pi)$.
}
\label{table:potentials}
\end{table}
For our case, the scalar annihilations can be affected by the
photon,
$Z$-boson an
Higgs boson induced potentials, collected
in tab.~\ref{table:potentials}.
In the broken phase of the electroweak symmetry,
and in a thermal environment of the early universe, some quantities become
temperature-dependent.
By following phenomenological prescriptions~\cite{Ghiglieri:2016xye,Kim:2016kxt},
one finds that
\begin{itemize}
  \item[(i)]
    the Higgs VEV $v$ and
    the Higgs mass depend on the temperature;
  \item[(ii)]
    the temporal part of the gauge bosons acquire thermal masses, {\em also} the photon;
  \item[(iii)]
    the weak mixing angle $\theta_{w}$,
    or Weinberg angle, evolves with the temperature.
\end{itemize}
We include these effects in the static potentials when estimating the Sommerfeld factors.
In appendix~\ref{sec:app_DM_2}
we collect the relevant definitions,
whereas we refer to~\cite{Ghiglieri:2016xye} for a more detailed discussion.

\begin{figure}[t]
\centering
\includegraphics[width=0.49\textwidth]{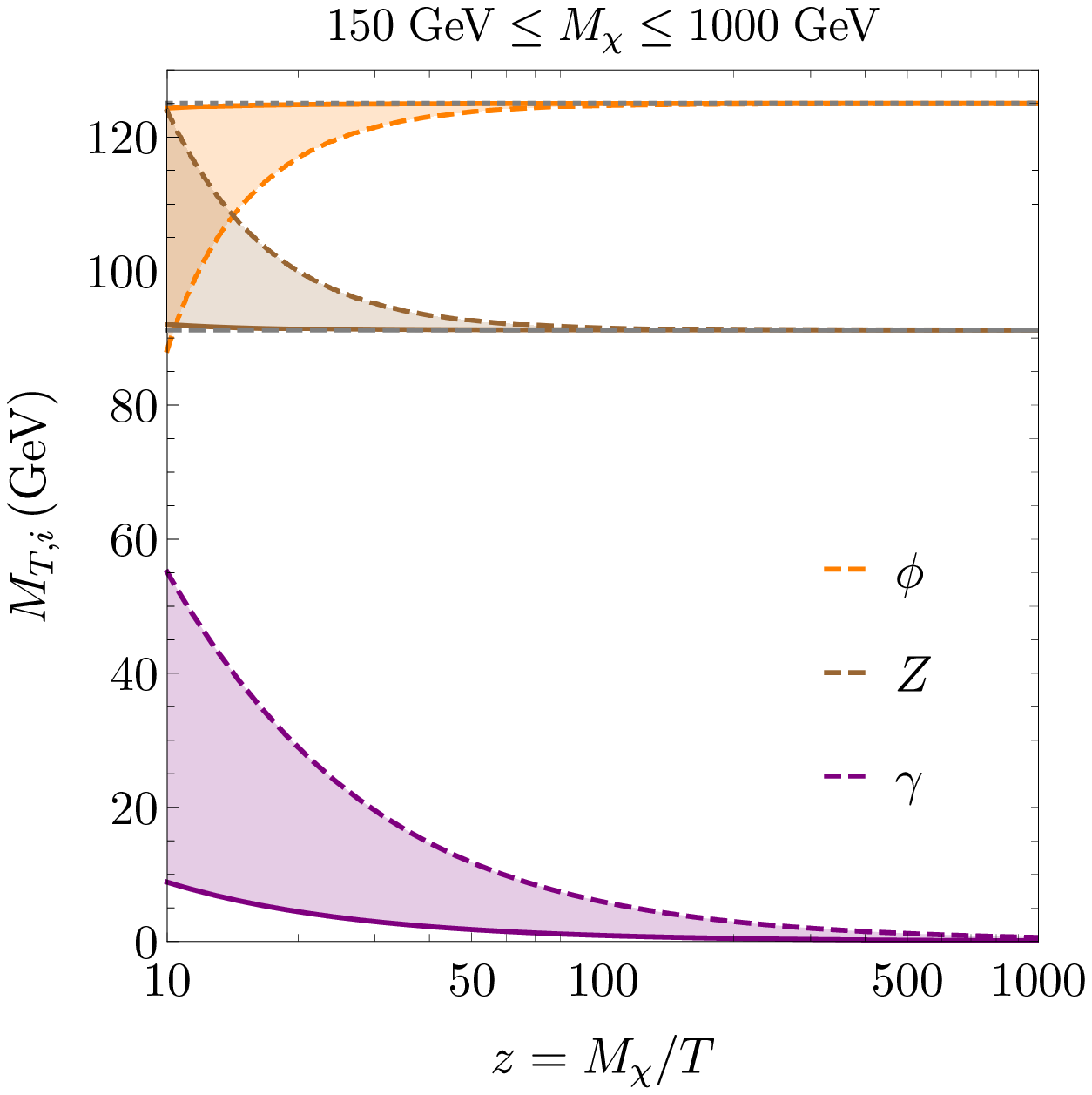}%
\hfill
\includegraphics[width=0.495\textwidth]{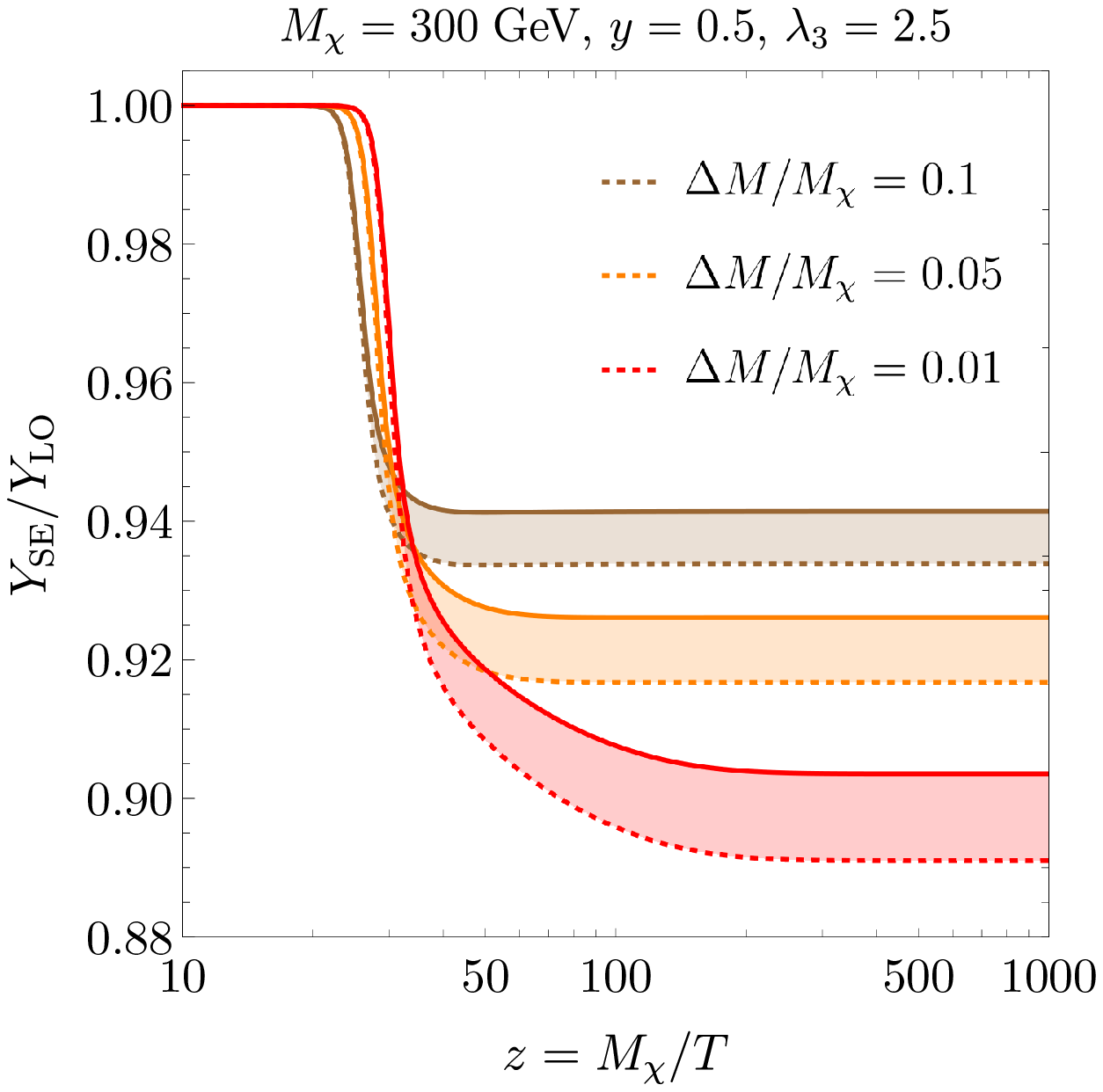}
\caption{%
  Left: Thermal masses for the gauge and Higgs bosons and their in-vacuum counterparts (grey lines).
  Right: Ratio of the DM yields with and without the Sommerfeld enhancement for
  the co-annihilating scalar particle.
  The bands show the impact of a Yukawa-like potential when
  a thermal mass of the photons is included (solid lines).
}
\label{fig:thermal_masses}
\end{figure}

We follow the approach of ref.~\cite{Iengo:2009ni}
to numerically extract the Sommerfeld enhancement due to the exchange
of a Higgs boson, photon, and $Z$-boson --
both for the $s$- and $p$-wave annihilations.
We also include temperature-dependent masses for the force carriers
that are indicated as
$M_{\T,\phi}$ in eq.~\eqref{eq:mphi:T}
and
$M_{\T,\gamma}$,
$M_{\T,\rmii{$Z$}}$ in eq.~\eqref{eq:mZgamma:T}, and
reproduce the values
$M_\phi=125.1$~GeV,
$M_\gamma=0$, and
$\MZ=91.2$~GeV
for
$T\to 0$.
In our treatment, also the photon induces a Yukawa-like potential, rather than a Coulomb potential, for sufficiently high temperatures corresponding to non-negligible thermal masses.
The temperature-dependent masses are shown
in fig.~\ref{fig:thermal_masses} (left).
Here we consider
the dark matter mass to vary between
$150~{\rm GeV} \leq M_\chi \leq 10^3~{\rm GeV}$,
and the solid (dashed) lines stand for the smallest (largest) masses and temperatures.
The grey dotted lines indicate the in-vacuum masses for the Higgs boson and
$Z$-boson.
Thermal masses differ by 5\% of the corresponding in-vacuum values at
the chemical freeze-out temperature
$T \simeq M_\chi/25$.
For the photon, a thermal mass can potentially be more relevant,
since at $T=0$ the photon is massless.
However, we find the Sommerfeld factor corresponding to a Yukawa-like potential, with
a finite photon mass, to be smaller by a few-per-cent than
the Sommerfeld factor computed in the Coulomb limit.
We crosscheck the numerical extraction of the Sommerfeld factors with
the formalism of~\cite{Kim:2016kxt}, that relies on determining
the spectral function of
the $\eta$-$\eta^\dagger$ pair, and
obtain compatible results.

The ratio between the dark matter yield
$Y_{\rmii{SE}}$($Y_{\rmii{LO}}$)
with(out) the Sommerfeld enhancement from the $\gamma$, $Z$ and Higgs boson exchange
is shown in fig.~\ref{fig:thermal_masses} (right).
The dominant Sommerfeld enhancement is induced by
the photon exchange in the mass range relevant for our work, namely
$M_\chi, M_\eta \lesssim 1$~TeV.
To show the effect of a finite (thermal) photon mass, we show the DM yields as obtained with the Sommerfeld factors in the Coulomb limit (dashed lines) and for the Yukawa-like potential (solid lines).
The difference between the two cases is small, with an effect of $\sim 1\%$ on
the yield ratio.
As expected, the screened potential corresponds to less effective annihilations and
solid lines exceed the dashed ones.
Our assessment shows that the $T=0$ treatment is well suited for
the relic density extraction in this model in the freeze-out scenario.
One can safely use the $T \to 0$ limit of the potentials
(see e.g.\ ref.~\cite{Bollig:2021psb} for the same model).
We
fixed the dark matter mass to
$M_\chi=300$~GeV,
considered three relative mass splittings
$\Delta M/M_\chi \in \{0.1,0.05,0.01\}$, and
fixed the Higgs-portal coupling to $\lambda_3=2.5$.
A non-vanishing scalar portal coupling induces many processes that add up to
the scalar-pair annihilations;
see cross sections in appendix~\ref{sec:app:DM}.
In agreement with the co-annihilation scenario~\cite{Edsjo:1997bg,Ellis:2014ipa,Garny:2015wea},
the smaller the splitting the larger the effect of the
$\eta$-$\eta^\dagger$ annihilations.
For this choice of the dark matter mass and the Yukawa coupling $y$,
the Sommerfeld enhancement reduces the dark matter abundance by
6--11\% depending on the mass splitting.
We find that varying the DM mass
$M_\chi \in [10^2, 10^3]$~GeV and
$y \in [0.1,2]$ changes
$Y_{\rmii{SE}}/Y_{\rmii{LO}}$ only slightly, whereas
decreasing $\lambda_3$ impacts the yield ratio stronger
(with a less important Sommerfeld enhancement).

The situation is more intricate for the bound-state effects.
First, the potentials are Yukawa-like and it is not obvious if they are sufficiently
long-ranged with respect to the energy scales at play.
A reasonable estimate can
be obtained by demanding that the screening length is not smaller than
the typical bound-state size, i.e.\
$1/M_X \gtrsim a_0$, where the Bohr radius is
$a_0=2/(M_\eta \alpha_X)$ with
$X=\gamma, Z, \phi$, and the corresponding fine structure constant is
$\alpha_X$.
More precisely, one can exploit the numerical evaluation that demands
for a ground state to form, that reads
$1/M_X \geq 0.84 a_0$~\cite{Rogers:1970xx}.
Then, for the $Z$-boson induced bound states, we obtain a lower bound
$M_\eta \gtrsim 70$~TeV, which is higher than the masses we are interested in
(our estimate compares well with the one detailed recently in ref.~\cite{Bollig:2021psb}).
For the Higgs boson, we find that even for the largest coupling that we allowed, 
$\lambda_3 =3.0$, the bound gives
$M_\eta \gtrsim 5$~TeV.
Therefore, only the photon exchange can sustain the formation and existence of bound states
in the relevant mass and temperature range.

The annihilation of a $\eta$-$\eta^\dagger$ pairs as bound states can efficiently occur
only if they are not dissociated by the interactions with the medium constituents.
Owing to some similarities with heavy quarkonium in medium, where
the bound-state dynamics is driven by gluo-dissociation~\cite{Kharzeev:1994pz} and
the dissociation by inelastic parton scattering~\cite{Grandchamp:2001pf},
the model at hand features similar processes involving photons.
See refs.~\cite{Petraki:2015hla,Biondini:2017ufr,Binder:2020efn,Biondini:2021ycj} for applications to dark matter.
The first dissociation process entails  a thermal photon hitting the
$\eta$-$\eta^\dagger$ pair
in a bound-state and, if sufficient energy is available,
breaking it into an unbound above-threshold pair.
The second dissociation process comes as a $2 \to 2$ scattering reaction, where
the particles in the thermal bath transfer energy/momenta to the heavy
$\eta$-$\eta^\dagger$ pair through a photon exchange,
turning a bound-state into an unbound one.
Since we have checked that a thermal mass for the photon gives
a practically negligible effect on the DM abundance when including it the Sommerfeld enhancement
(see right panel in fig.~\ref{fig:thermal_masses}),
we remain in the Coulombic regime and treat the photon as massless.
This approximation enables us to adapt former derivations for
the bound-state formation cross section~\cite{Petraki:2015hla}, from which
the dissociation rate $\Gamma_{\rmi{bsd}}$ can be obtained via
the Milne relation~\cite{LL};%
\footnote{
  Results in~\cite{Petraki:2015hla} can be used by
  changing accordingly the coupling between
  scalar $\eta$ and photon, $g \to \go c_w$.
}
see appendix~\ref{sec:app:DM} for details on the rates and
Boltzmann equation for bound states.
We find that the bound-state effects are fairly small for
$M_\eta \lesssim 1$~TeV and an electroweak coupling strength
$\alpha_\gamma \sim \mathcal{O}(10^{-2})$.
In fact, we obtain a correction of 1--2\% with respect to
the yield where
the Sommerfeld enhancement is already accounted for.
This is consistent with
the analysis in ref.~\cite{vonHarling:2014kha}, and confirmed by
the studies~\cite{Petraki:2015hla,Biondini:2017ufr,Binder:2020efn}.

%%%%%%%%%%%%%%%%%%%%%%%%%%%%% SUBSUBSECTION %%%%%%%%%%%%%%%%%%%%%%%%%%%%%%%%%%%%
%
\subsubsection{Numerical results and parameter space}

The results for the model parameter space compatible with
the observed dark matter energy density
$\Omega_{\rmii{DM}} h^2 \big \vert_{\rmi{obs.}} = 0.1200 \pm 0.0012$~\cite{Planck:2018nkj}
are shown in this section.
The predicted DM energy density depends on the model parameters, namely
$\left\lbrace M_\chi, M_\eta, y, \lambda_3, \lambda_2 \right\rbrace$.
At the order we are working, the
$\eta$ self-coupling
$\lambda_2$ enters the extraction of the DM energy density only through
the renormalisation group equations (RGE), that
provide the values of all couplings at a given energy scale;%
\footnote{
  The specific value of $\lambda_2$ is more important for the EWPT instead.
}
see eqs.~\eqref{eq:rge:g1}--\eqref{eq:rge:y}.
We consider
different and complementary ways to visualise the parameter space compatible with
$\Omega_{\rmii{DM}} h^2 \big \vert_{\rmi{obs.}}$.

\begin{figure}[t!]
\centering
\includegraphics[width=0.47\textwidth]{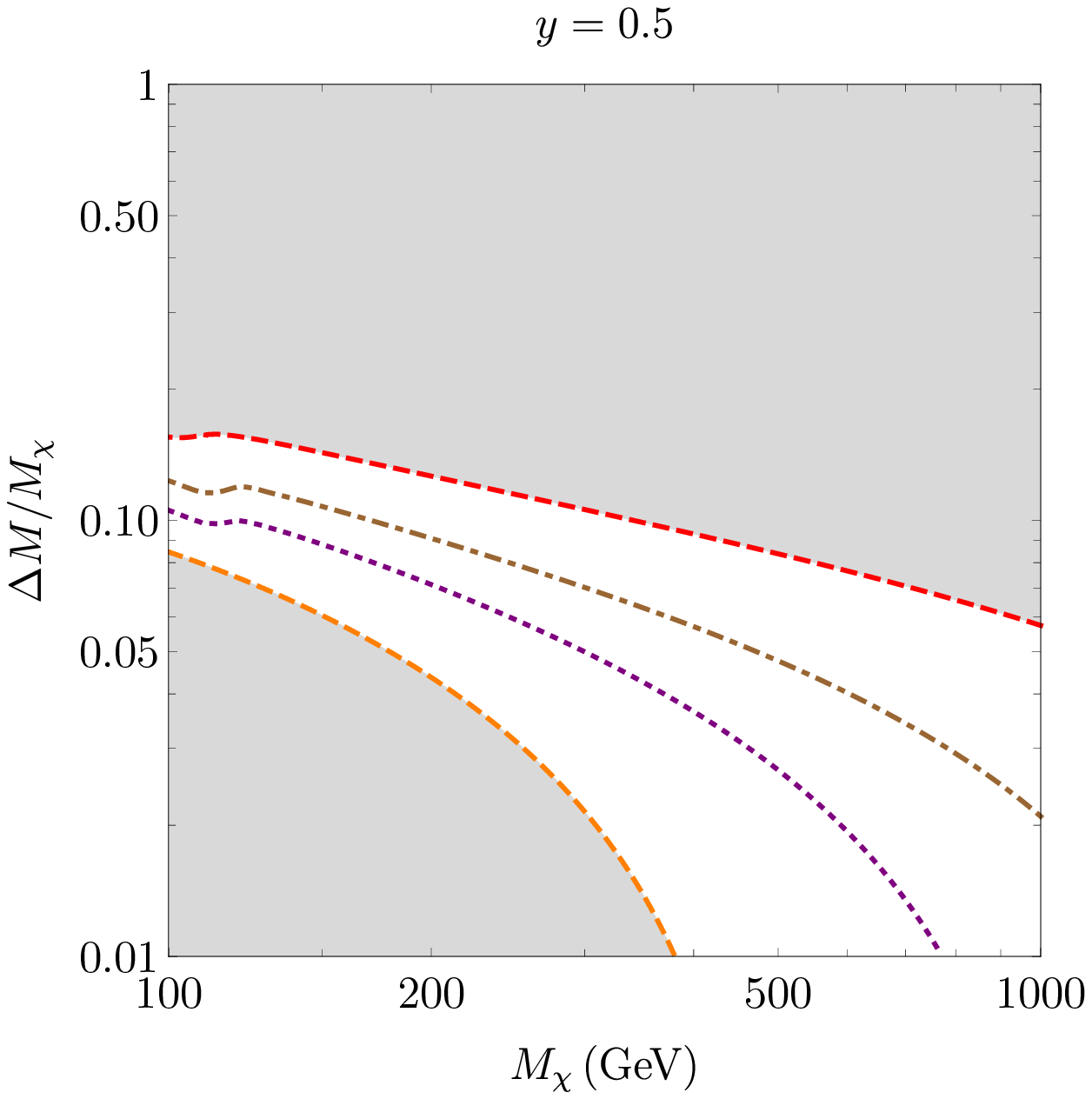}%
\hspace{0.3 cm}
\includegraphics[width=0.47\textwidth]{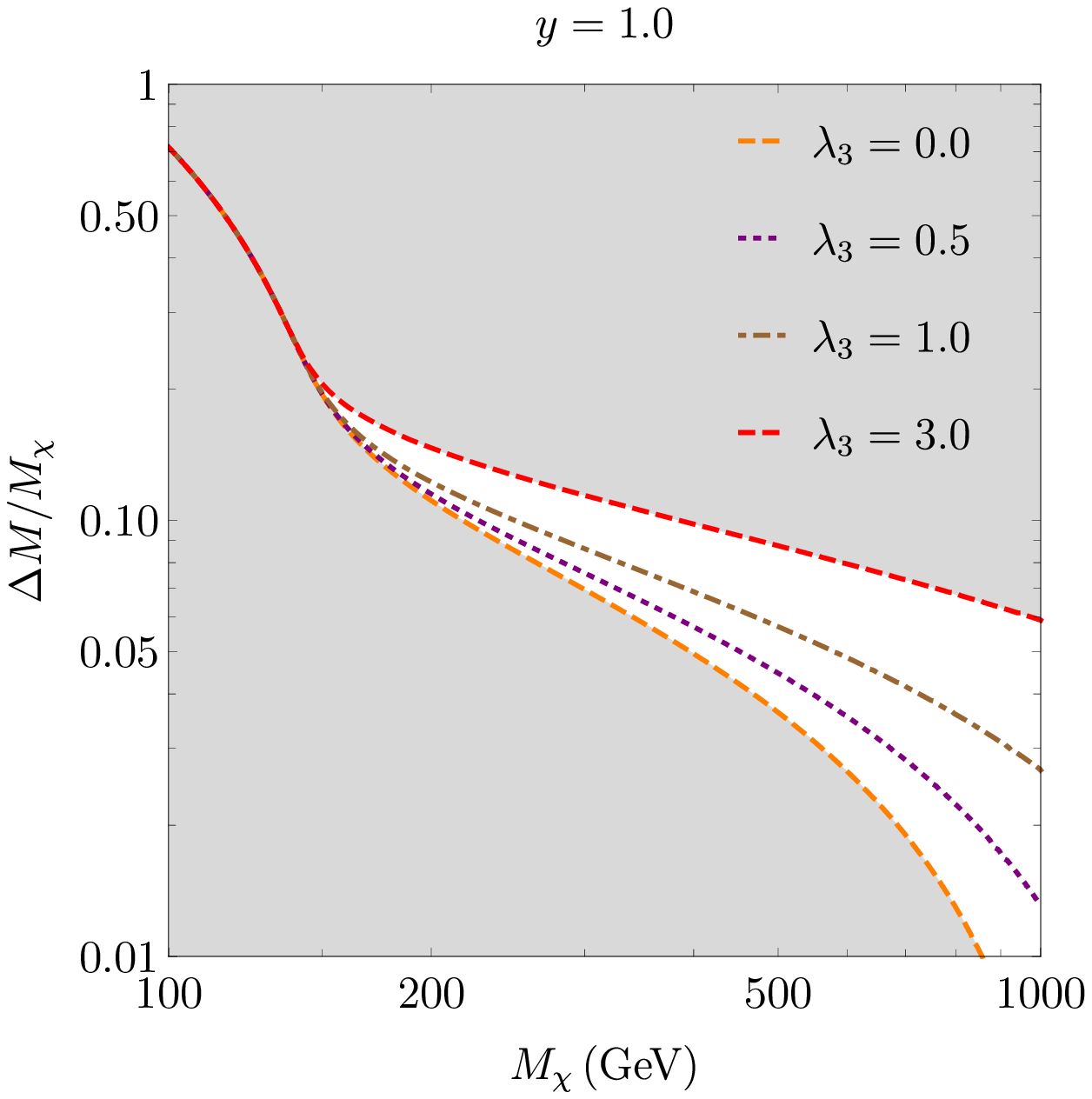}
\caption{%
  Parameter space in the $(M_\chi, \Delta M /M_\chi)$-plane compatible with
  the observed DM energy density as
  white area comprised between the
  orange~dashed ($\lambda_3=0.0$) and
  red~dashed ($\lambda_3=3.0$) curves.
}
\label{fig:param_DM_1}
\end{figure}
A first visualisation of the parameter space as
in fig.~\ref{fig:param_DM_1} was already adopted for
the family of simplified models to which our model Lagrangian belongs,
see e.g.~\cite{Garny:2015wea,Biondini:2018ovz,Biondini:2019int}.
To obtain the curves shown in fig.~\ref{fig:param_DM_1}, we fix $y$ and $\lambda_2$, we trade $M_\eta$ with the relative mass splitting, and take
$(M_\chi, \Delta M / M_\chi)$ as free parameters.
We consider different values for the portal coupling $\lambda_3$.
The effect of non-vanishing $\lambda_3$ allows for larger mass splittings
since the overall cross section~\eqref{co_cross} is larger due to many additional annihilation processes enabled by the coupling $\lambda_3$.
Moreover, co-annihilations are more relevant for
smaller Yukawa couplings $y$
(compare panels in fig.~\ref{fig:param_DM_1}).
This effect arises for smaller dark matter masses and
traces back to the relative importance of
the various contributions to the cross section.
The smaller $y$ the larger the relative importance of
the pair annihilation of scalar pairs, that features
$y$-independent annihilation channels.

\begin{figure}[t]
\centering
\includegraphics[width=0.465\textwidth]{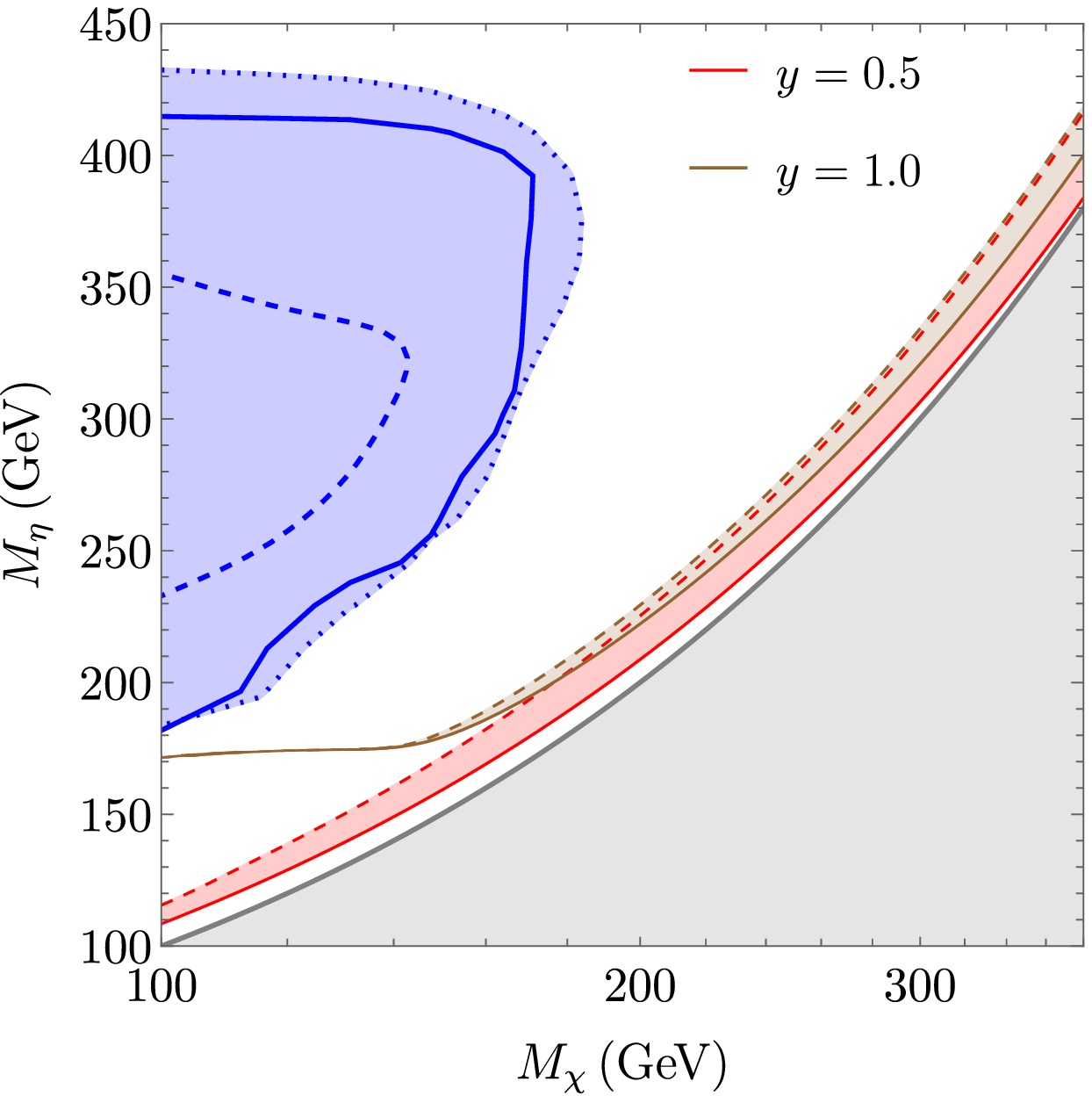}%
\hfill
\includegraphics[width=0.48\textwidth]{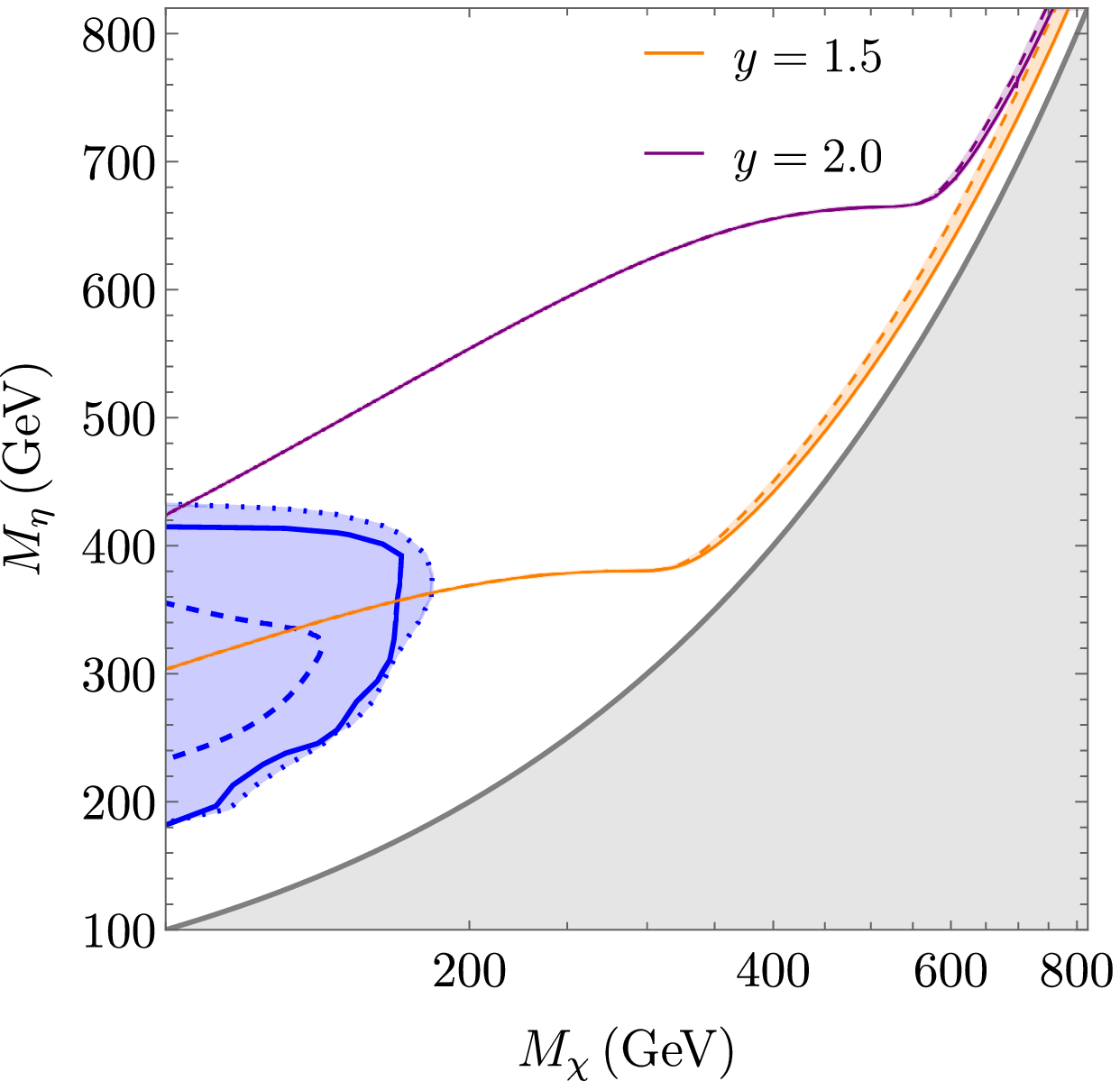}
\caption{%
  Coloured bands reproduce the observed energy density in the parameter space
  $(M_\chi,M_\eta)$; at
  $y=0.5$ and $y=1.0$ (left), as well as
  $y=1.5$ and $y=2.0$ (right).
  The solid (dashed) line corresponds to
  $\lambda_3=0.0$ ($\lambda_3=3.0$).
  The shaded grey area corresponds to $M_\chi< M_\eta$, which is
  not a viable option for the model.
  The collider-excluded regions for the different lepton flavours are
  the shaded blue areas from the ATLAS searches~\cite{ATLAS:2019gti,ATLAS:2019lff}
  for the
  electron (solid),
  muon (dotted), and
  tau (dashed).
  For Majorana fermion masses $M_\chi > 140 \; (180)$~GeV
  when coupling to taus (muons),
  experimental limits are absent for $M_\eta$.
  }
\label{fig:param_DM_2}
\end{figure}
As a second option, we visualise
the curves reproducing the observed DM energy density
in the $(M_\chi,M_\eta)$-plane
in fig.~\ref{fig:param_DM_2}.
To ensure the stability of the dark fermion,
the grey area $M_\chi > M_\eta$ is not allowed in the model.
The left panel includes two choices for the Yukawa coupling
$y=0.5$ and
$y=1.0$
with corresponding bands
$\lambda_3 \in [0.0,3.0]$
($\lambda_3=0$ for solid lines,
$\lambda_3=3.0$ for dashed lines).
The larger effect of
co-annihilations and
$\eta$-pair annihilations for a smaller Yukawa coupling $y$ is also visible and
results in a wider
$y=0.5$ band (red) than the
$y=1.0$ one (brown).
The right panel shows corresponding bands that reproduce
$\Omega_{\rmii{DM}} h^2 \big \vert_{\rmi{obs.}}$
at larger
$y=1.5,2.0$.
The sharp transition to the co-annihilation strip is visible as
the transition from a line to a band, and is delayed to
larger $M_\chi$ for
larger $y$.
We included the ATLAS exclusion limit at 95\% confidence level for
right-handed slepton searches at the LHC, in particular
for the stau (dashed)~\cite{ATLAS:2019gti} and
for the smuon (dotted) and selectron (solid)~\cite{ATLAS:2019lff}
as excluded shaded~blue regions in the parameter space.
These constraints also apply to the simplified model since
the experimental limits are obtained from the Drell-Yan production of the scalars, that
decay promptly into a dark fermion and a lepton.%
\footnote{
  As long as $y$ is not very small, its precise value is irrelevant since the decay process
  $\eta\to\chi \, e$ is a prompt decay anyway.
  Tiny couplings, compatible with a freeze-in production mechanism, would instead
  give different signatures involving long-lived particles
  (see e.g.~\cite{CMS:2013czn,ATLAS:2017tny}).
  Moreover, the production of scalar pairs from an off-shell Higgs, as induced by
  the portal coupling $\lambda_3 \neq 0$ and not considered in
  the experimental analyses~\cite{ATLAS:2019gti,ATLAS:2019lff}, would account for
  a small correction (few per cent) to
  the Drell-Yan processes at the current LHC energies and
  slepton masses {\em viz.}\ $M_\eta$;
  see ref.~\cite{Hessler:2014ssa}.
}

\begin{figure}[t]
\centering
\includegraphics[width=0.48\textwidth]{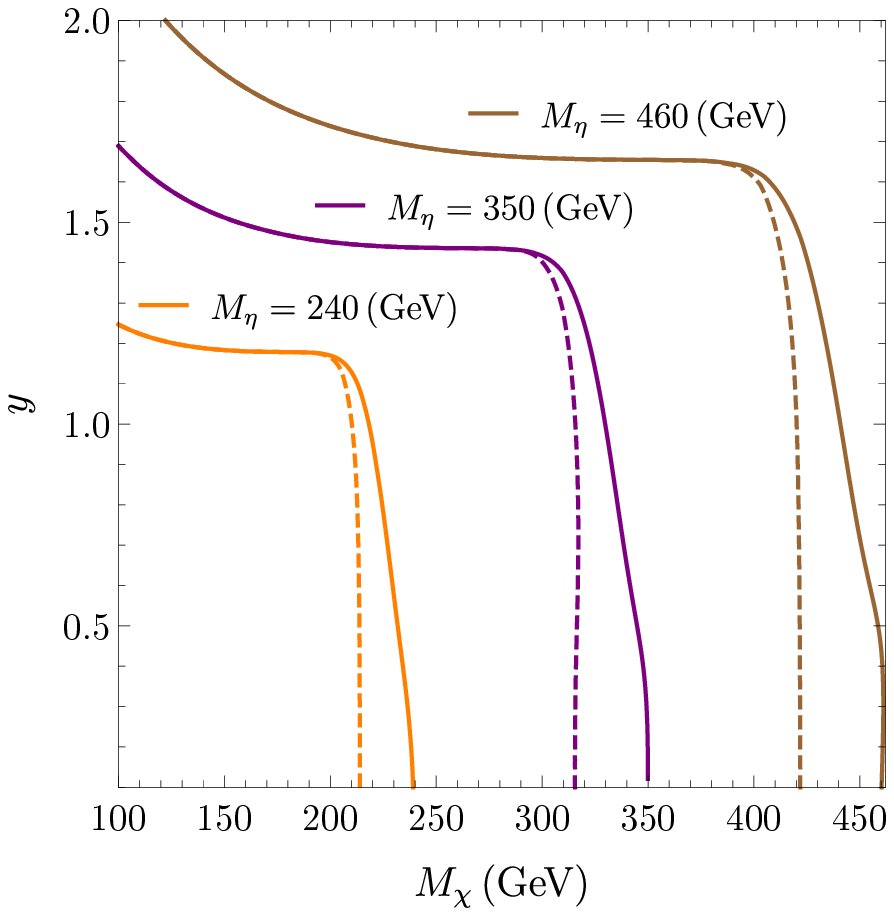}%
\hfill
\includegraphics[width=0.485\textwidth]{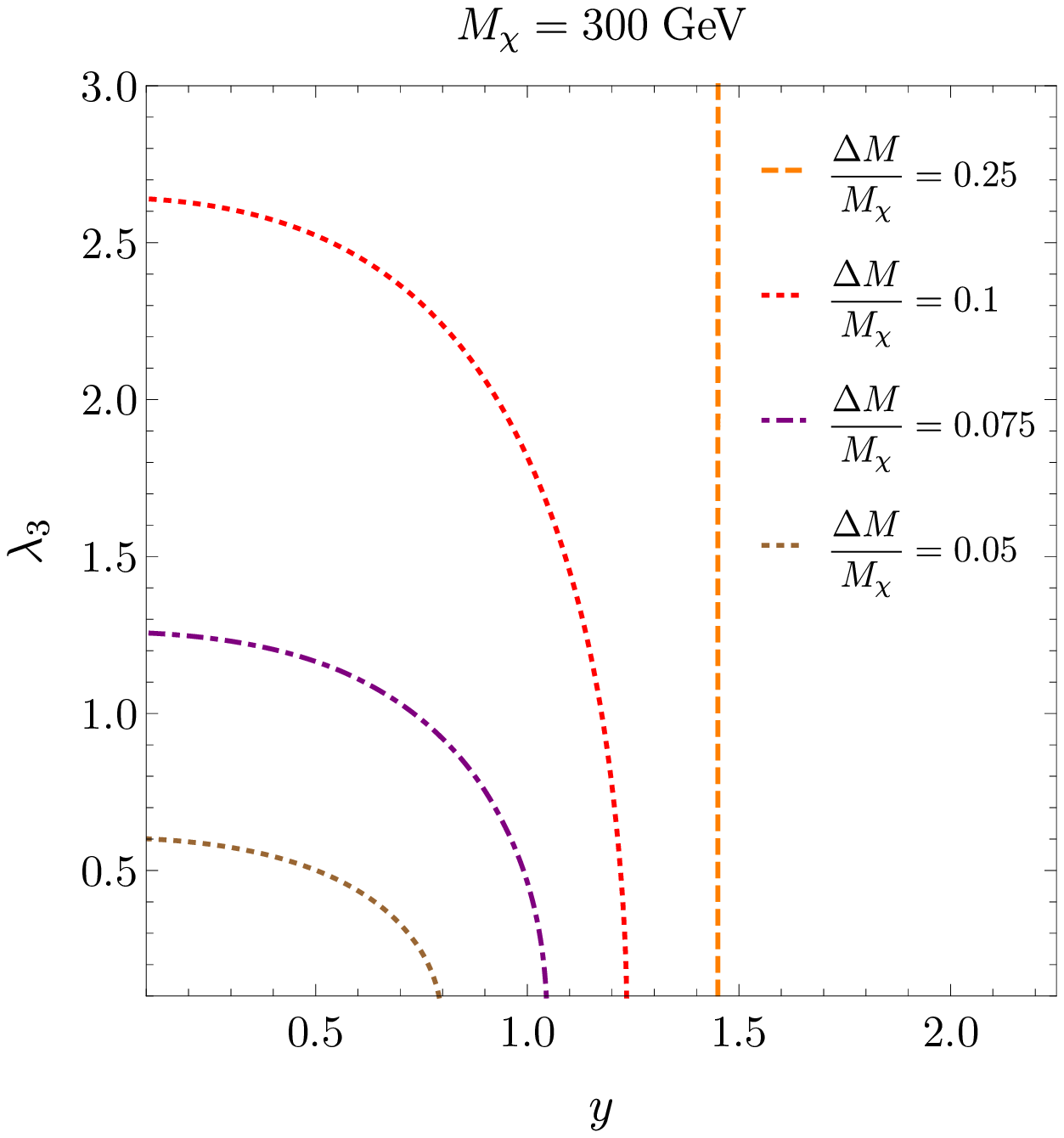}
\caption{%
  Left:
  Curves reproduce the observed energy density in the parameter space $(M_\chi,y)$.
  Three benchmark values for
  the scalar mass $M_\eta\in\{240,350,460\}$~GeV are considered at
  $\lambda_3 = 3.0$ (dashed) and
  $\lambda_3 = 0$ (solid).
  Right:
  Curves in the parameter space $(y,\lambda_3)$ for
  the DM mass $M_\chi=300$~GeV and different relative mass splittings
  $\Delta M/M_\chi$.
  }
\label{fig:param_DM_3}
\end{figure}
Two additional visualisations of the parameter space compatible with the observed energy density are offered in fig.~\ref{fig:param_DM_3}.
The left panel focuses on
the $(M_\chi,y)$-plane for
three different values of the scalar mass
$M_\eta\in\{240,350,460\}$~GeV
for
$\lambda_3 = 3.0$ (dashed) and
$\lambda_3 = 0$ (solid).
The behaviour of the curves can be understood by recalling the general form of the cross section for the scalar annihilations. On the one hand, for large values of the scalar-portal coupling, many contributions to the cross section
$\langle \sigma_{\eta \eta^\dagger} \vrel \rangle$
are active and one can allow for large mass splittings.
On the other hand, for $\lambda_3=0$ one has to
require smaller mass splitting to compensate for a smaller cross section
$\langle \sigma_{\eta \eta^\dagger} \vrel \rangle$ such that
the two masses $M_\chi$ and $M_\eta$ are almost degenerate.
The right panel of fig.~\ref{fig:param_DM_3} shows the contours for
$\Omega_{\rmii{DM}} h^2 \big \vert_{\rmi{obs.}}$
in the $(y, \lambda_3)$-plane at fixed dark matter mass
$M_\chi=300$~GeV.
For the relative mass splitting
$\Delta M / M_\chi=0.25$, beyond the co-annihilation strip, there is no dependence on
$\lambda_3$ because of a suppressed scalar population at the time of freeze-out and
later stages.
Conversely, by progressively decreasing the splitting,
the Yukawa and scalar-portal couplings become intertwined and
the observed energy density can be realised by
tuning one of the two couplings larger while decreasing the other.

%%%%%%%%%%%%%%%%%%%%%%%%%%%%% SUBSECTION %%%%%%%%%%%%%%%%%%%%%%%%%%%%%%%%%%%%
%
\subsection{Freeze-in}
\label{sec:freeze_in}

There is a two-fold motivation to not
pursue a detailed analysis for
the freeze-in parallel to the freeze-out mechanism.
First, we want to assess
the effect of a fermionic state coupled to the scalar sector, which undergoes
a phase transition.
Typical values of the Yukawa coupling $y$ required for freeze-in production would
completely decouple the fermion $\chi$ from the thermodynamics of the phase transition.
This is at variance with $y \sim \mathcal{O}(1)$ for the freeze-out scenario.
Second, and most importantly,
a reliable extraction of the parameter space of
the model compatible with the observed DM energy density is highly non-trivial.
In the following, we highlight further aspects in contrast with the freeze-out case.

The freeze-in production occurs in
a complementary temperature regime than freeze-out, namely
$T \gtrsim M_\eta, M_\chi$.
The latter is understood as the in-vacuum physical masses.
In this model, and rather in general for freeze-in produced DM, dark particles are generated through
the decays of a heavier accompanying state, here
$\eta \to \chi e$, and $2 \to 2$
scatterings that may involve SM particles.
If one assumes a vanishing initial abundance for the DM,
the tiny couplings with other fields prevent DM to ever reach chemical equilibrium.
Hence, dark matter particles only appear in the final
state of the relevant processes, and their abundance builds up and increases over the thermal
history.
The production takes place over a wide temperature range,
that includes
$T \gg M_\eta,M_\chi$.
Hence, thermal effects can be relevant.

The impact of thermal masses on freeze-in produced dark matter has been studied recently
by focusing on decay processes that would be forbidden at zero temperature.
Instead, such processes are realised in a
thermal environment~\cite{Baker:2016xzo,Baker:2017zwx,Dvorkin:2019zdi,Darme:2019wpd,Biondini:2020ric}.
Moreover, for the quark-philic model where the DM is coupled to a quark via
a QCD-charged scalar, the contribution to the dark matter production rate from
multiple soft scatterings at high temperature, oftentimes called the
Landau-Pomeranchuk-Migdal (LPM) effect~\cite{Landau:1953gr,Landau:1953um,Migdal:1956tc}, can drastically change the DM production rate depending on
the relative mass splitting $\Delta M /M_\chi$~\cite{Biondini:2020ric}.%
\footnote{
  Mass splittings of order unity lead to $\mathcal{O}(1)$ effects on the DM energy density.
  By demanding
  $\Delta M/M_\chi \gtrsim 10$,
  corrections of $\mathcal{O}(10\%)$ to the energy density can occur.
  Multiple scatterings with particles in the thermal bath were
  extensively studied in the neutrino production rate and
  applied
  to leptogenesis~\cite{Anisimov:2010gy,Besak:2012qm,Ghisoiu:2014mha,Ghiglieri:2016xye}.
}
Such resummations capture the quantum mechanical interference of collective plasma phenomena in a collinear kinematic regime at high temperature.
The main result is that effective $2 \to 1$ processes occur and enhance the production of the DM particle for $T \gg M_\chi$.
In the present model one has
$\eta \to \chi e$,
$e \to \eta \chi$, and
$\eta e \to \chi$.

At temperatures higher than the electroweak (crossover) transition,
the dynamics of the scalar $\eta$ can be intricate.
The scalar may also undergo a phase transition, and
broken and symmetric phases can alternate along the thermal history
(see two-step versus one-step transition in fig.~\ref{fig:multi_step}).
This induces a strong temperature dependence of the physical mass of $\eta$ and
its interactions with other particles in the plasma.
Hence, an accurate treatment of the scalar thermodynamics needs to be interfaced with
the DM production along the thermal history to compute decays and scattering processes on
a solid basis.
See e.g.~\cite{Bringmann:2021sth} for an implementation of these aspects for
a scalar singlet DM coupled to the SM Higgs boson within
a more phenomenological treatment of the EWPT.

For tiny $y$ couplings,
two sources contribute to the overall DM energy density
in this model~\cite{Garny:2018ali,Junius:2019dci}.
In addition to the freeze-in mechanism, that dominates at temperatures
$T \gtrsim M_\eta$,
instead the super-WIMP mechanism~\cite{Feng:2003xh,Feng:2003uy} occurs
much later in the thermal history at $T \ll M_\eta$.
During the latter mechanism,
the final abundance of $\eta$ particles is fixed by
freeze-out dynamics, and dark matter is produced in
the subsequent $\eta$ decay process
$\eta \to \chi e$.
Such a decay process will become efficient much later than
the chemical freeze-out due to the minuscule coupling $y \ll 1$.
The observed dark matter energy density is then given by 
\begin{equation}
\label{energy_density_tot}
    (\Omega_{\rmii{DM}}h^2)_{\rmi{obs.}}= 
    (\Omega_{\rmii{DM}}h^2)_{\rmi{freeze-in}}
  + (\Omega_{\rmii{DM}}h^2)_{\rmi{super-WIMP}}
  \;.
\end{equation}

We postpone extracting the DM energy density including LPM resummation,
the effect of thermal masses,
together with the interplay of the thermodynamics of the $\eta$ scalar to future studies.
For recent investigations of
freeze-in production,
super-WIMP, and
conversion-driven freeze-out of
the model considered in our work,
see ref.~\cite{Junius:2019dci,Bollig:2021psb}.
There the above-mentioned effects have not been included.

%%%%%%%%%%%%%%%%%%%%%%%%%%%%% SECTION %%%%%%%%%%%%%%%%%%%%%%%%%%%%%%%%%%%%
%
\section{Overlapping parameter space for phase transition and dark matter}
\label{sec:interplay}

This section investigates the joint
model parameter space responsible for both
the observed DM energy density and the EWPT.
Thus, we explore to what extent the model allows,
{\em at the same time}, for
the observed energy density and
a strong first-order phase transition.
The main results are visualised in two ways.

The first option focuses on the
$(M_\eta,\lambda_3)$-plane.
There the ranges that mark a strong first-order phase transition are easier to understand (cf.\ sec.~\ref{sec:ewpt}).
\begin{figure}[t!]
\centering
\includegraphics[width=0.5\textwidth]{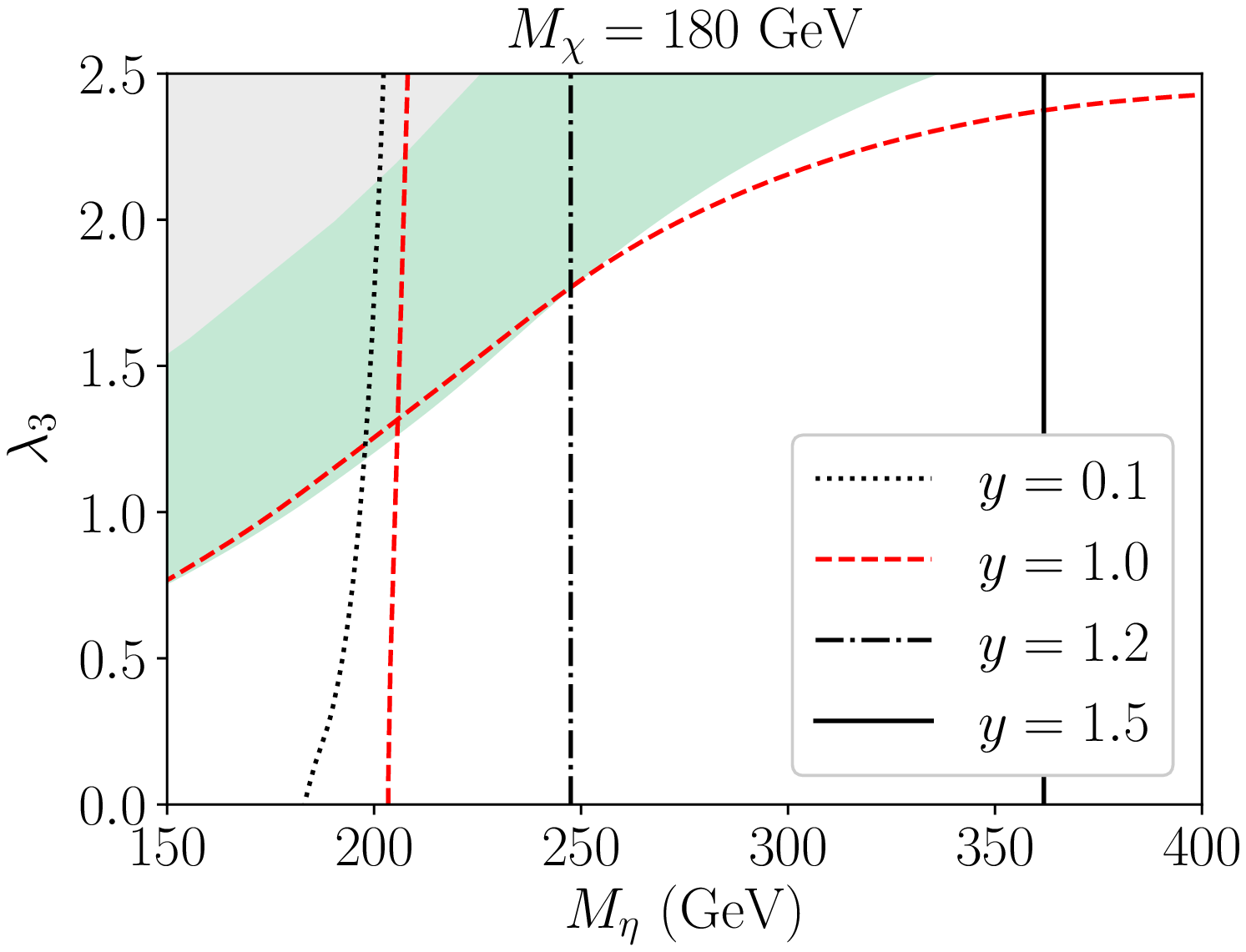}%
\includegraphics[width=0.5\textwidth]{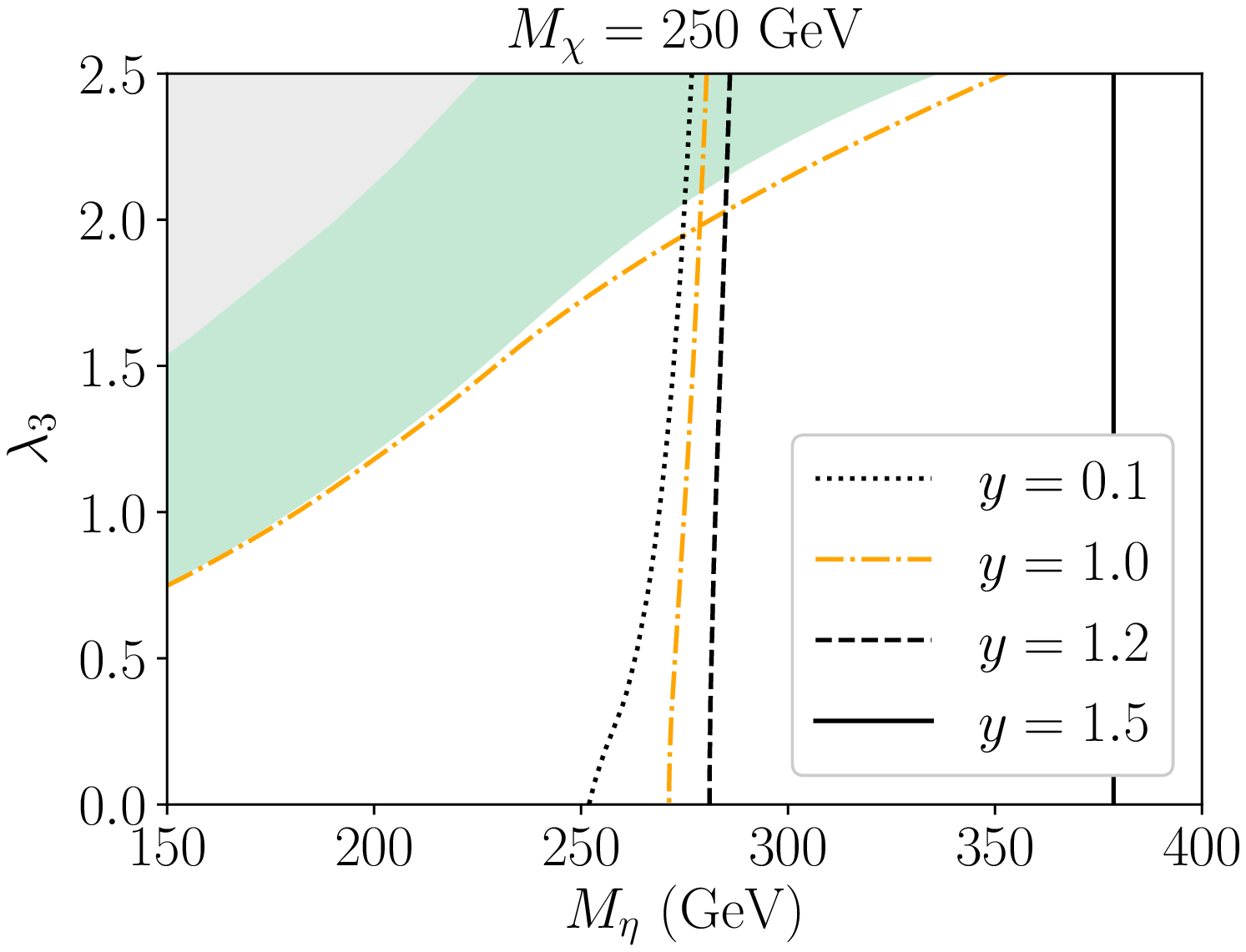}%
\caption{%
  Overlay of contours for the observed DM abundance
  $\Omega_{\rmii{DM}}h^2 \vert_{\rmi{obs.}}$ (vertical lines) in the
  $(M_\eta,\lambda_3)$-plane, for
  $M_\chi = 180$ GeV (left) and
  $M_\chi = 250$ GeV (right).
  The green band features a strong phase transition for $y=0$
  (cf.\ fig.~\ref{fig:EWPT-y-Mchi}).
  For $y=1$, we contour
  $v_{3,{\rm c}}/\sqrt{T_{{\rm c},\phi}} = 1$ for
  $M_\chi=180$~GeV (red, dashed) and
  $M_\chi=250$ GeV (orange, dash-dotted).
  For $y=1$, above the intersection point of the
  red~dashed and
  orange~dash-dotted lines,
  a strong phase transition is compatible with the observed DM abundance.
}
\label{fig:phases_comp_1}
\end{figure}
Figure~\ref{fig:phases_comp_1} provides the findings for
two different hypotheses of the dark matter fermion mass
$M_\chi=180$~GeV (left) and
$M_\chi=250$~GeV (right).
Such values evade the present-day collider searches for each lepton flavour
(cf.\ fig.~\ref{fig:param_DM_2}).
For the EWPT, we
show the metastable region (grey) and
reference the FOPT region
at $y=0$ as green as in fig.~\ref{fig:EWPT-y-Mchi}.
Then, we contour
$v_{3,{\rm c}}/\sqrt{T_{{\rm c},\phi}}=1$
corresponding to a non-vanishing dark-matter interaction with the scalar sector for 
$y=1.0$,
$M_\chi=180$~GeV (red, dashed) and
$y=1.0$,
$M_\chi=250$~GeV (orange, dash-dotted). 
Since the trend extends also to larger $y$-values (cf.\ sec.~\ref{sec:ewpt}),
we merely provide the benchmark line for $y=1.0$.

For dark matter, the
dotted,
dash-dotted,
dashed, and
solid contours in
the $(M_\eta,\lambda_3)$-plane reproduce
$\Omega_{\rmii{DM}}h^2 \vert_{\rmi{obs.}}=0.1200$.
The dependence between the two variables is rather mild and is 
progressively lost when increasing the Yukawa coupling $y$, which results in
a straight vertical line (solid, $y=1.5$) in both panels.
Small values of $y$
render $\chi\chi$ annihilations poorly effective (cf.\ sec.~\ref{sec:dark_matter}). Hence, the scalar pair annihilations, which are $\lambda_3$-dependent, drive the energy density and demand small mass splittings. This
results in values of $M_\eta$ that tend to be close to
$M_\chi=180$~GeV and
$M_\chi=250$~GeV for small $y$ and $\lambda_3$.
As long as $y$ increases,
the $\chi \chi$ annihilations become more important, and one can allow for
larger $M_\eta$ masses.
For large enough $y$,
the dependence on the scalar annihilations, and then on $\lambda_3$ is lost,
as one may see from the straight vertical lines for $y=1.5$.

\begin{figure}[t!]
\centering
\includegraphics[width=0.5\textwidth]{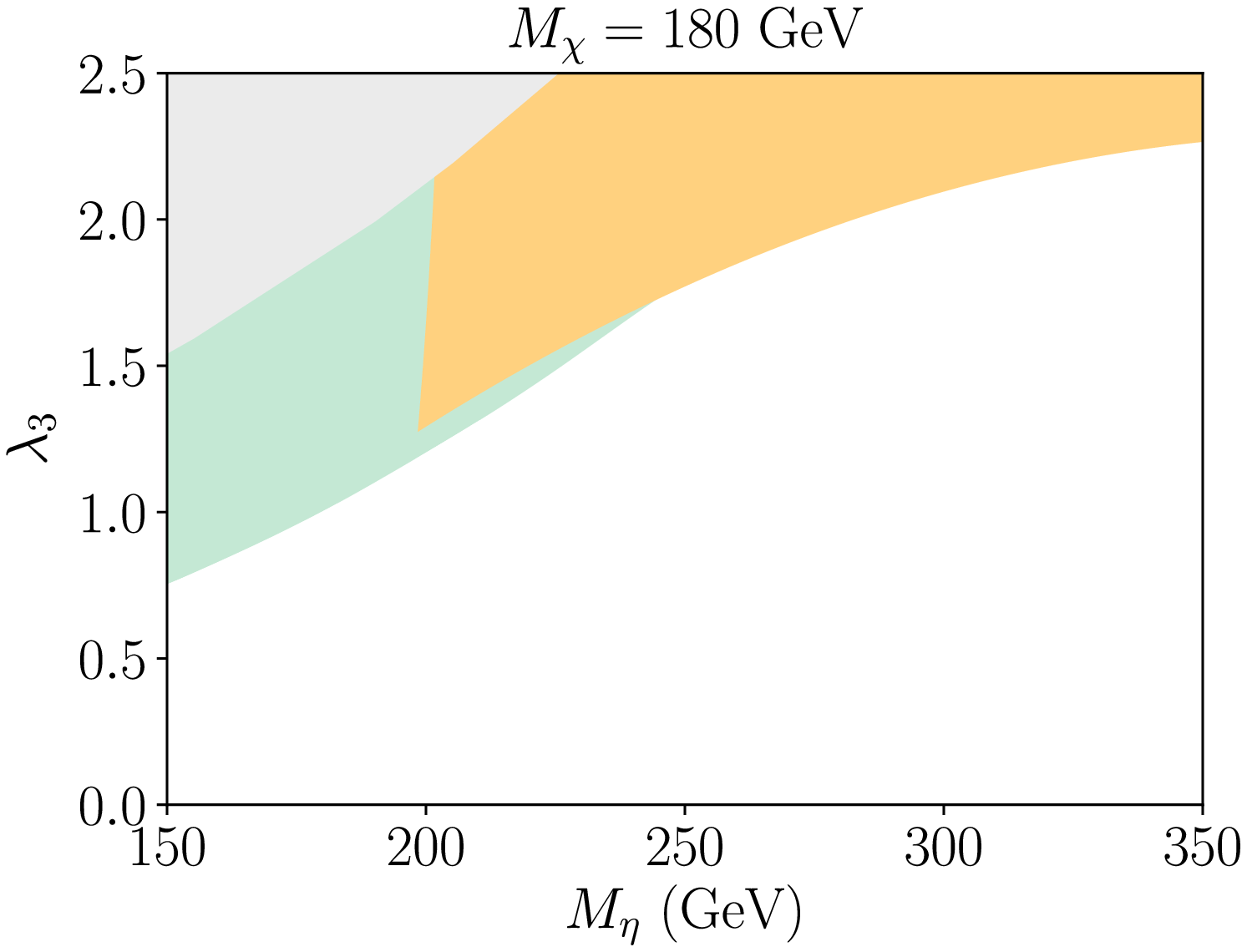}%
\includegraphics[width=0.5\textwidth]{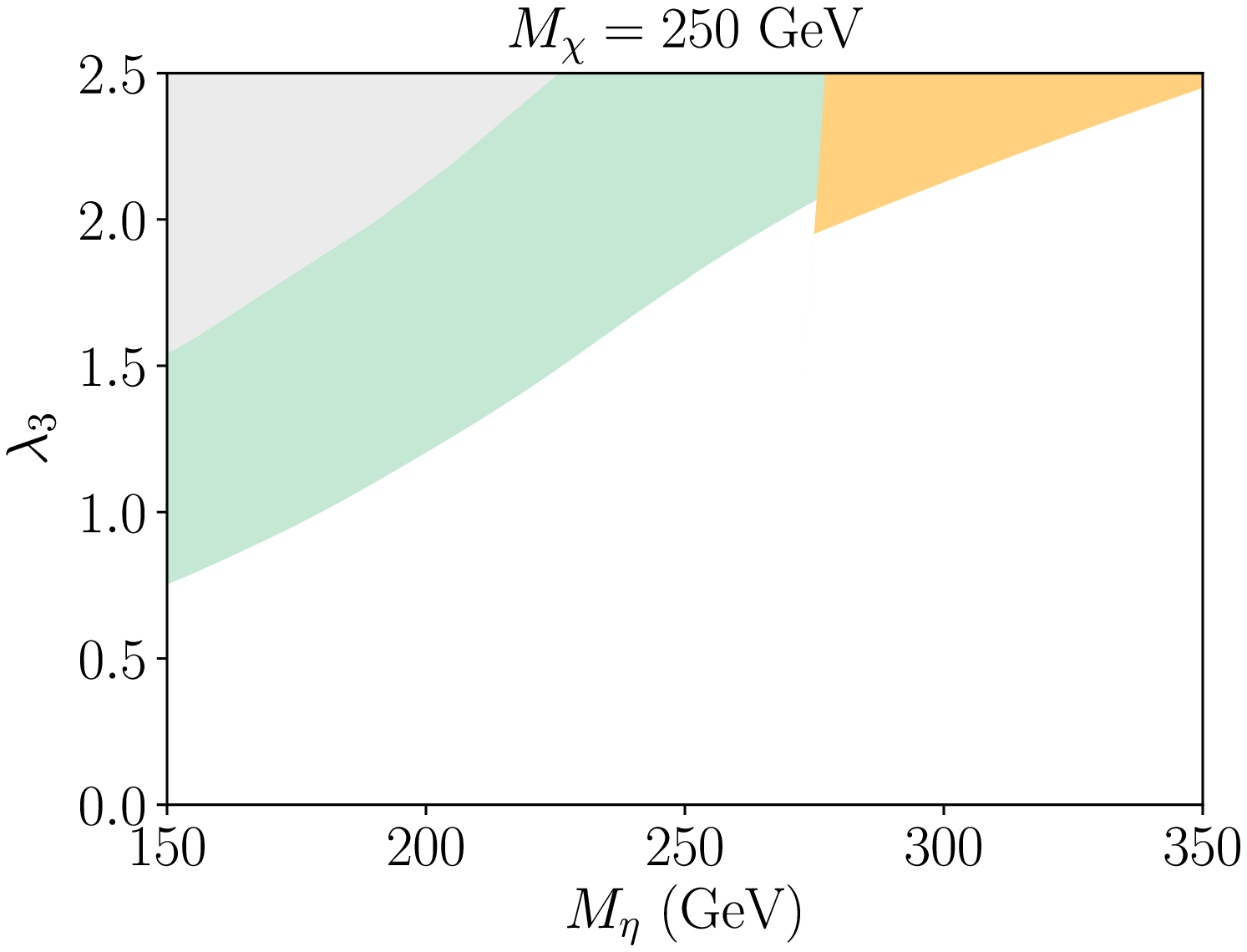}%
\caption{%
  Overlapping regions (orange) of
  strong FOPT and
  observed dark matter energy density for
  fixed $M_\chi$ and
  $0.1 < y < 1.5$.
  The $y=0$ FOPT green band
  is adopted from
  fig.~\ref{fig:EWPT-vanilla}.
  For smaller $M_\eta$ values,
  the orange band is cut based on the smallest used value $y=0.1$.
}
\label{fig:interplay}
\end{figure}

The bottom line of this comparison is the following.
Some parameter regions of the model can account for
the observed DM energy density
{\em and}
a strong first-order phase transition.
The overlapping region widens for smaller dark fermion masses which
also lies in the parameter space where
the perturbative assessment for the phase transition is more reliable.
Moreover, taking $y <0.1$ does not change the black~dotted curve
since the DM energy density is dominantly fixed by the scalar pair-annihilations already
at $y=0.1$.
Thus,
the scalar mass $M_\eta$
is bound from below by the black~dotted line
whereas increasing $y>1.5$ would push $M_\eta \gtrsim 400$
loosing entirely the connection with the EWPT.
The plot
in fig.~\ref{fig:interplay},
shows the overlapping region (orange) for varying $y$ and
is further discussed in the conclusions below.

\begin{figure}[t!]
\centering
\includegraphics[width=0.7\textwidth]{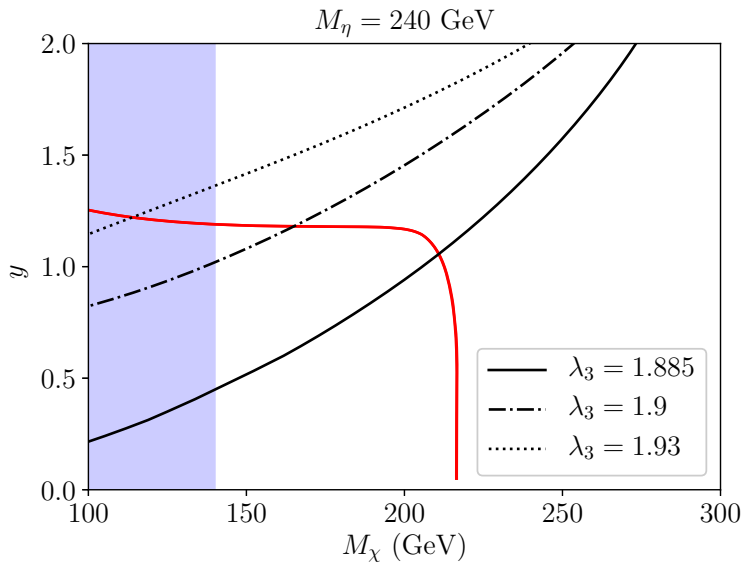}
\caption{%
  Overlap of strong phase transition and DM energy density in
  the $(M_\chi,y)$-plane for
  $M_\eta=240$~GeV and
  $\lambda_2=1.25$.
  Black lines contour the
  $v_{3,{\rm c}}/\sqrt{T_{{\rm c},\phi}} = 1$ for different values of
  $\lambda_3=\{1.885,1.9,1.93\}$.
  The red narrow band
  depicts the observed DM energy density which
  changes indiscernibly between the different values of $\lambda_3$.
  The blue shaded region is excluded by the collider searches, and we show
  the most stringent case due to the coupling to
  a muon ($M_\chi > 140$~GeV).
  Unlike in other plots contouring
  $v_{3,{\rm c}}/\sqrt{T_{{\rm c},\phi}}$,
  the transition strength does not increase towards the upper left corner.
  The behaviour of such contours is more complicated in the $(M_\chi,y)$-plane,
  which we do not visualise here to maintain clarity.
}
\label{fig:phases_comp_2}
\end{figure}
The second visualisation focuses on the
$(M_\chi,y)$-plane in fig.~\ref{fig:phases_comp_2}
at fixed $M_\eta=240$~GeV.
The three black curves (solid, dash-dotted, dotted)
contour
$v_{3,{\rm c}}/\sqrt{T_{{\rm c},\phi}}=1$,
that signal a strong phase transition,
for three (slightly) different portal couplings $\lambda_3$.
A large sensitivity to a small change of $\lambda_3$
at one-per-cent level in the $(M_\chi,y)$-plane is clearly visible.
This reflects the non-trivial behaviour in
the complementary parameter space $(M_\eta,\lambda_3)$ upon changing
the dark fermion mass and Yukawa coupling $y$ (cf.\ fig.~\ref{fig:EWPT-y-Mchi}).

The situation is rather the opposite for the dark matter energy density.
In the same fig.~\ref{fig:phases_comp_2},
all three values of $\lambda_{3}$ collapse onto
the red narrow band and correspond to
$\Omega_{\rmii{DM}} h^2 \vert_{\rmi{obs.}}$.
This is largely consistent with a per-cent change in scalar portal coupling,
that can only slightly affect the annihilation cross section and
the corresponding extraction of the energy density.
Hence, in this complementary visualisation, one may appreciate how differently
the first-order phase transition and DM energy density depend on the model parameters.
We also included the experimental constraint on the DM mass
$M_\chi> 140$~GeV for the case of the coupling of $\chi$ and $\eta$ with a muon.
This limit is
most stringent for $M_\eta=240$~GeV among the three lepton flavours
(see fig.~\ref{fig:param_DM_2}).
In this light, one can exclude the case $\lambda_3=1.93$ for 
the simultaneous event of a strong phase transition and the observed DM relic density. 
A similar analysis can be repeated for other choices of parameters,
to explore possible values for portal couplings.

%%%%%%%%%%%%%%%%%%%%%%%%%%%%% SECTION %%%%%%%%%%%%%%%%%%%%%%%%%%%%%%%%%%%%
%
\section{Conclusions and outlook}
\label{sec:outlook}

In this article, we investigated
the coexistence of a strong electroweak phase transition and
the observed dark matter abundance using
state-of-the-art methodology for the combined analyses,
in perturbation theory.
The phase transition analysis is performed within
the 3d EFT which includes complete NLO thermal resummations via dimensional reduction.
For the dark matter energy density via freeze-out, we included the effect of thermal masses, Sommerfeld enhancement and bound-state effects.

The model considered in the article
is a $t$-channel mediator dark matter model, that belongs to a next-to-minimal class of simplified models
offering a rich phenomenology at collider, direct and indirect searches.
This setup goes beyond the minimal option of a DM singlet scalar coupled to
the SM Higgs boson.
Concretely, the model features
a dark matter Majorana fermion
that is a singlet under the SM gauge group, and
a scalar particle that mediates between the visible and dark matter sector.
Gauge-invariant and renormalisable operators can be built from
the Majorana DM fermion,
the scalar mediator and
the SM chiral fermions.
We considered interactions with right-handed leptons.
The scalar mediator can couple directly to
the Higgs boson and, therefore,
affect the electroweak phase transition.
As the main result, our study
discerns regions of parameters space where both
a strong phase transition and
the observed DM energy density allow for one another.

Quantitatively we find that a strong first-order phase transition can be realised for
a complex scalar mass
$150~{\rm GeV} \lesssim M_\eta \lesssim 340~{\rm GeV}$ and
a scalar portal coupling
$0.75 \lesssim  \lambda_3 \lesssim 2.5$, in the limit of a decoupled DM fermion.
Next, we assessed the effect of the Yukawa coupling and the fermionic degree of freedom.
The overall effect on the FOPT diagram $(M_\eta,\lambda_3)$ is mild
since the Majorana fermion does not interact directly with the SM Higgs boson.
However, the interplay between
the Yukawa coupling and
the DM mass non-trivially changes the parameter space compatible with
a first-order transition.
Different combinations of $y$ and $M_\chi$ enlarge or narrow
the region compatible with a FOPT.

For extracting the dark matter energy density,
we focused on the freeze-out production mechanism.
To find accurately the parameter space consistent with observations,
one needs to include the effects of co-annihilations of the accompanying scalar mediator.  For sufficiently small mass splittings, the scalar pair annihilations are relevant and
we have included the processes that are induced by the interaction between
the scalar $\eta$ and the SM Higgs boson.
Moreover, since the scalar mediator interacts with the $Z$-boson and the photon, the pair annihilations are affected by non-perturbative effects, namely Sommerfeld enhancement and bound-state formation.
The main contribution to both effects is due to the interaction between the scalar $\eta$ and the photon for masses
$M_\eta \lesssim 1$~TeV.
By combing the two effects, we find an impact on the DM energy density that ranges between 1--10\% depending on
the relative mass splitting
$(M_\eta-M_\chi)/M_\chi$ and
the scalar portal coupling $\lambda_3$.
We restricted the mass range
$M_\chi,M_\eta \lesssim 1$~TeV because larger values are not interesting for the EWPT.
More importantly,
this mass range maintains the phase transition and DM production as
two separate events along the cosmological history, as
the freeze-out temperature is below
the critical temperature of the phase transition.

The main result of our study is shown in
figs.~\ref{fig:phases_comp_1} and \ref{fig:interplay}.
The overlapping region of a correct DM energy density and FOPT strongly depends on
the Majorana fermion mass $M_\chi$ and
the Yukawa coupling $y$.
Upon increasing $M_\chi$,
one has to also take larger values of $M_\eta$
(the DM model requires $M_\chi < M_\eta$), and hence
the perturbative upper bound
$M_\eta \lesssim 350$~GeV is soon saturated.
The same holds when increasing the value of the Yukawa coupling to $y \approx 1$.
We find that the value $y \simeq 0.1$ provides the lowest possible scalar masses.
There is no effect in decreasing $y$ further since
the DM energy density is governed by
the scalar pair annihilations for the mass range of interest for $y \simeq 0.1$.
When choosing the smallest DM fermion mass that avoids entirely the collider constrains,
we find that
the observed energy density and
a strong electroweak phase transition can be realised in the present model for
$
140~{\rm GeV} < M_\chi \lesssim
300~{\rm GeV}$
($
180~{\rm GeV} < M_\chi \lesssim
300~{\rm GeV}$)
for coupling to taus (muons).
Moreover, one may look at the interplay between a FOPT and DM in
the parameter space $(M_\chi,y)$ in fig.~\ref{fig:phases_comp_2}.
Here,
a rather sharp difference on the dependence of the scalar portal coupling appears.
While very slight changes in $\lambda_3$ sensibly change
the $v_{c}/T_{\rmi{c},\phi} =1$ condition,
they leave the DM energy density unaffected.
Most importantly,
the collider constraints can exclude values of $\lambda_3$ that allow for a strong EWPT.

The present analysis can be extended into various directions.
First, the freeze-in production deserves attention in the light of
null searches of WIMP-like DM particles.
The DM could be merely interacting feebly with the visible sector and
there could still exist compelling constraints for $t$-channel models from
long-lived particles at colliders
(in our case the lifetime of the produced $\eta$ would be long due to a tiny $y$).
For the freeze-in mechanism,
additional complexity in our investigation would be introduced
by relaxing the assumption that both DM production and EWPT occurred
as separate events in the thermal history of the universe.
Thus, non-trivial cross influences of both events are conceivable but
would require theoretical tools to extend to broader temperature ranges.
Furthermore,
different realisations of the present model where
the Majorana fermion is coupled to a left-handed fermion
are worth exploring
as in e.g.~\cite{Garny:2015wea,Cline:2017qpe}.
The corresponding mediator would then be charged under
the ${\rm SU(2)}_\rmii{L}$ gauge group.

Second,
it is conceivable to implement the pipeline connecting
the thermodynamics of a strong first-order phase transition
with
the production of a stochastic gravitational wave spectrum.
Such an implementation can be realised within the 3d EFT
to account for NLO thermal resummations
as in~\cite{Croon:2020cgk,Gould:2021oba}.
Furthermore, the EFT picture allows for consistently computing
the bubble nucleation rate and
nucleation temperature~\cite{Gould:2021ccf} as well as
including higher order corrections~\cite{Ekstedt:2021kyx,Ekstedt:2022tqk}.
Such a computation goes beyond the scope of this article and is left for future studies.
Nevertheless, we can expect that a strong enough GW signature for
LISA-generation interferometers could be generated in a subregion of
two-step phase transition regions --
in analogy to a recent analysis~\cite{Friedrich:2022cak}.

Finally, we foresee a link between
the simplified model approach for dark matter and
exploring possible new-physics affecting the electroweak phase transitions.
There has been extensive effort in classifying DM models,
that exploit the Higgs portal as the main actor in connecting the visible to a dark sector
(see e.g.\ reviews~\cite{Abdallah:2015ter,Arcadi:2019lka}).
A similar framework could help to estimate
the effects on the EWPT due to new physics coupled to
the SM Higgs boson.
In the same spirit of a simplified model approach for DM,
one can classify new-physics models depending on
the particle that couples to the Higgs boson, such as scalars, gauge bosons, or fermions.
The latter may share additional interactions
with the SM degrees of freedom or
with a dark sector or
both.
One may still capture the relevant effect on the thermodynamics of the phase transition
once a minimal set of fields and couplings is given.
This way, the impact on the EWPT can be assessed
without necessarily relying on a fully-fledged UV theory, as we
explored in our study.

%%%%%%%%%%%%%%%%%%%%%%%%%%%%% SECTION %%%%%%%%%%%%%%%%%%%%%%%%%%%%%%%%%%%%
%
\section*{Acknowledgments}

The work of Simone Biondini is supported by the Swiss National Science Foundation under the Ambizione grant PZ00P2\_185783.
Philipp Schicho was supported
by the European Research Council, grant no.~725369, and
by the Academy of Finland, grant no.~1322507.
The work of Tuomas V.~I.~Tenkanen has been supported in part
by the National Science Foundation of China grant no.~19Z103010239.
The authors are grateful for
Stefan Vogl for useful discussions on the collider limits on the simplified model, and for
Tommi Tenkanen for comments on the manuscript.

%%%%%%%%%%%%%%%%%%%%%%%%%%%%% APPENDIX %%%%%%%%%%%%%%%%%%%%%%%%%%%%%%%%%%%%
%
\appendix
\renewcommand{\thesection}{\Alph{section}}
\renewcommand{\thesubsection}{\Alph{section}.\arabic{subsection}}
\renewcommand{\theequation}{\Alph{section}.\arabic{equation}}

%%%%%%%%%%%%%%%%%%%%%%%%%%%%% SECTION %%%%%%%%%%%%%%%%%%%%%%%%%%%%%%%%%%%%
%
\section{Dimensional reduction and thermal effective potential}
\label{sec:model:match}

This appendix collects
renormalisation group equations (RGE),
the matching relations of the model defined in eq.~\eqref{eq:lag:4d} to its
dimensionally reduced three-dimensional effective theory in eq.~\eqref{eq:lag:3d},
and
the thermal effective potential computed within the EFT.

%%%%%%%%%%%%%%%%%%%%%%%%%%%%% SUBSECTION %%%%%%%%%%%%%%%%%%%%%%%%%%%%%%%%%%%%
%
\subsection{Renormalisation and one-loop beta functions}
\label{sec:rge}

The renormalisation group equations listed below
are associated with the parameters of the model in eq.~\eqref{eq:lag:4d} and
encode their running with respect to
the \MSbar{} renormalisation scale $\LamD$ via the beta functions.
To this end, we use
\begin{equation}
\label{eq:rge:g1}
t \equiv \ln\bar{\mu}^2
\;,
\end{equation}
and find
at one-loop level
\begin{align}
\partial_{t}^{ }
\go^2 &=
    \beta_{\rmii{SM}}(\go^2)
  + \frac{1}{(4\pi)^2} \Big( \frac{1}{12} Y_\eta^{2} \go^4 \Big)
  \;, \\
\partial_{t}^{ }
\gt^2 &=
    \beta_{\rmii{SM}}(\gt^2)
  \;, \\[2mm]
\partial_{t}^{ }
\gY^2 &=
    \beta_{\rmii{SM}}(\gY^2)
  \;, \\[2mm]
\partial_{t}^{ }
\mu_{\phi}^{2} &=
    \beta_{\rmii{SM}}(\mu^2_\phi)
  + \frac{1}{(4\pi)^2} \Bigl( \lambda_{3}^{ } \mu_\eta^2 \Bigr)
  \;, \\
\partial_{t}^{ }
\mu_{\eta}^{2} &=
  \frac{1}{(4\pi)^2} \Big(
      2 \lambda_{3}^{ } \mu^2_\phi
    - 2 |y|^2 \mu^2_\chi
    + \Bigl(4 \lambda_{2}^{ } - \frac{3}{4} Y_\eta^{2} \go^2 + |y|^2\Bigr) \mu^2_\eta
  \Big)
  \;, \\
\partial_{t}^{ }
\mu_{\chi}^{2} &=
  \frac{1}{(4\pi)^2} |y|^2 \mu_{\chi}^2
  \;, \\
\partial_{t}^{ }
\lambda_{1} &=
    \beta_{\rmii{SM}}(\lambda_{1})
  + \frac{1}{(4\pi)^2} \Bigl( \frac{1}{2} \lambda_{3}^2 \Bigr)
  \;, \\
\partial_{t}^{ }
\lambda_{2}^{ } &=
  \frac{1}{(4\pi)^2} \Big(
      \lambda_{3}^2
    + 10 \lambda_{2}^2
    - \frac{3}{2} Y_\eta^2 \go^2 \lambda_{2}^{ }
    + \frac{3}{16} Y_\eta^4 \go^4
    + (2 \lambda_{2}^{ } - |y|^2)|y|^2
  \Big)
  \;, \\
\partial_{t}^{ }
\lambda_{3}^{ } &=
  \frac{1}{(4\pi)^2} \Big(
    \lambda_{3} \Big[
        2 \lambda_{3}^{ }
      + 4 \lambda_{2}^{ }
      - \frac{3}{4}(3 \gt^2 + (Y_\phi^{2} + Y_\eta^{2}) \go^2)
      + \Nc^{ }\gY^2
      + |y|^2
      + 6\lambda_{1} \Big]
  \nn &
  \hphantom{=\frac{1}{(4\pi)^2} \Big(}
  + \frac{3}{8} Y^2_\eta\Ys^2 \go^4 \Big)
  \;, \\
\label{eq:rge:y}
\partial_{t}^{ }
|y|^2 &=
  \frac{1}{(4\pi)^2}\Bigl(
    2 |y|^4
  - \frac{3}{4} |y|^2 \go^2
      Y_\eta^2
  \Bigr)
  \;.
\end{align}
A generalisation of the beta function for $|y|$ can be found
in~\cite{Luo:2002ti}.
The functions $\beta_{\rmii{SM}}$ are pure SM contributions,
that can be read e.g.~from~\cite{Brauner:2016fla}.
At the accuracy of a NLO dimensional reduction,
the strong coupling $\gs$
merely enters the thermal mass for
the Higgs doublet at two-loop order
and can be fixed to
$\gs(\MZ) = 1.48409$~\cite{Workman:2022ynf}.

%%%%%%%%%%%%%%%%%%%%%%%%%%%%% SUBSUBSECTION %%%%%%%%%%%%%%%%%%%%%%%%%%%%%%%%%%%%
%
\subsubsection*{Background field dependent mass eigenvalues}

Using the scalar field parameterisation in eq.~\eqref{eq:phi:eta:param},
the mass eigenvalues in terms of generic background fields $v$ and $x$ read
\begin{align}
\MW^2 &= \frac{1}{4}\gt^2 v^2
  \;, \\
M_{\pm}^{2} &= \frac{1}{8} \biggl(
    (\gt^2 + \go^2)v^2
  + Y_\eta^{2} \go^2 x^2
  \nn &
  \hphantom{{}= \frac{1}{8} \biggl( (\gt^2 + \go^2)v^2}
  \pm \sqrt{
    \gt^4 v^4
    + 2 \gt^2 \go^2 v^2 (v^2-Y_\eta^{2} x^2)
    + \go^4 (v^2 + Y_\eta^{2} x^2)^2 }
  \biggr)
  \;,
\end{align}
for the gauge fields.
The $W$-mass is double degenerate, and
$M^{}_+$($M^{ }_-$) is the eigenvalue for the $Z$-boson (photon)
that reduces to
the SM expressions for vanishing $x$.
For scalars, 
\begin{align}
\mG^2 &=
      \mu^2_\phi
    + \lambda_{1} v^2
    + \frac{1}{2} \lambda_{3} x^2
  \;, \\
\mA^2 &=
      \mu^2_\eta
    + \lambda_{2} x^2
    + \frac{1}{2} \lambda_{3} v^2
    \;, \\
m^2_{\pm} &=
      \frac{1}{2} (\mu^2_\phi + \mu^2_\eta )
    + \Bigl(\frac{3}{2} \lambda_{1} + \frac{1}{4} \lambda_{3} \Bigr) v^2
    + \Bigl(\frac{3}{2} \lambda_{2} + \frac{1}{4} \lambda_{3} \Bigr) x^2
    \nn &
    \pm \Bigl(
        (v^4 + 14 v^2 x^2 + x^4) \lambda_{3}^2
      + 4 \lambda_{3}^{ }(v^2-x^2) (
          3 \lambda_{1}^{ } v^2
        + 3 \lambda_{2}^{ } x^2
        - \mu_\phi^{2}
        + \mu_\eta^{2}
        ) 
    \nn &
      + 4 (
          3 \lambda_{1}^{ } v^2
        + 3 \lambda_{2}^{ } x^2
        - \mu_\phi^{2}
        + \mu_\eta^{2}
        )^2 
    \Bigr)^{\frac{1}{2}}
    \;,
\end{align}
where
the Goldstone mass eigenvalue $\mG^{2}$ is triple degenerate, and
the top quark has
the mass eigenvalue
\begin{align}
m^2_t &= \frac{1}{2} \gY^2 v^2
\;.
\end{align}
For other SM fermions,
the Yukawa couplings are assumed to vanish,
as their effect is negligible for EWPT thermodynamics~\cite{Kajantie:1995dw}.

%%%%%%%%%%%%%%%%%%%%%%%%%%%%% SUBSUBSECTION %%%%%%%%%%%%%%%%%%%%%%%%%%%%%%%%%%%%
%
\subsubsection*{Relation of \MSbar{} and input parameters}

At zero temperature,
the vacuum-expectation-value for
the singlet is assumed to vanish.
As a result of
setting the background field $x=0$, we identify 
$\mA = m_+ \equiv M_\eta$.
The Goldstone mass eigenvalue vanishes at the electroweak minimum  
\begin{align}
v_0 = \sqrt{\frac{4\MW^2}{g^2_0}}
  \;,
\end{align}
where
we install a shorthand notation
$g^2_0 \equiv 4 \sqrt{2} G_{\rmi{F}}^{ } \MW^2$, and
the reduced Fermi constant
$G_{\rmi{F}} = 1.1663787 \times 10^{-5}~{\rm GeV}^{-2}$.
In terms of the masses
$$
\bigl\{ \MW, \MZ, M_t \bigr\} = \bigl\{
80.379~{\rm GeV},
91.1876~{\rm GeV},
172.76~{\rm GeV} 
\bigr\}\;,
$$
the gauge and Yukawa couplings are
\begin{align}
\gt^2 &= g^2_0
  \;,&
\go^2 &= g^2_0\Big( \frac{\MZ^2}{\MW^2} - 1 \Big)
  \;,& 
\gY^2 &= \frac{1}{2} g^2_0 \frac{M^2_t}{\MW^2}
  \;.
\end{align}
By inverting the scalar mass eigenvalues,
we get
\begin{align}
\lambda_{1}^{ } &= \frac{1}{2} \frac{M^2_\phi}{v^2_0}
  \;,&
\mu^2_\phi &= -\frac{1}{2} M^2_\phi
  \;,&
\mu^2_\eta &= M^2_\eta - \frac{1}{2} \lambda_{3}^{ } v^2_0
  \;.
\end{align}
Here, the Higgs mass
$M_\phi = 125.1$~GeV, 
$\lambda_{2}$, 
$\lambda_{3}$, and 
the unknown mass $M_\eta$ 
are treated as free input parameters. 
We assume that the Higgs is identified with the lighter eigenstate,
i.e.\
$M^2_\phi = m^2_-$, and
the singlet related states have identical masses
$M^2_\eta \equiv \mA^2 = m^2_+$.
The above relations are valid at LO, and receive loop corrections
that could be included
at one-loop, or NLO
along the lines of~\cite{Kajantie:1995dw,Laine:2017hdk,Kainulainen:2019kyp,Niemi:2021qvp}.
However, we do not consider these corrections here.
The dark matter mass parameter
has a trivial LO relation
\begin{align}
\mu^2_{\chi} = M^2_{\chi}
  \;,
\end{align}
with its physical mass $M_{\chi}$.

For the tree-level potential ($V$) at zero temperature to be bounded from below, 
the parameters have to satisfy~\cite{Ivanov:2018jmz}
\begin{align}
\lambda_{1},
\lambda_{2} > 0
  \;,\qquad
\lambda_{3} + 2 \sqrt{\lambda_{1} \lambda_{2}} > 0
  \;.
\end{align}
The Higgs and singlet phases are described by 
\begin{align}
(v,x) \stackrel{\text{Higgs}}{=} 
    \biggl(
      i\sqrt{\frac{\mu^2_\phi}{\lambda_1}}, 0
    \biggr)
  \;, \qquad
(v,x) \stackrel{\text{singlet}}{=}
    \biggl(
      0, i\sqrt{\frac{\mu^2_\eta}{\lambda_{2}}}
    \biggr)
  \;.
\end{align}
These solutions for the background fields extremise 
the tree-level potential, and we identify
the Higgs phase as the zero-temperature electroweak minimum.
There are also other solutions for extrema, that are not minima of the potential.
For a two-variable function, 
the extremising condition for the minima is
\begin{align}
    \underbrace{\frac{\partial^2 V}{\partial v^2}}_{>0}
    \frac{\partial^2 V}{\partial x^2}
  - \Big( \frac{\partial^2 V}{\partial v \partial x} \Big)^2 > 0
  \;.
\end{align}
If both Higgs and singlet minima coexist at the same parameter point,
we require a global Higgs minimum.

%%%%%%%%%%%%%%%%%%%%%%%%%%%%% SUBSECTION %%%%%%%%%%%%%%%%%%%%%%%%%%%%%%%%%%%%
%
\subsection{Parameters of the 3d EFT}
\label{sec:eft:3d}

The effective parameters of the dimensionally reduced theory are collected below.
Our independent computation here also agrees with
the output of {\tt DRalgo}~\cite{Ekstedt:2022bff}.
To aid compactness,
we define a shorthand notation
\begin{align}
L_{b} &\equiv
    2 \ln\Big( \frac{\mu}{T} \Big)
  - 2 \Big( \ln(4\pi) - \gammaE \Big)
  \;, \quad
L_{f} \equiv L_{b} + 4\ln2
    \;, \\
c &=
  \frac{1}{2} \Bigl(
    \ln \Big( \frac{8\pi}{9} \Big)
    + (\ln\zeta_{2})'
  - 2 \gammaE \Bigr)
  \;,
\end{align}
where 
$\gammaE$ is the Euler-Mascheroni constant, 
$\zeta_{s}=\zeta(s)$ for $\re\,(s) > 1$ is the Riemann zeta function, and
$(\ln \zeta_s)'=\zeta'(s)/\zeta(s)$.
In the high-temperature expansion,%
\footnote{
  The high-temperature expansion is applied to all mass parameters
  $\mu_\phi, \mu_\eta, \mu_\chi \sim g T$.
  Later in this section,
  we discuss the results without high-temperature expansion for
  the Majorana fermion mass parameter.
}
and given that the number of fermion generations
$\nf=3$, 
the matching relations are
\begin{align}
g_{1,3}^{2} &=
  \go^2T\bigg[
    1
    - \frac{\go^2}{(4\pi)^2}\frac{1}{6}\Big(
        L_{b}\Big[\Ys^{2} +\frac{1}{2}Y_{\eta}^{2}\Big]
      + L_{f}Y_{\rmi{2f}}^{ }\,\nf^{ }
  \Big)\bigg]
\;,\\
g_{2,3}^2 &=
  \gt^2T\bigg[
    1
  + \frac{\gt^2}{(4\pi)^2}\Big(
      \frac{43}{6}L_b
    + \frac{2}{3}
    - \frac{(\Nc+1)\nf}{3}L_f
  \Big)\bigg]
\;,\\
\label{eq:mD1}
\mDi{\rmii{1}}^{2} &=
  \Big( \mDi{\rmii{1}}^{2} \Big)_{\rmii{SM}}
  + T^2 \go^{2}\bigg[
      \frac{1}{12}
    + \frac{1}{(4\pi)^2} \frac{\mu_{\eta}^{2}}{T^2}
  \bigg]Y_{\eta}^{2}
  \nn &
  + \frac{T^{2}}{(4\pi)^2}\go^{2}\bigg[
    -\bigg(
        \frac{L_{b}-7}{144}Y_{\eta}^{4}
      + \frac{L_{b}+2}{36}\Ys Y_{\eta}^{2}
      + \frac{(L_{b} + 4L_{f} - 2)Y_{\rmi{2f}}^{ }Y_{\eta}^{2}}{288}\nf
      \bigg)\go^{2}
  \nn &
  \hphantom{{}+ \frac{T^{2}}{(4\pi)^2}\go^{2}\bigg[}
    + \frac{1}{6} \lambda_{3}^{ }\big[\Ys^{2} + Y_{\eta}^{2}\big]
    + \frac{1}{3} \lambda_{2}^{ }Y_{\eta}^{2}
    - \frac{Y_{\eta}^{2}}{24} |y|^2
  \bigg]
\;,\\
\label{eq:mD2}
\mDi{\rmii{2}}^{2} &=
  \Big( \mDi{\rmii{2}}^{2} \Big)_{\rmii{SM}}
  + \frac{T^2}{(4\pi)^2}\frac{1}{6} \gt^{2} \lambda_{3}^{ }
\;,\\
\label{eq:m:phi:2l}
\mu_{\phi,3}^{2} &=
    \Big( \mu^2_{\phi,3} \Big)_{\rmii{SM}}
  + \frac{T^2}{12} \lambda_{3}^{ }
  - \frac{L_b}{(4\pi)^2} \lambda_{3}^{ } \mu^2_\eta
  \nn &
  + T^{2}\frac{Y_{\eta}^{2}}{(4\pi)^{2}}\Big(
      \frac{5}{288}\,\go^{4}\,\Ys^{2}
    + \frac{1}{24}\,\go^{2}\,\lambda_{3}
    \Big)
  + T^{2}L_{f}\frac{\lambda_{3}}{(4\pi)^{2}}\frac{1}{12}\Bigl(
      \frac{1}{2}\,|y|^{2}
    - \Nc\,\gY^{2}
    \Bigr)
  \nn &
  - T^{2}L_{b}\frac{1}{(4\pi)^{2}}\bigg(
      \frac{7}{192}\,\go^{4}\,Y_{\eta}^{2}\Ys^{2}
  \nn &
  \hphantom{- T^{2}L_{b}\frac{1}{(4\pi)^{2}}\bigg(}
      + \lambda_{3}\Big(
        \frac{5}{12}\,\lambda_{3}^{ }
      + \frac{\lambda_{1}}{2}
      + \frac{\lambda_{2}}{3}\,
      + \frac{1}{8}\,|y|^{2}
      - \frac{1}{16}\go^{2}\bigl(Y_{\phi}^{2} + Y_{\eta}^{2}\bigr)
      - \frac{3}{16}\,\gt^{2}
      \Big)
      \bigg)
  \nn &
  + T^{2}\,\clog\frac{1}{(4\pi)^{2}}\bigg(
    - \lambda_{3}^{2}
    + \frac{1}{2}\,\go^{2}\,\lambda_{3}^{ }\,Y_{\eta}^{2}
    - \frac{1}{16}\,\go^{4}\,Y_{\eta}^{2}\Ys^{2}
    \bigg)
%  \nn &
%  + \epsilon^{-1}T^{2}\frac{1}{(4\pi)^{2}}\bigg(
%     \frac{1}{4}\,\lambda_{3}^{2}
%    - \frac{1}{8}\,\go^{2}\,\lambda_{3}^{ }\,Y_{\eta}^{2}
%    + \frac{1}{64}\,\go^{4}\,Y_{\eta}^{2}\Ys^{2}
%  \bigg)
  \;, \\
\lambda_{1,3} &=
  \Big( \lambda_{1,3} \Big)_{\rmii{SM}}
  + T \bigg[ - \frac{L_b}{(4\pi)^2} \frac{1}{2} \lambda_{3}^2 \bigg]
\;,
\end{align}
where
$\Nc=3$ is the number of colours.
The momentum-dependent parts of the renormalised 2-point correlation functions
yield
\begin{align}
\hat{\Pi}'_{A_{0}^{a}A_{0}^{b}} &=
\frac{\gt^{2}}{(4\pi)^2} \bigg(
   3 
  + \frac{(\Nc+1)}{3}\nf(L_f-1)
  + \Big(\xi_2 - \frac{25}{6} \Big) L_b
  - 2 \xi_2 \bigg)
    \;, \\
\hat{\Pi}'_{A_{r}^{a}A_{s}^{b}} &=
\frac{\gt^{2}}{(4\pi)^2} \bigg(
  - \frac{2}{3}
  + \frac{(\Nc+1)}{3} \nf L_f
  + \Big(\xi_2 - \frac{25}{6} \Big) L_b \bigg)
    \;, \\
\hat{\Pi}'_{B_{0}B_{0}} &=
\frac{\go^{2}}{(4\pi)^2} \frac{1}{6}\Big(
    \Big[\Ys^{2} +\frac{1}{2}Y_{\eta}^{2}\Big](L_{b}+2)
    + Y_{\rmi{2f}}^{ }\,\nf^{ }(L_f-1)
  \Big)
  \;, \\
\hat{\Pi}'_{B_{r}B_{s}} &=
\frac{\go^{2}}{(4\pi)^2}\frac{1}{6}\Big(
       \Big[\Ys^{2} +\frac{1}{2}Y_{\eta}^{2}\Big]L_b
      + Y_{\rmi{2f}}^{ }\,\nf^{ } L_f
  \Big)
  \;, \\
\label{eq:phi:2pt:1l:1d}
\hat{\Pi}'_{\phi^\dagger \phi} &=
    \frac{1}{(4\pi)^2} \bigg(
      - \frac{L_b}{4} \Big( 3(3-\xi_2)\gt^{2} + (3-\xi_1)\go^{2}\Ys^{2}\Big)
      + \Nc L_f \gY^{2} \bigg)
    \;, \\
\hat{\Pi}'_{\eta^{\dagger}\eta} &=
    \frac{1}{(4\pi)^2} \bigg(
      - \frac{L_b}{4} (3-\xi_1)\go^{2}Y_{\eta}^{2}
      + L_f |y|^{2} \bigg)
\;,
\end{align}
where
$\xi_1$ is the ${\rm U(1)}_\rmii{Y}$ and
$\xi_2$ is the ${\rm SU(2)}_\rmii{L}$
gauge-fixing parameter.
The remaining matching relations for
the thermal mass of the complex
${\rm SU(2)}_\rmii{L}$ singlet and
its quartic couplings
take the form
\begin{align}
\label{eq:m:eta:2l}
\mu_{\eta,3}^{2} &=
  \mu^2_{\eta}
  + T^2 \Big(
      \frac{1}{6} \lambda_{3}
    + \frac{1}{3}\lambda_{2}
    + \frac{Y_\eta^{2}}{16} \go^2
    + \frac{1}{12}|y|^2
  \Big)
  \nn &
  + \frac{1}{(4\pi)^2} \Big(
      L_f (2\mu_\chi^2 - \mu_{\eta}^{2}) |y|^2
    - L_b\Big(2\mu^2_\phi \lambda_{3}
      + \mu_{\eta}^{2}\Big[
          4 \lambda_{2}
        - \frac{3}{4}\go^{2}Y_{\eta}^{2}
      \Big]\Big)
    \Big)
    \nn &
    + T^{2}\frac{1}{(4\pi)^{2}}\bigg(
      - \frac{\go^{4}}{72} Y_{\eta}^{2}\Bigl(
          Y_{\eta}^{2}
        - \frac{5}{2}\,Y_{\phi}^{2}
        - \frac{1}{4}\,\nf^{ }\,Y_{\rmi{2f}}^{ }
      \Bigr)
      + \frac{1}{6}\,\go^{2}\Bigl(
          \lambda_{2}^{ }\,Y_{\eta}^{2}
        + \frac{1}{2}\,\lambda_{3}^{ }\,Y_{\phi}^{2}
      \Bigr)
    \nn &
    \hphantom{{}+ T^{2}\frac{1}{(4\pi)^{2}}\bigg(}
      + \frac{1}{4}\,\gt^{2}\,\lambda_{3}^{ }
      - \frac{1}{48}|y|^{2}\,\go^{2}\,\Ye^{2}
    \bigg)
    \nn &
    + T^{2}L_{b}\frac{1}{(4\pi)^{2}}\bigg(
      - \frac{\go^{4}}{96} Y_{\eta}^{2}\Big(
          \frac{13}{2}\,Y_{\eta}^{2}
        + 7 Y_{\phi}^{2}
        + \frac{3}{2}\,\nf^{ }Y_{\rmi{2f}}^{ }
      \Big)
    \nn &
    \hphantom{{}+ T^{2}L_{b}\frac{1}{(4\pi)^{2}}\bigg(}
      + \lambda_{3}\Big(
        - \frac{2}{3}\bigl(\lambda_{2}^{ } + \lambda_{3}^{ }\bigr)
        - \lambda_{1}
        + \frac{1}{8}\,\go^{2}\bigl(Y_{\eta}^{2} + Y_{\phi}^{2}\bigr)
        + \frac{3}{8}\,\gt^{2}
        - \frac{\Nc}{4} \gY^{2}
      \Big)
    \nn &
    \hphantom{{}+ T^{2}L_{b}\frac{1}{(4\pi)^{2}}\bigg(}
      - \lambda_{2}^{ }\Bigl(
          \frac{10}{3} \lambda_{2}^{ }
        - \frac{1}{2} \go^{2}\,Y_{\eta}^{2}
        + \frac{1}{2} |y|^{2}
      \Bigr)
      + \frac{1}{12}\,|y|^{4}
      + \frac{1}{96}\bigl(6Y_{\eta}^{2}+\Ye^{2}\bigr)|y|^{2}\,\go^{2}
      \bigg)
    \nn &
    + T^{2}L_{f}\frac{1}{(4\pi)^{2}}\bigg(
        \frac{1}{192}\,\nf^{ }\,\go^{4}\,Y_{\eta}^{2}Y_{\rmi{2f}}^{ }
      - \frac{1}{96}\bigl(6Y_{\eta}^{2}-5\Ye^{2}\bigr)|y|^{2}\,\go^{2}
    \nn &
    \hphantom{{}+ T^{2}L_{f}\frac{1}{(4\pi)^{2}}\bigg(}
      + \frac{\Nc}{12} \gY^{2}\,\lambda_{3}^{ }
      - \frac{1}{6} |y|^{2}\bigl(\lambda_{2}^{ } + \lambda_{3}^{ }\bigr)
      + \frac{1}{12}\,|y|^{4}
    \bigg)
    \nn &
    + T^{2}\,\clog\frac{1}{(4\pi)^{2}}\bigg(
      - \frac{3}{8}\,\go^{4}\,Y_{\eta}^{4}
      - \frac{1}{8}\,\go^{4}\,Y_{\eta}^{2}\,Y_{\phi}^{2}
      + \go^{2}\bigl(
          \lambda_{3}^{ }\,Y_{\phi}^{2}
        + 2 \lambda_{2}^{ }\,Y_{\eta}^{2}
      \bigr)
    \nn &
    \hphantom{{}+ T^{2}\,\clog\frac{1}{(4\pi)^{2}}\bigg(}
      - 2 \lambda_{3}^{2}
      - 8 \lambda_{2}^{2}
      + 3 \gt^{2}\,\lambda_{3}^{ }
    \bigg)
%    \nn &
%    - \epsilon^{-1}T^{2}\frac{1}{(4\pi)^{2}}\frac{1}{4}\bigg(
%      - \frac{3}{8}\,\go^{4}\,Y_{\eta}^{4}
%      - \frac{1}{8}\,\go^{4}\,Y_{\eta}^{2}\,Y_{\phi}^{2}
%      + \go^{2}\bigl(
%          \lambda_{3}^{ }\,Y_{\phi}^{2}
%        + 2 \lambda_{2}^{ }\,Y_{\eta}^{2}
%      \bigr)
%    \nn &
%    \hphantom{{}- \epsilon^{-1}T^{2}\frac{1}{(4\pi)^{2}}\frac{1}{4}\bigg(}
%      - 2 \lambda_{3}^{2}
%      - 8 \lambda_{2}^{2}
%      + 3 \gt^{2}\,\lambda_{3}^{ }
%    \bigg)
    \;, \\
\lambda_{2,3} &= T \bigg[ \lambda_2 + \frac{1}{(4\pi)^2} \bigg(
      \frac{Y_\eta^{4}}{8} \go^4
    + L_f \Big( |y|^2 - 2\lambda_{2} \Big)|y|^{2}
    \nn &
    \hphantom{{}= T \bigg[ \lambda_{2} + \frac{1}{(4\pi)^2} \bigg( \frac{Y_\eta^{4}}{8} \go^4}
    - L_b \Big(
        \lambda_{3}^2
      + 10 \lambda_{2}^2
      + \frac{3}{16} Y_\eta^{4} \go^4
      - \frac{3}{2} Y_\eta^{2} \go^2 \lambda_{2}^{ }
      \Big)
    \bigg)
    \bigg]
    \;, \\
\lambda_{3,3} &= T \bigg[ \lambda_{3} + \frac{1}{(4\pi)^2} \bigg(
    \frac{Y_\eta^{2}\Ys^{2}}{4} \go^4
  - L_b \Big(
      2 \lambda_{3}^2
    + 4 \lambda_{3} \lambda_{2}
    + \frac{3}{8} Y_\eta^{2}Y_{\phi}^{2} \go^4
    + 6 \lambda_{1} \lambda_{3}
  \nonumber \\ &
  \hphantom{=T\bigg[\lambda_{3}+\frac{1}{(4\pi)^2}\bigg(}
    - \frac{3}{4}\go^{2} \lambda_{3}(\Ys^2+Y_{\eta}^2)
    - \frac{9}{4}\gt^{2} \lambda_{3}
  \Big) 
  - L_f \lambda_{3}\big(\Nc\gY^2 + |y|^2\big)
  \bigg)
  \bigg]
  \;, \\
h_{3}'&=\frac{\go^2 T}{2}\bigg[
  \Ys^{2}
  + \frac{1}{(4\pi)^2}\bigg(
      \frac{3g^2}{2}\Ys^{2}
      - \frac{1}{12}\Big(
        2 (L_b-1)\Ys^{2}
        + (L_b+2)Y_{\eta}^{2}
        + 2(L_f-1)Y_{\rmi{2f}}^{ }\,\nf^{ }
      \Big)\go^2 \Ys^{2}
  \nn &\hp{{}=\frac{\go^2 T}{4}\bigg[\Ys^{2}}
      - 2(\Yq^{2}+\Yu^{2})\Nc\gY^{2}
      + 12\lambda_{1}^{ }\Ys^{2}
      + 2\lambda_{3}^{ }Y_{\eta}^{2}
    \bigg)
  \bigg]
\;,\\
\rho_{3}' &= \frac{\go^2 T}{2} \bigg[
  Y_{\eta}^{2}
  + \frac{1}{(4\pi)^2}\bigg(
      \frac{1}{12}\Big(
        2 (L_b+2)\Ys^{2}
        + (L_b-4)Y_{\eta}^{2}
        - 2(L_f-1)Y_{\rmi{2f}}^{ }\,\nf^{ }
      \Big)\go^{2} Y_{\eta}^{2}
  \nn &\hphantom{{}=\frac{\go^2 T}{2}\bigg[Y_{\eta}^{2}}
      + \big(
          (\Ye^{2}-Y_{\eta}^{2})L_f
        - 2\Ye^{2}
      \big)|y|^2
      + 8\lambda_{2}^{ }Y_{\eta}^{2}
      + 4\lambda_{3}^{ }\Ys^{2}
    \bigg)
  \bigg]
\;,
\end{align}
for which we abbreviate recurring sums as
\begin{align}
  \sum_{f} \Yf^{4} \equiv
  Y_{\rmi{4f}}^{ } &=
    \Bigl[ (\Ye^{4} + 2\Yl^{4}) + \Nc(\Yu^{4} + \Yd^{4} + 2\Yq^{4})\Bigr]
  = \frac{2}{81}(729 + 137\Nc)
  = \frac{760}{27}
  \;,\\
  \sum_{f} \Yf^{2} \equiv
  Y_{\rmi{2f}}^{ } &=
    \Bigl[ (\Ye^{2} + 2\Yl^{2}) + \Nc(\Yu^{2} + \Yd^{2} + 2\Yq^{2})\Bigr]
  = \frac{2}{9}(27 + 11\Nc)
  = \frac{40}{3}
  \;,
\end{align}
with the corresponding hypercharges
$\Ye,\Yl,\Yu,\Yd,\Yq$ collected in~\cite{Brauner:2016fla}.

The novel ultrasoft matching relations besides from
the SM~\cite{Brauner:2016fla}
are
\begin{align}
\bar{\mu}_{\eta,3}^{2} &=
    \mu_{\eta,3}^{2}
  - \frac{1}{8\pi}
    \rho_{3}' \mDi{\rmii{1}}^{ }
\;,\\
\bar{\lambda}_{2,3}^{ } &=
    \lambda_{2,3}^{ }
  - \frac{1}{32\pi}
    \frac{\rho_{3}'^{2}}{\mDi{\rmii{1}}^{ }}
\;,\\
\bar{\lambda}_{3,3}^{ } &=
    \lambda_{3,3}^{ }
  - \frac{1}{16\pi}
    \frac{h_{3}'\rho_{3}'}{\mDi{\rmii{1}}^{ }}
\;.
\end{align}

%%%%%%%%%%%%%%%%%%%%%%%%%%%%% SUBSUBSECTION %%%%%%%%%%%%%%%%%%%%%%%%%%%%%%%%%%%%
%
\subsubsection*{One-loop thermal functions for the Majorana fermion}

The assumption of high-temperature expansion is relaxed for
the Majorana fermion mass parameter,
at one-loop level.
A full two-loop treatment
is relegated to future work
and would be required for full NLO dimensional reduction. 

The one-loop fermionic master integrals can be recast into
a vacuum part and
a thermal integral $\widetilde{Z}^{\T}$
that can be evaluated numerically
\begin{align}
\widetilde{Z}_{s;i}^{\alpha} &= 
  \Tint{\{P\}} \frac{p_n^\alpha}{[P^2 + m_{i}^{2}]^{s}}
  \;, 
  &
\widetilde{Z}_{s;i}^{0} = 
\widetilde{Z}_{s;i}^{ } &= 
  \Tint{\{P\}} \frac{1}{[P^2 + m_{i}^{2}]^{s}} =
    I_{s;i}^{4}
  + \widetilde{Z}_{s;i}^{\T}
  \;,
\end{align}
where
the four-momenta
$P\equiv(p_n^{ },\vec{p})$ and 
$p_n$ is a Matsubara frequency.
The curly brackets indicate the fermionic nature of the thermal sums and
that $p_n$ is fermionic, i.e.\
$p_n = (2n+1)\pi T$ with $n\in\mathbb{Z}$.
The $d$-dimensional integral measure is
\begin{equation}
\label{eq:measure}
  \int_{\vec{p}} \equiv \int \frac{{\rm d}^d \vec{p}}{(2\pi)^d} =
  \frac{2}{(4\pi)^2\Gamma(\frac{d}{2})} \int_{0}^{\infty} {\rm d}p\,p^{d-1}
  \;.
\end{equation}
The term $I_{s;i}^{4}$ is the 4d vacuum integral
in $d\to d+1 =4-2\epsilon$
\begin{equation}
  I_{s;i}^{d} \equiv
  \Bigl(\frac{\LamD^{2}e^\gammaE}{4\pi}\Bigr)^{\epsilon}
  \int_{\vec{p}}
  \frac{1}{(P^2+m_{i}^{2})^\alpha} =
  \Bigl(\frac{\LamD^{2}e^\gammaE}{4\pi}\Bigr)^{\epsilon}
  \frac{[m_{i}^{2}]^{\frac{d}{2}-s}}{(4\pi)^\frac{d}{2}}
  \frac{\Gamma\bigl(s-\frac{d}{2}\bigr)}{\Gamma(s)}
  \;.
\end{equation}

The high-temperature expansion ($m_i \ll T$) of
the master integrals
that contribute to the fermionic sector of the matching relations
at one-loop level
in $d=3-2\epsilon$ is
\begin{eqnarray}
  \widetilde{Z}_{1;i}^{ } &=&
  I_{1;i}^{4}
+ \widetilde{Z}_{1;i}^{\T} 
  \nn &=&
  - \frac{m_{i}^{2}}{(4\pi)^2}\Bigl[
      \frac{1}{\epsilon}
    + \ln\Bigl(\frac{\LamD^2}{m_{i}^{2}}\Bigr)
    + 1
    \Bigr]
    + \widetilde{Z}_{1;i}^{\T}
    + \mathcal{O}(\epsilon^{ })
  \nn &\stackrel{m_i \ll T}{=}&
  - \frac{T^2}{24}
  - \frac{m_{i}^2}{(4\pi)^2}\Bigl[\frac{1}{\epsilon} + L_f^{ }\Bigr] 
  + \mathcal{O}\Bigl(\frac{m_{i}^4}{T^2},\epsilon^{ }\Bigr)
  \;,\\[2mm]
  \widetilde{Z}_{2;i}^{ } &=&
  I_{2;i}^{4}
+ \widetilde{Z}_{2;i}^{\T} 
  \nn &=&
    \frac{1}{(4\pi)^2}\Bigl[
      \frac{1}{\epsilon}
    + \ln\Bigl(\frac{\LamD^2}{m_{i}^{2}}\Bigr)
    \Bigr]
    + \widetilde{Z}_{2;i}^{\T}
    + \mathcal{O}(\epsilon^{ })
  \nn &\stackrel{m_i \ll T}{=}&
    \frac{1}{(4\pi)^2}\Bigl[\frac{1}{\epsilon} + L_f^{ }\Bigr]
  + \mathcal{O}\Bigl(\frac{m_{i}^2}{T^2},\epsilon^{ }\Bigr)
  \;,\\[2mm]
  \widetilde{Z}_{1;i}^{2} &=& 
    - \frac{m_{i}^{2}}{d+1} I_{1;i}^{4}
    - \widetilde{Z}_{1;i}^{2,\T}
    - m_{i}^{2}\widetilde{Z}_{1;i}^{\T}
    \nn &=&
    \frac{m_{i}^{4}}{4(4\pi)^2}\Bigl[
      \frac{1}{\epsilon}
    + \ln\Bigl(\frac{\LamD^2}{m_{i}^{2}}\Bigr)
    + \frac{3}{2}
    \Bigr]
    - \widetilde{Z}_{1;i}^{2,\T}
    + \mathcal{O}(\epsilon^{ })
  \nn &\stackrel{m_i \ll T}{=}&
    \frac{7(4\pi)^2}{3840}T^4
  - \frac{T^2 m_i^{2}}{48}
  + \frac{m_i^{4}}{4(4\pi)^2}\Bigl[
      \frac{1}{\epsilon}
      + L_f^{ }
      + 2
    \Bigr]
  + \mathcal{O}\Bigl(\frac{m_{i}^6}{T^2},\epsilon^{ }\Bigr)
  \;,
\end{eqnarray}
where
the explicit fermionic thermal integrals up to
$\mathcal{O}(\epsilon^0)$ are
\begin{align}
\widetilde{Z}_{1;i}^{\T} &=
  - \int_{\vec{p}} \frac{\nF(E_{p,i})}{E_{p,i}}
  \;, \\
\widetilde{Z}_{2;i}^{\T} &=
  - \frac{1}{2}\int_{\vec{p}} \frac{\nF(E_{p,i})}{p^2 E_{p,i}}
  \;, \\
\widetilde{Z}_{1;i}^{2,\T} &=
  - \int_{\vec{p}} \frac{p^2\nF(E_{p,i})}{E_{p,i}}
  \;,
\end{align}
with
$E_{p,i} \equiv \sqrt{p^2 + m_{i}^2}$,
$m_{i}$ a general mass, and
the Fermi distribution function
$\nF(E) \equiv 1/[\mbox{exp}(E/T) +1]$. 

Below we give the corresponding Majorana fermionic sector of
the dimensional reduction matching relations.
Therein the full one-loop dependence on the thermal integrals
for the Majorana fermion
$\widetilde{Z}_{s;\chi}^{\alpha,\T}$ is installed:
\begin{align}
\hat{\Pi}'_{\eta^{\dagger}\eta} &\supset
  -|y|^{2}
  \Bigl[
      \frac{2}{\mu_{\chi}^{2}}\widetilde{Z}_{1;\chi}^{\T}
    + \frac{4}{3}\frac{1}{\mu_{\chi}^{4}}\Bigl(
          \widetilde{Z}_{1;\chi}^{2,\T}
        - \widetilde{Z}_{1;0}^{2,\T}
    \Bigr)
    - \frac{1}{(4\pi)^2}\Bigl(
        \ln\Bigl(\frac{\LamD^2}{\mu_{\chi}^{2}}\Bigr)
      + \frac{1}{2}
    \Bigr)
  \Bigr]
\;,\\
\label{eq:m:eta:fer}
\mu_{\eta,3}^{2} &\supset
  - |y|^{2}\Bigl[
    2\widetilde{Z}_{1;\chi}^{\T}\Bigl(
      1
      - \frac{\mu_{\eta}^{2}}{\mu_{\chi}^{2}}
    \Bigr)
    - \frac{4}{3}\frac{\mu_{\eta}^{2}}{\mu_{\chi}^{4}}\Big(
        \widetilde{Z}_{1;\chi}^{2,\T}
      - \widetilde{Z}_{1;0}^{2,\T}
    \Bigr)
    \nn &
    \hphantom{{}\supset
  - |y|^{2}\Bigl[
    2\widetilde{Z}_{1;\chi}^{\T}\Bigl(
      1
      - \frac{\mu_{\eta}^{2}}{\mu_{\chi}^{2}}
    \Bigr)}
    - \frac{\mu_{\chi}^{2}}{(4\pi)^2}\Bigl(
      \ln\Bigl(\frac{\LamD^2}{\mu_{\chi}^{2}}\Bigr)\Bigl(
       2
      - \frac{\mu_{\eta}^{2}}{\mu_{\chi}^{2}}
      \Bigr)
    + 2
    - \frac{1}{2}\frac{\mu_{\eta}^{2}}{\mu_{\chi}^{2}}
    \Bigr)
  \Bigr]
\;,\\
\lambda_{2,3}
  &\supset
    T |y|^{4}\Bigl[
    \widetilde{Z}_{2;\chi}^{\T}
    + \frac{1}{(4\pi)^2}
      \ln\Bigl(\frac{\LamD^2}{\mu_{\chi}^{2}}\Bigr)
  \Bigr]
  \nn &
  + 2 T |y|^{2}\lambda_{2}^{ }\Bigl[
      \frac{2}{\mu_{\chi}^{2}}\widetilde{Z}_{1;\chi}^{\T}
    + \frac{4}{3}\frac{1}{\mu_{\chi}^{4}}\Bigl(
      \widetilde{Z}_{1;\chi}^{2,\T}
    - \widetilde{Z}_{1;0}^{2,\T}
    \Bigr)
    - \frac{1}{(4\pi)^2}\Bigl(
        \ln\Bigl(\frac{\LamD^2}{\mu_{\chi}^{2}}\Bigr)
      + \frac{1}{2}
    \Bigr)
  \Bigr]
  \;,\\
\lambda_{3,3} &\supset
    T |y|^{2}\lambda_{3}^{ }\Bigl[
      \frac{2}{\mu_{\chi}^{2}}\widetilde{Z}_{1;\chi}^{\T}
    + \frac{4}{3}\frac{1}{\mu_{\chi}^{4}}\Bigl(
      \widetilde{Z}_{1;\chi}^{2,\T}
    - \widetilde{Z}_{1;0}^{2,\T}
    \Bigr)
    - \frac{1}{(4\pi)^2}\Bigl(
        \ln\Bigl(\frac{\LamD^2}{\mu_{\chi}^{2}}\Bigr)
      + \frac{1}{2}
    \Bigr)
  \Bigr]
  \;,\\
\rho_{3}' &\supset
  -\frac{\go^2 |y|^{2} T}{2} \bigg[
    (\Ye^{2}-Y_{\eta}^{2})
    \Bigl[
      \frac{2}{\mu_{\chi}^{2}}\widetilde{Z}_{1;\chi}^{\T}
    + \frac{4}{3}\frac{1}{\mu_{\chi}^{4}}\Bigl(
      \widetilde{Z}_{1;\chi}^{2,\T}
    - \widetilde{Z}_{1;0}^{2,\T}
    \Bigr)
    - \frac{1}{(4\pi)^2}\Bigl(
        \ln\Bigl(\frac{\LamD^2}{\mu_{\chi}^{2}}\Bigr)
      + \frac{1}{2}
    \Bigr)
  \Bigr]
  \nn &
  \hphantom{{}\supset\frac{\go^2 |y|^{2} T}{2} \bigg[}
  + \frac{2\Ye^{2}}{(4\pi)^2}
  \bigg]
\;.
\end{align}
Here,
$\widetilde{Z}_{s;0}^{\alpha,\T}$ is
the zero-mass version of the corresponding thermal integral.
In the limit $\mu_{\chi} \ll T$, we recover 
the matching relations as stated in the beginning of
appendix~\ref{sec:eft:3d}.
One subtlety is
the vanishing 
$\widetilde{Z}_{1;\chi}^{\T}$ term in eq.~\eqref{eq:m:eta:fer}
in the limit $\mu_\chi \to \mu_\eta$.
In that case, 
the high-temperature LO term $\frac{T^2|y|^{2}}{12}$ is produced from
the momentum-dependent part of the $\eta$ correlator.
This is a NLO contribution.
Since LO and NLO are then of the same order,
$\mu_\chi$ has to be at least soft or parametrically larger than 
a soft $\mu_\eta$ for a high-temperature expansion to be valid.

For fig.~\ref{fig:EWPT-y-Mchi},
we verified that there is no qualitative difference for the
$v_{3,{\rm c}}/\sqrt{T_{{\rm c},\phi}}$ contours
whether
using dimensional reduction matching relations with
a generic $\mu_{\chi}$ or
a soft $\mu_{\chi}\sim gT$ with high-temperature expansion.
However,
using the high-temperature expansion could compromise the accuracy for
large physical $M_\chi$
when determining the phase transition thermodynamics --
such as the phase transition
strength and
inverse duration (cf.\ eg.~\cite{Croon:2020cgk}). 
We leave such an investigation for future work.

%%%%%%%%%%%%%%%%%%%%%%%%%%%%% SUBSUBSECTION %%%%%%%%%%%%%%%%%%%%%%%%%%%%%%%%%%%%
%
\subsubsection*{Thermal effective potential within 3d EFT}

The effective potential at one-loop order within the 3d EFT reads
\begin{align}
V^{\rmi{3d}}_{\rmi{eff}} &=
    V^{ }_{\rmi{tree}}
  + V^{ }_{\rmi{1-loop}}
  \;.  
\end{align}
At tree-level
\begin{align}
  V_{\rmi{tree}} &=
      \frac{1}{2} \mu^2_{\phi,3} v^2_3
    + \frac{1}{2} \mu^2_{\eta,3} x^2_3
    + \frac{1}{4} \lambda_{1,3}^{ } v^4_3
    + \frac{1}{4} \lambda_{2,3}^{ } x^4_3
    + \frac{1}{4} \lambda_{3,3}^{ } v^2_3 x^2_3
    \;.
\end{align}
At one-loop level,
the effective potential 
can be written in terms of the master integral 
in
general dimensions and
explicit $d=3-2\epsilon$ dimensions
\begin{align}
\label{eq:1loop-master-d}
J_{d}(m^2) &\equiv \frac{1}{2} \int_{\vec{p}} \ln(p^2 + m^2) =
  -\frac{1}{2}
  \Big( \frac{\Lamd^2 e^\gammaE}{4\pi} \Big)^\epsilon
  \frac{[m^2]^\frac{d}{2}}{(4\pi)^{\frac{d}{2}}} \frac{\Gamma(-\frac{d}{2})}{\Gamma(1)}
  \;,\\
\label{eq:1loop-master-3d}
J_{3}(m^2) &=
    - \frac{[m^2]^{\frac{3}{2}}}{12 \pi}
    + \mathcal{O}(\epsilon)
  \;.
\end{align}
In Landau gauge,
\begin{align}
V_{\rmi{1-loop}} &= (d-1) \Big(
  2 J_3^{ }(m^2_{\rmii{$W$},3})
  + J_3^{ }(M^2_{+,3})
  + J_3^{ }(M^2_{-,3}) \Big)
  \nn &
  + 3 J_3^{ }(m^2_{\rmii{$G$},3}) 
  + J_3^{ }(m^2_{\rmii{$A$},3})
  + J_3^{ }(m^2_{+,3})
  + J_3^{ }(m^2_{-,3})
  \;,
\end{align}
where
an additional subscript highlights that
the mass eigenvalues are functions of the 3d EFT parameters.
Since the integral $J_3$ is UV-finite,
one can directly set $d\to 3$.
Despite computing the effective potential
at one-loop level within the 3d EFT, it still includes all
NLO hard thermal contributions via dimensional reduction, such as
two-loop thermal masses~\cite{Schicho:2022wty}.

%%%%%%%%%%%%%%%%%%%%%%%%%%%%% SECTION %%%%%%%%%%%%%%%%%%%%%%%%%%%%%%%%%%%%
%
\section{Dark matter relic density}
\label{sec:app:DM}

This appendix details the cross section
$\sigma_{ij} v$ that enters eq.~\eqref{co_cross} of sec.~\ref{sec:dark_matter}, and
accounts for the annihilation processes that drive
the dark matter energy density in the freeze-out scenario.
Here $\vrel$ is the relative velocity of the annihilating particles~\cite{Gondolo:1990dk}.
In the following,
we display the leading terms in the velocity expansion and work up to
$\mathcal{O}(\vrel^2)$.
For the processes that feature a velocity-independent leading term,
we omit the (often lengthy) expressions of the sub-leading $\vrel^2$ contributions,
despite including them in the numerical calculation.
Since
$M_e/M_\chi \lesssim 10^{-2}$ for
$M_\chi>100$~GeV
the lepton mass is not included when calculating cross sections as
it induces minuscule corrections~\cite{Garny:2015wea}
for cross sections that comprise light fermion masses.
We further detail the extraction of
the Sommerfeld factors and bound-state effects for the $\eta$ particles.

%%%%%%%%%%%%%%%%%%%%%%%%%%%%% SUBSECTION %%%%%%%%%%%%%%%%%%%%%%%%%%%%%%%%%%%%
%
\subsection{Cross sections}

Three classes of processes contribute to the cross section
$\sigma_{ij} v$ in eq.~\eqref{co_cross};
the corresponding diagrams are collected in fig.~\ref{fig:xsec:2to2}.
\begin{figure}[t!]
\begin{align}
  \mathcal{M}_{\eqref{chi_chi_ann}} &=
      \VtxvT(\Lqu,\Lqum,\Lqum,\Luq,\Lscii)
    + \VtxvU(\Lqu,\Lqum,\Lqum,\Luq,\Lscii)
  \;,
  &&
  \nn[2mm]
  \mathcal{M}_{\eqref{chi_chi_l_g},\eqref{chi_eta_ann}} &=
    \VtxvT(\Lgli,\Lscii,\Lqum,\Luq,\Lscii)
  + \VtxvS(\Lgli,\Lscii,\Lqum,\Luq,\Lqu)
  \;,
  \nn[2mm]
  \mathcal{M}_{\eqref{eta_eta_ll}} &=
    \VtxvT(\Luq,\Lscii,\Lscii,\Luq,\Lqum)
  + \VtxvU(\Luq,\Lscii,\Lscii,\Luq,\Lqum)
  \;,
  &&
  \nn[2mm]
  \mathcal{M}_{\eqref{eta_etabar_l_lbar}} &=
    \VtxvT(\Luq,\Lscii,\Lcsii,\Lqu,\Lqum)
  + \VtxvS(\Luq,\Lscii,\Lcsii,\Lqu,\Lgli)
  + \VtxvS(\Luq,\Lscii,\Lcsii,\Lqu,\Lxx)
  \;,
  \nn[2mm]
  \mathcal{M}_{\eqref{eta_etabar_gg}} &=
    \Vtxvo(\Lgli,\Lscii,\Lcsii,\Lgli)
  + \VtxvT(\Lgli,\Lscii,\Lcsii,\Lgli,\Lscii)
  + \VtxvU(\Lgli,\Lscii,\Lcsii,\Lgli,\Lscii)
  \;,
  &&
  \nn[2mm]
  \mathcal{M}_{\eqref{eta_etabar_ZZ}} &=
    \Vtxvo(\Lgli,\Lscii,\Lcsii,\Lgli)
  + \VtxvT(\Lgli,\Lscii,\Lcsii,\Lgli,\Lscii)
  + \VtxvU(\Lgli,\Lscii,\Lcsii,\Lgli,\Lscii)
  + \VtxvS(\Lgli,\Lscii,\Lcsii,\Lgli,\Lxx)
  \;,
  &&
  \nn[2mm]
  \mathcal{M}_{\eqref{eta_etabar_WW}} &=
    \Vtxvo(\Lgli,\Lscii,\Lcsii,\Lgli)
  + \VtxvS(\Lgli,\Lscii,\Lcsii,\Lgli,\Lgli)
  + \VtxvS(\Lgli,\Lscii,\Lcsii,\Lgli,\Lxx)
  \;,
  &&
  \nn[2mm]
  \mathcal{M}_{\eqref{eta_etabar_Zg}} &=
    \Vtxvo(\Lxx,\Lscii,\Lcsii,\Lgli)
  \;,
  &&
  \nn[2mm]
  \mathcal{M}_{\eqref{eta_etabar_Zh}} &=
    \VtxvS(\Lxx,\Lscii,\Lcsii,\Lgli,\Lgli)
  \;,
  &&
  \nn[2mm]
  \mathcal{M}_{\eqref{eta_etabar_t_tbar}} &=
    \VtxvS(\Luq,\Lscii,\Lcsii,\Lqu,\Lgli)
  + \VtxvS(\Luq,\Lscii,\Lcsii,\Lqu,\Lxx)
  \;,
  &&
  \nn[2mm]
  \mathcal{M}_{\eqref{eta_etabar_hh}} &=
    \Vtxvo(\Lxx,\Lscii,\Lcsii,\Lxx)
  + \VtxvT(\Lxx,\Lscii,\Lcsii,\Lxx,\Lxx)
  + \VtxvU(\Lxx,\Lscii,\Lcsii,\Lxx,\Lxx)
  + \VtxvS(\Lxx,\Lscii,\Lcsii,\Lxx,\Lxx)
  \;.\nonumber
\end{align}
\caption{%
  $2\to 2$
  pair annihilation processes contributing to the cross sections
  in eqs.~\eqref{chi_chi_ann}--\eqref{eta_etabar_hh}.
  Crossed diagrams are not shown.
  The dark fermion $\chi$ is displayed by a double-solid line,
  the complex scalar $\eta$ by an arrowed double-dashed line,
  the SM lepton $e$ and top quark $t$ by arrowed solid lines,
  $W^\pm$ and $Z$ bosons by wiggly lines, and
  Higgs bosons by a dashed line.
  A sum over polarisations is kept implicit for $W^\pm$ and $Z$ bosons.
  }
\label{fig:xsec:2to2}
\end{figure}
Since gauge bosons are involved,
the diagrams are calculated in
a general covariant $R_\xi$ gauge, and
we explicitly verified gauge invariance.
Conjugate processes are not displayed in the following since they have
the same cross section.

For the Majorana fermion pair annihilation
merely one process contributes, namely
$\chi \chi \to \bar{e} e$.
We reproduce the known result in the literature and
up to $\mathcal{O}(\vrel^2)$
it reads~\cite{Garny:2015wea}
\begin{align}
\label{chi_chi_ann}
\sigma \vrel (\chi \chi \to e \bar{e}) &=
  \frac{|y|^4}{48 \pi}
  \frac{M_\chi^2 (M_\chi^4+M_\eta^4)}{(M_\chi^2+M_\eta^2)^4} \vrel^2
  \;.
\end{align}
Next,
one needs to find the co-annihilation processes, where
a Majorana fermion and a scalar antiparticle enter as incoming states.
The corresponding cross sections are 
\begin{align}
\label{chi_chi_l_g}
\sigma \vrel (\chi \eta^\dagger \to e \, \gamma)  &=
  \frac{|y|^2\go^2 c_w^2 Y_\eta^2}{64 \pi M_\eta(M_\eta+M_\chi)} 
  \;,\\[1mm]
\label{chi_eta_ann}
\sigma \vrel(\chi \eta^\dagger \to e \, Z) &=
  \frac{|y|^2\go^2 s_w^2 Y_\eta^2}{128 \pi}
  \sqrt{1-\frac{\MZ^2}{4 M_\eta^2}}
  \nn &\times
  \frac{2 M_\eta^2 (M_\eta+M_\chi)^4 + 2 \MZ^4 M_\eta^2 -\MZ^2(M_\eta+M_\chi)^2 (9 M_\eta^2+2 M_\eta M_\chi +M_\chi^2)}{M_\eta^3 (M_\eta+M_\chi)^3((M_\eta + M_\chi)^2-\MZ^2)}  
  \;.
\end{align}
Finally, we find the processes for
$\eta \eta$ and
$\eta \eta^\dagger$ annihilation processes. 
In the result for e.g.\
$\eta \eta^\dagger \to W^+ W^-$, we 
account for the sum of all polarisations.
We include the coupling between the Higgs boson with
the top quark,
whereas we neglect the contribution from the other SM fermions due to
the much smaller Yukawa couplings.
The resulting cross sections read
\begin{align}
\label{eta_eta_ll}
\sigma \vrel(\eta \eta \to \bar{e} \, \bar{e}) &=
  \frac{|y|^4}{6 \pi} \frac{M_\chi^2}{(M_\chi+M_\eta)^4}
  \;,\\[1mm]
\label{eta_etabar_l_lbar} 
\sigma \vrel(\eta \eta^\dagger \to e \bar{e} ) &= \frac{\vrel^2}{48 \pi M_\eta^2}
  \biggl[
    \frac{\go^2 Y_\eta Y_e}{8} \biggl(
        c_w^2
      - \frac{4 \, s_w^2  M_\eta^2}{4 M_\eta^2-\MZ^2}
    \biggr)
    + \frac{|y|^2 M_\eta^2}{(M_\eta^2+M_\chi^2)}
  \biggr]^2
  \;,\\[1mm]
\label{eta_etabar_gg}
\sigma \vrel (\eta \eta^\dagger \to \gamma \gamma ) &=
  \frac{\go^4 Y_\eta^4 c_w^4}{128 \pi M_\eta^2}
  \;,\\[1mm]
\label{eta_etabar_ZZ}
\sigma \vrel (\eta \eta^\dagger \to ZZ) &=
  \frac{1}{64 \pi M_\eta^2}
  \biggl[ \frac{\go^4Y_\eta^4 s_w^4}{2}
  + \lambda_3^2 \biggl( 1+ \frac{2 M_\phi^2}{4 M_\eta^2-M^2_\phi}\biggr)^2 \biggr] 
  \sqrt{1-\frac{\MZ^2}{M_\eta^2}}
  \;,\\[1mm]
\label{eta_etabar_WW}
\sigma \vrel (\eta \eta^\dagger \to W^+ W^-) &=
  \frac{ \lambda_3^2}{32 \pi M_\eta^2}
  \biggl( 1+ \frac{2 M_\phi^2}{4 M_\eta^2-M^2_\phi}\biggr)^2
  \sqrt{1-\frac{\MW^2}{M_\eta^2}}
  \;,\\[1mm]
\label{eta_etabar_Zg}
\sigma \vrel (\eta \eta^\dagger \to Z \gamma)  &= 
  \frac{c_w^2 s_w^2 \go^4 Y_\eta^4}{32 \pi M_\eta^2}
  \biggl( 1- \frac{\MZ^2}{4 M_\eta^2} \biggr)
  \;,\\[1mm]
\label{eta_etabar_Zh}
\sigma \vrel (\eta \eta^\dagger \to Z \phi)  &= 
  \frac{\go^4 Y_\eta^2}{12288 \pi M_\eta^2}
  \sqrt{\left(1- \frac{(M_\phi-\MZ)^2}{4 M_\eta^2}\right)
  \left(1- \frac{(M_\phi+\MZ)^2}{4 M_\eta^2}\right)}
  \\ &\times
  \frac{(4 M_\eta^2-\MZ^2)^3-4 M_\eta^2 M_\phi^2 (8 M_\eta^2-M_\phi^2)+M_\phi^2 \MZ^2 (2 \MZ^2-M_\phi^2)}{(4 M_\eta^2-\MZ^2)^3}
  \;,\nn[1mm]
\label{eta_etabar_t_tbar}
\sigma \vrel (\eta\eta^\dagger \to t \bar{t}) &=
  \frac{3 \lambda_3^2}{128 \pi}
  \frac{M_t^2}{(4 M_\eta^2-M_\phi^2)^2}
  \biggl( 1- \frac{M_t^2}{M_\eta^2}\biggr)^{3/2}
  \;,\\[1mm]
\label{eta_etabar_hh}
\sigma \vrel(\eta\eta^\dagger \to \phi \phi ) &=
  \frac{ \lambda_3^2}{32 \pi M_\eta^2}
  \biggl( 1+ \frac{6 M_\phi^2}{4 M_\eta^2-M^2_\phi}\biggr)^2
  \biggl(1-\frac{M_\phi^2}{M_\eta^2}\biggr)^{1/2}
  \;,
\end{align}
where
the weak mixing angle or Weinberg angle at $T=0$ reads
\begin{equation}
\label{Weinberg_T0}
  \sin{ \left( 2 \theta_w \right)}  = \frac{\go\gt}{\go^2+\gt^2}
  \;,
\end{equation}
and is abbreviated via
$c_{w} \equiv \cos(\theta_w)$ and
$s_{w} \equiv \sin(\theta_w)$. 
In our work, we considered
$100~{\rm GeV} < M_\chi < M_\eta$ and
this makes including the decay width of the Higgs boson irrelevant in
the $s$-channel diagrams,
that would instead be necessary when $2 M_\eta \simeq M_\phi$;
see e.g.\ ref.~\cite{Cline:2013gha}. 

%%%%%%%%%%%%%%%%%%%%%%%%%%%%%%%%%%%%%%%%%%%%%%%%%%%%%%%%%%%%%%%%%%%%%%%%%%%%%%%
\subsection{Sommerfeld enhancement and bound-state effects}
\label{sec:app_DM_2}

This section collects
the main formulae and ingredients that
were used to estimate and include non-perturbative effects for the scalar annihilations.
At variance with the Majorana DM fermions,
the scalar particles interact with the SM gauge and Higgs bosons.

In the freeze-out scenario, the plasma temperature is much below
the mass scale of the annihilating states.
Hence, the scalar particles are moving slowly and they can undergo several interactions before
annihilating into light Standard Model particles.
Multiple exchanges of photons lead to
the Sommerfeld effect that increases (decreases) the annihilation rate for
an attractive (repulsive) potential experienced by
the heavy pair in an above-threshold scattering state~\cite{Sommerfeld:1931,Hisano:2004ds}.
The same interaction leads
to the formation of bound states, the below-threshold counterpart of
the Sommerfeld effect for negative energy two-particle states.
The formation of bound states and their decays into
light degrees of freedom (pairs of photons) open up an additional depletion channel for the scalar particle.
As a consequence, in the co-annihilation regime,
this can affect the overall DM relic density.

For obtaining the Sommerfeld factors,
the main ingredients are
the potentials experienced by the annihilating pairs.
In our case,
we are interested in the combinations
$\eta \eta$,
$\eta^\dagger \eta^\dagger$,
$\eta \eta^\dagger$.
A scalar-mediated potential exchange,
here due to the Higgs boson,
is always attractive.
Contrarily,
a vector-induced exchange,
here due to $Z$-boson and photon, is
repulsive for
$\eta \eta$ and
$\eta^\dagger \eta^\dagger$, and
attractive for
$\eta \eta^\dagger$.
When deriving
the static potentials between the heavy scalar pair,
we included
masses and the weak mixing angle
at finite-temperature.
The following relations have to be understood as
a phenomenological recipe to include
finite-temperature effects, and are strictly valid in
the Hard Thermal Loop (HTL) approximation of
the relevant self-energies~\cite{Kim:2016kxt,Biondini:2017ufr}.
By combining the tree-level effects from
the Higgs mechanism with
the finite-temperature self-energies,
one finds for
the Higgs thermal mass~\cite{Kim:2016kxt}
\begin{equation}
\label{eq:mphi:T}
   M_{\T,\phi}^{2} =
   2 \lambda_1^{ } v_\T^2
   \;,\quad
   v_\T^2  = \frac{1}{\lambda_1} \left[
      \frac{M_\phi^2}{2}
    - \frac{(\go^2 + 3\gt^2 + 8\lambda_1^{ } + 4\gY^2)}{8}
    \right]
  \;.
\end{equation}
The corresponding potential reads
(see also~\cite{Biondini:2018ovz}, and
\cite{Bollig:2021psb} for the $T=0$ limit)
\begin{equation}
  \mathcal{V}_\phi =
    - \frac{\lambda_3^2 v_{\T}^2}{16\pi M_\eta^2} \frac{e^{- M_{\T,\phi} r }}{r}
  \;.
\end{equation}

The vector potentials contain the Debye mass parameters
$\mDi{\rmii{1}}$ for the ${\rm U(1)}_\rmii{Y}$ and
$\mDi{\rmii{2}}$ for the ${\rm SU(2)}_\rmii{L}$ SM gauge group, introduced
in eqs.~\eqref{eq:mD1} and \eqref{eq:mD2}
of appendix~\ref{sec:model:match}.
Here,
using their one-loop expression suffices.
The neutral gauge mass parameters are~\cite{Ghiglieri:2016xye,Kim:2016kxt}
\begin{align}
  M_{\T,\rmii{$Z$}} &= M_{\T,+}
 \;, \quad
 M_{\T,\gamma} = M_{\T,-}
 \;,\nn
\label{eq:mZgamma:T}
M_{\T,\pm} &=
  \frac{1}{2} \biggl\{
      \MZ^2 
      + \mDi{\rmii{1}}^{2}
      + \mDi{\rmii{2}}^{2} \pm
      \sqrt{\sin^2{ \left( 2 \theta_w \right)} \MZ^4
      + \bigl( \cos{ \left( 2 \theta_w \right)} \MZ^2
        +  \mDi{\rmii{2}}^{2}
        - \mDi{\rmii{1}}^{2}  \bigr)^2 }
    \biggr\}
 \;.
\end{align}
In eqs.~\eqref{eq:mZgamma:T} and \eqref{Weinberg_T},
$\MZ$ has to be understood as
a temperature-dependent mass on its own, due to
the Higgs VEV in eq.~\eqref{eq:mphi:T}, and it reads
$\MZ=v_\T \sqrt{\go^2+\gt^2}/2$.
We do not introduce additional labelling to distinguish it from
the $T=0$ value in the main body.
The attractive static potential due to
the $Z$-boson and
the photon reads
(one can understand it just originating from the $B_\mu$ exchange)
\begin{equation}
  \mathcal{V}_{B}(r)= -\frac{g_1^2}{4 \pi} \left( \frac{Y_\eta}{2}\right)^2
  \left[
      \tilde{c}_{w}^{2} \frac{e^{- M_{\T,\gamma} r} }{r}
    + \tilde{s}_{w}^{2} \frac{e^{- M_{\T,\rmii{$Z$}} r} }{r}
    \right] \, ,
  \label{B_potential}
\end{equation}
where
$\tilde{c}_{w} = \cos(\tilde{\theta}_w)$ and
$\tilde{s}_{w} = \sin(\tilde{\theta}_w)$ as abbreviated
in tab.~\ref{table:potentials}, with
the finite-temperature mixing angle that reads
\begin{eqnarray}
\label{Weinberg_T}
   \sin{ ( 2 \tilde{\theta}_w )} = \frac{ \sin{ \left( 2 \theta_w \right)}}{\sqrt{\sin^2{ \left( 2 \theta_w \right)} \MZ^4
   + \left( \cos{ \left( 2 \theta_w \right)} \MZ^2 +  \mDi{\rmii{2}}^{2} - \mDi{\rmii{1}}^{2}  \right)^2 }}
   \,.
\end{eqnarray}
In the potential list
in tab.~\ref{table:potentials},
we split
the $Z$-boson and photon contributions in eq.~\eqref{B_potential}.

For the particle-particle annihilation, only one process contributes
in eq.~\eqref{eta_eta_ll},
and the thermally averaged cross section reads
(the same applies for the complex conjugate process)
\begin{eqnarray}
  \langle \sigma_{\eta \eta} \, \vrel \rangle =
  \sigma \vrel(\eta \eta \to \bar{e} \, \bar{e}) \langle
  \mathcal{S}_0 (\zeta_\gamma, \zeta_{\rmii{$Z$}}, \zeta_{\rmii{$H$}}) \rangle
  \;.
\end{eqnarray}
Here,
$\mathcal{S}_0(\zeta_\gamma, \zeta_{\rmii{$Z$}}, \zeta_{\rmii{$H$}})$
is the Sommerfeld factor as extracted by
the repulsive photon and $Z$-boson potentials and the attractive Higgs potential, where
$\zeta_X  \equiv \alpha_X /\vrel$.
The symbol
$\mathcal{S}_{l}$ is used for
the Sommerfeld factor of the $\eta \eta$
(and $\eta^\dagger \eta^\dagger$) pairs, in contrast with
$S_{l}$ that is reserved for the particle-antiparticle pair, where
$l$ is the orbital angular momentum of the relative motion.
The Sommerfeld factor is extracted according to the techniques detailed
in~\cite{Iengo:2009ni}, and
is thermally averaged according to~\cite{Gondolo:1990dk,Feng:2010zp}.
The same holds for the conjugate process
$\eta^\dagger \eta^\dagger \to e \, e$.
The impact of the Sommerfeld factor on this mixed attractive-repulsive channel is practically negligible,
resulting in
$\mathcal{S}_0(\zeta_\gamma, \zeta_{\rmii{$Z$}}, \zeta_{\rmii{$H$}}) \approx1$
for the relevant parameter space.

When considering the particle-antiparticle annihilations, the thermally averaged cross section can be written as~\cite{Ellis:2015vaa}
\begin{eqnarray}
\label{Cross_section_eff}
   \langle\sigma^{\rmi{eff}}_{\eta \eta^\dagger} \,\vrel^{ } \rangle  =
   \langle \sigma_{\eta \eta^\dagger} \, \vrel^{ } \rangle
   + \sum_{n} \langle \sigma^n_{\rmi{bsf}} \, \vrel^{ } \rangle \, \frac{\Gamma_{\rmi{ann}}^n}{\Gamma_{\rmi{ann}}^n+\Gamma_{\rmi{bsd}}^n}
   \;,
\end{eqnarray}
where
$\langle \sigma_{\eta \eta^\dagger} \, \vrel \rangle$
is understood as
the sum of all annihilation processes as listed in
eqs.~\eqref{eta_etabar_l_lbar}--\eqref{eta_etabar_hh} and
weighted by the corresponding
$s$- and $p$-wave Sommerfeld factors $S_0$ and $S_1$.
The thermal average for the annihilation cross section is implemented according to standard definitions;
see e.g.~\cite{Feng:2010zp,vonHarling:2014kha}.
The Sommerfeld factors
$S_0$ and
$S_1$ are computed according to the strategy in~\cite{Iengo:2009ni}.
For attractive potentials, these factors moderately enhance
the cross section and reduce the corresponding dark matter abundance
(cf.\ right panel in fig.~\ref{fig:thermal_masses}).
The effect of a finite thermal mass of the photon is minuscule, and
is therefore not included in estimating bound-state effects.

The additional, summed terms contain
the thermally averaged bound-state formation cross section
$\langle \sigma^n_{\rmi{bsf}} \, \vrel^{ } \rangle$,
the bound-state decay width
$\Gamma_{\rmi{ann}}^n$, and
the bound-state thermal width
$\Gamma_{\rmi{bsd}}^n$, the latter accounting for the dissociation process.
The combination of the decay width and dissociation width
$\Gamma_{\rmi{ann}}^n$/($\Gamma_{\rmi{ann}}^n + \Gamma_{\rmi{bsd}}^n)$,
determines the efficiency of DM annihilations via bound states.
One typically has to wait until the temperature, that sets the scale for
the energy of the light particles that hit the bound states,
is of the order of
the binding energy of the bound states or smaller.
In this regime, bound states are not ionised and can decay into lighter particles
(here a pair of photons) and deplete the number of dark matter particles.

The bound state formation cross section can be
obtained from~\cite{Petraki:2015hla} and
adapted to our coupling between the scalar $\eta$ and the photon.
Only the photon is sufficiently lighter than the scalar $\eta$ to induce
bound-state formation; see sec.~\ref{sec:dark_matter}.
We reiterate that here the photon is treated as massless, and no thermal mass is included.
This allows for using Coulombic wave functions that enter
the calculation of the relevant cross sections and widths.
For the ground-state formation
$| 1 0 0 \rangle \equiv | 1 S \rangle$
it reads
\begin{equation}
  \sigma^{1S}_{\rmi{bsf}} \, \vrel^{ } = \frac{\go^4 c_w^4 \, 2^6}{ 3 M_\eta^2} \frac{\zeta_\gamma^5}{(1+\zeta_\gamma^2)^2}  \frac{e^{- 4 \zeta_\gamma \arccot \zeta_\gamma}}{1-e^{- 2 \pi \zeta_\gamma}}
    \;.
\end{equation}
The dissociation rate
$\Gamma_{\rmi{bsd}}^n$ can be inferred from the dissociation cross section $\sigma_{\rmi{bsd}}$, obtained via
the Milne relation~\cite{vonHarling:2014kha}, or from
the self-energy of the bound state in
a potential non-relativistic effective theory~\cite{Biondini:2021ycj}.
Finally, the decay width into a pair of photons is
\begin{equation}
  \Gamma_{\rmi{ann}}^{1S} = \frac{M_\eta \go^5 c_w^5}{4}
  \;.
\end{equation}
When estimating bound-state effects,
we merely include the ground state
($n=1$) in the sum~\eqref{Cross_section_eff}.
As the leading effect,
it is often the choice adopted in former studies.
However, the effect of excited bound states
was recently investigated~\cite{Binder:2021vfo,Garny:2021qsr,Biondini:2021ycj}, and
goes beyond the scope of this work.

{\small
%%%%%%%%%%%%%%%%%%%%%%%%% BIBLIO %%%%%%%%%%%%%%%%%%%%%%%%%%%%%%%%%%%%%%%%%
%
%\bibliographystyle{utphys}
%\bibliography{ref}

\begin{thebibliography}{999}

\bibitem{Kuzmin:1985mm}
V.~A. Kuzmin, V.~A. Rubakov, and M.~E. Shaposhnikov, {\em{On the Anomalous
  Electroweak Baryon Number Nonconservation in the Early Universe},}
  \href{http://dx.doi.org/10.1016/0370-2693(85)91028-7}{Phys. Lett. B {\bf 155}
  (1985) 36}.

\bibitem{Shaposhnikov:1987tw}
M.~E. Shaposhnikov, {\em{Baryon Asymmetry of the Universe in Standard
  Electroweak Theory},}
  \href{http://dx.doi.org/10.1016/0550-3213(87)90127-1}{Nucl. Phys. B {\bf 287}
  (1987) 757}.

\bibitem{Morrissey:2012db}
D.~E. Morrissey and M.~J. Ramsey-Musolf, {\em{Electroweak baryogenesis},}
  \href{http://dx.doi.org/10.1088/1367-2630/14/12/125003}{New J. Phys. {\bf 14}
  (2012) 125003} [\href{http://arxiv.org/abs/1206.2942}{{\ttfamily
  1206.2942}}].

\bibitem{Ramsey-Musolf:2019lsf}
M.~J. Ramsey-Musolf, {\em{The electroweak phase transition: a collider
  target},} \href{http://dx.doi.org/10.1007/JHEP09(2020)179}{JHEP {\bf 09}
  (2020) 179} [\href{http://arxiv.org/abs/1912.07189}{{\ttfamily 1912.07189}}].

\bibitem{LISA:2017pwj}
{\bfseries LISA} Collaboration, P.~Amaro-Seoane {\em et~al.}, {\em{Laser
  Interferometer Space Antenna},}
  [\href{http://arxiv.org/abs/1702.00786}{{\ttfamily 1702.00786}}].

\bibitem{Kawamura:2006up}
S.~Kawamura {\em et~al.}, {\em{The Japanese space gravitational wave antenna
  DECIGO},} \href{http://dx.doi.org/10.1088/0264-9381/23/8/S17}{Class. Quant.
  Grav. {\bf 23} (2006) S125}.

\bibitem{Guo:2018npi}
W.-H. Ruan, Z.-K. Guo, R.-G. Cai, and Y.-Z. Zhang, {\em{Taiji program:
  Gravitational-wave sources},}
  \href{http://dx.doi.org/10.1142/S0217751X2050075X}{Int. J. Mod. Phys. A {\bf
  35} (2020) 2050075} [\href{http://arxiv.org/abs/1807.09495}{{\ttfamily
  1807.09495}}].

\bibitem{Harry:2006fi}
G.~M. Harry, P.~Fritschel, D.~A. Shaddock, W.~Folkner, and E.~S. Phinney,
  {\em{Laser interferometry for the big bang observer},}
  \href{http://dx.doi.org/10.1088/0264-9381/23/15/008}{Class. Quant. Grav. {\bf
  23} (2006) 4887}.

\bibitem{Caprini:2019egz}
C.~Caprini {\em et~al.}, {\em{Detecting gravitational waves from cosmological
  phase transitions with LISA: an update},}
  \href{http://dx.doi.org/10.1088/1475-7516/2020/03/024}{JCAP {\bf 03} (2020)
  024} [\href{http://arxiv.org/abs/1910.13125}{{\ttfamily 1910.13125}}].

\bibitem{Barger:2008jx}
V.~Barger, P.~Langacker, M.~McCaskey, M.~Ramsey-Musolf, and G.~Shaughnessy,
  {\em{Complex Singlet Extension of the Standard Model},}
  \href{http://dx.doi.org/10.1103/PhysRevD.79.015018}{Phys. Rev. D {\bf 79}
  (2009) 015018} [\href{http://arxiv.org/abs/0811.0393}{{\ttfamily
  0811.0393}}].

\bibitem{Barger:2010yn}
V.~Barger, M.~McCaskey, and G.~Shaughnessy, {\em{Complex Scalar Dark Matter
  vis-{\`a}-vis CoGeNT, DAMA/LIBRA and XENON100},}
  \href{http://dx.doi.org/10.1103/PhysRevD.82.035019}{Phys. Rev. D {\bf 82}
  (2010) 035019} [\href{http://arxiv.org/abs/1005.3328}{{\ttfamily
  1005.3328}}].

\bibitem{Espinosa:2011ax}
J.~R. Espinosa, T.~Konstandin, and F.~Riva, {\em{Strong Electroweak Phase
  Transitions in the Standard Model with a Singlet},}
  \href{http://dx.doi.org/10.1016/j.nuclphysb.2011.09.010}{Nucl. Phys. B {\bf
  854} (2012) 592} [\href{http://arxiv.org/abs/1107.5441}{{\ttfamily
  1107.5441}}].

\bibitem{Ahriche:2012ei}
A.~Ahriche and S.~Nasri, {\em{Light Dark Matter, Light Higgs and the
  Electroweak Phase Transition},}
  \href{http://dx.doi.org/10.1103/PhysRevD.85.093007}{Phys. Rev. D {\bf 85}
  (2012) 093007} [\href{http://arxiv.org/abs/1201.4614}{{\ttfamily
  1201.4614}}].

\bibitem{Chowdhury:2011ga}
T.~A. Chowdhury, M.~Nemevsek, G.~Senjanovic, and Y.~Zhang, {\em{Dark Matter as
  the Trigger of Strong Electroweak Phase Transition},}
  \href{http://dx.doi.org/10.1088/1475-7516/2012/02/029}{JCAP {\bf 02} (2012)
  029} [\href{http://arxiv.org/abs/1110.5334}{{\ttfamily 1110.5334}}].

\bibitem{Borah:2012pu}
D.~Borah and J.~M. Cline, {\em{Inert Doublet Dark Matter with Strong
  Electroweak Phase Transition},}
  \href{http://dx.doi.org/10.1103/PhysRevD.86.055001}{Phys. Rev. D {\bf 86}
  (2012) 055001} [\href{http://arxiv.org/abs/1204.4722}{{\ttfamily
  1204.4722}}].

\bibitem{Gonderinger:2012rd}
M.~Gonderinger, H.~Lim, and M.~J. Ramsey-Musolf, {\em{Complex Scalar Singlet
  Dark Matter: Vacuum Stability and Phenomenology},}
  \href{http://dx.doi.org/10.1103/PhysRevD.86.043511}{Phys. Rev. D {\bf 86}
  (2012) 043511} [\href{http://arxiv.org/abs/1202.1316}{{\ttfamily
  1202.1316}}].

\bibitem{Gil:2012ya}
G.~Gil, P.~Chankowski, and M.~Krawczyk, {\em{Inert Dark Matter and Strong
  Electroweak Phase Transition},}
  \href{http://dx.doi.org/10.1016/j.physletb.2012.09.052}{Phys. Lett. B {\bf
  717} (2012) 396} [\href{http://arxiv.org/abs/1207.0084}{{\ttfamily
  1207.0084}}].

\bibitem{Cline:2012hg}
J.~M. Cline and K.~Kainulainen, {\em{Electroweak baryogenesis and dark matter
  from a singlet Higgs},}
  \href{http://dx.doi.org/10.1088/1475-7516/2013/01/012}{JCAP {\bf 01} (2013)
  012} [\href{http://arxiv.org/abs/1210.4196}{{\ttfamily 1210.4196}}].

\bibitem{Cline:2017qpe}
J.~M. Cline, K.~Kainulainen, and D.~Tucker-Smith, {\em{Electroweak baryogenesis
  from a dark sector},}
  \href{http://dx.doi.org/10.1103/PhysRevD.95.115006}{Phys. Rev. D {\bf 95}
  (2017) 115006} [\href{http://arxiv.org/abs/1702.08909}{{\ttfamily
  1702.08909}}].

\bibitem{Cho:2021itv}
G.-C. Cho, C.~Idegawa, and E.~Senaha, {\em{Electroweak phase transition in a
  complex singlet extension of the Standard Model with degenerate scalars},}
  \href{http://dx.doi.org/10.1016/j.physletb.2021.136787}{Phys. Lett. B {\bf
  823} (2021) 136787} [\href{http://arxiv.org/abs/2105.11830}{{\ttfamily
  2105.11830}}].

\bibitem{Alanne:2014bra}
T.~Alanne, K.~Tuominen, and V.~Vaskonen, {\em{Strong phase transition, dark
  matter and vacuum stability from simple hidden sectors},}
  \href{http://dx.doi.org/10.1016/j.nuclphysb.2014.11.001}{Nucl. Phys. B {\bf
  889} (2014) 692} [\href{http://arxiv.org/abs/1407.0688}{{\ttfamily
  1407.0688}}].

\bibitem{Basler:2020nrq}
P.~Basler, M.~M\"uhlleitner, and J.~M\"uller, {\em{BSMPT v2 a tool for the
  electroweak phase transition and the baryon asymmetry of the universe in
  extended Higgs Sectors},}
  \href{http://dx.doi.org/10.1016/j.cpc.2021.108124}{Comput. Phys. Commun. {\bf
  269} (2021) 108124} [\href{http://arxiv.org/abs/2007.01725}{{\ttfamily
  2007.01725}}].

\bibitem{Jiang:2015cwa}
M.~Jiang, L.~Bian, W.~Huang, and J.~Shu, {\em{Impact of a complex singlet:
  Electroweak baryogenesis and dark matter},}
  \href{http://dx.doi.org/10.1103/PhysRevD.93.065032}{Phys. Rev. D {\bf 93}
  (2016) 065032} [\href{http://arxiv.org/abs/1502.07574}{{\ttfamily
  1502.07574}}].

\bibitem{Chiang:2017nmu}
C.-W. Chiang, M.~J. Ramsey-Musolf, and E.~Senaha, {\em{Standard Model with a
  Complex Scalar Singlet: Cosmological Implications and Theoretical
  Considerations},} \href{http://dx.doi.org/10.1103/PhysRevD.97.015005}{Phys.
  Rev. D {\bf 97} (2018) 015005}
  [\href{http://arxiv.org/abs/1707.09960}{{\ttfamily 1707.09960}}].

\bibitem{Chen:2019ebq}
N.~Chen, T.~Li, Y.~Wu, and L.~Bian, {\em{Complementarity of the future $e^+
  e^-$ colliders and gravitational waves in the probe of complex singlet
  extension to the standard model},}
  \href{http://dx.doi.org/10.1103/PhysRevD.101.075047}{Phys. Rev. D {\bf 101}
  (2020) 075047} [\href{http://arxiv.org/abs/1911.05579}{{\ttfamily
  1911.05579}}].

\bibitem{Ghorbani:2018yfr}
K.~Ghorbani and P.~H. Ghorbani, {\em{Strongly First-Order Phase Transition in
  Real Singlet Scalar Dark Matter Model},}
  \href{http://dx.doi.org/10.1088/1361-6471/ab4823}{J. Phys. G {\bf 47} (2020)
  015201} [\href{http://arxiv.org/abs/1804.05798}{{\ttfamily 1804.05798}}].

\bibitem{Ghorbani:2019itr}
K.~Ghorbani and P.~H. Ghorbani, {\em{A Simultaneous Study of Dark Matter and
  Phase Transition: Two-Scalar Scenario},}
  \href{http://dx.doi.org/10.1007/JHEP12(2019)077}{JHEP {\bf 12} (2019) 077}
  [\href{http://arxiv.org/abs/1906.01823}{{\ttfamily 1906.01823}}].

\bibitem{Ertas:2021xeh}
F.~Ertas, F.~Kahlhoefer, and C.~Tasillo, {\em{Turn up the volume: listening to
  phase transitions in hot dark sectors},}
  \href{http://dx.doi.org/10.1088/1475-7516/2022/02/014}{JCAP {\bf 02} (2022)
  014} [\href{http://arxiv.org/abs/2109.06208}{{\ttfamily 2109.06208}}].

\bibitem{Kumar:2011np}
P.~Kumar and E.~Ponton, {\em{Electroweak Baryogenesis and Dark Matter with an
  approximate R-symmetry},}
  \href{http://dx.doi.org/10.1007/JHEP11(2011)037}{JHEP {\bf 11} (2011) 037}
  [\href{http://arxiv.org/abs/1107.1719}{{\ttfamily 1107.1719}}].

\bibitem{Kozaczuk:2011vr}
J.~Kozaczuk and S.~Profumo, {\em{Closing in on Supersymmetric Electroweak
  Baryogenesis with Dark Matter Searches and the Large Hadron Collider},}
  \href{http://dx.doi.org/10.1088/1475-7516/2011/11/031}{JCAP {\bf 11} (2011)
  031} [\href{http://arxiv.org/abs/1108.0393}{{\ttfamily 1108.0393}}].

\bibitem{Carena:2011jy}
M.~Carena, N.~R. Shah, and C.~E.~M. Wagner, {\em{Light Dark Matter and the
  Electroweak Phase Transition in the NMSSM},}
  \href{http://dx.doi.org/10.1103/PhysRevD.85.036003}{Phys. Rev. D {\bf 85}
  (2012) 036003} [\href{http://arxiv.org/abs/1110.4378}{{\ttfamily
  1110.4378}}].

\bibitem{Espinosa:2011eu}
J.~R. Espinosa, B.~Gripaios, T.~Konstandin, and F.~Riva, {\em{Electroweak
  Baryogenesis in Non-minimal Composite Higgs Models},}
  \href{http://dx.doi.org/10.1088/1475-7516/2012/01/012}{JCAP {\bf 01} (2012)
  012} [\href{http://arxiv.org/abs/1110.2876}{{\ttfamily 1110.2876}}].

\bibitem{Chala:2016ykx}
M.~Chala, G.~Nardini, and I.~Sobolev, {\em{Unified explanation for dark matter
  and electroweak baryogenesis with direct detection and gravitational wave
  signatures},} \href{http://dx.doi.org/10.1103/PhysRevD.94.055006}{Phys. Rev.
  D {\bf 94} (2016) 055006} [\href{http://arxiv.org/abs/1605.08663}{{\ttfamily
  1605.08663}}].

\bibitem{Ghorbani:2017jls}
P.~H. Ghorbani, {\em{Electroweak Baryogenesis and Dark Matter via a
  Pseudoscalar vs. Scalar},}
  \href{http://dx.doi.org/10.1007/JHEP08(2017)058}{JHEP {\bf 08} (2017) 058}
  [\href{http://arxiv.org/abs/1703.06506}{{\ttfamily 1703.06506}}].

\bibitem{Liu:2021mhn}
J.~Liu, X.-P. Wang, and K.-P. Xie, {\em{Searching for lepton portal dark matter
  with colliders and gravitational waves},}
  \href{http://dx.doi.org/10.1007/JHEP06(2021)149}{JHEP {\bf 06} (2021) 149}
  [\href{http://arxiv.org/abs/2104.06421}{{\ttfamily 2104.06421}}].

\bibitem{Gould:2021oba}
O.~Gould and T.~V.~I. Tenkanen, {\em{On the perturbative expansion at high
  temperature and implications for cosmological phase transitions},}
  \href{http://dx.doi.org/10.1007/JHEP06(2021)069}{JHEP {\bf 06} (2021) 069}
  [\href{http://arxiv.org/abs/2104.04399}{{\ttfamily 2104.04399}}].

\bibitem{Quiros:1999jp}
M.~Quiros, {\em{Finite temperature field theory and phase transitions},} in
  {\em {ICTP Summer School in High-Energy Physics and Cosmology}},
  pp.~187--259, 1, 1999 [\href{http://arxiv.org/abs/hep-ph/9901312}{{\ttfamily
  hep-ph/9901312}}].

\bibitem{Croon:2020cgk}
D.~Croon, O.~Gould, P.~Schicho, T.~V.~I. Tenkanen, and G.~White,
  {\em{Theoretical uncertainties for cosmological first-order phase
  transitions},} \href{http://dx.doi.org/10.1007/JHEP04(2021)055}{JHEP {\bf 04}
  (2021) 055} [\href{http://arxiv.org/abs/2009.10080}{{\ttfamily 2009.10080}}].

\bibitem{Sommerfeld:1931}
A.~Sommerfeld, {\em{Über die Beugung und Bremsung der Elektronen},}
  \href{http://dx.doi.org/https://doi.org/10.1002/andp.19314030302}{{Annalen
  der Physik} {\bf 403} (1931) 257}.

\bibitem{Sakharov:1991pia}
A.~D. Sakharov, {\em{Interaction of an Electron and Positron in Pair
  Production},} \href{http://dx.doi.org/10.1070/PU1991v034n05ABEH002492}{Zh.
  Eksp. Teor. Fiz. {\bf 18} (1948) 631}.

\bibitem{Hisano:2004ds}
J.~Hisano, S.~Matsumoto, M.~M. Nojiri, and O.~Saito, {\em{Non-perturbative
  effect on dark matter annihilation and gamma ray signature from galactic
  center},} \href{http://dx.doi.org/10.1103/PhysRevD.71.063528}{Phys. Rev. D
  {\bf 71} (2005) 063528}
  [\href{http://arxiv.org/abs/hep-ph/0412403}{{\ttfamily hep-ph/0412403}}].

\bibitem{vonHarling:2014kha}
B.~von Harling and K.~Petraki, {\em{Bound-state formation for thermal relic
  dark matter and unitarity},}
  \href{http://dx.doi.org/10.1088/1475-7516/2014/12/033}{JCAP {\bf 12} (2014)
  033} [\href{http://arxiv.org/abs/1407.7874}{{\ttfamily 1407.7874}}].

\bibitem{An:2013xka}
H.~An, L.-T. Wang, and H.~Zhang, {\em{Dark matter with $t$-channel mediator: a
  simple step beyond contact interaction},}
  \href{http://dx.doi.org/10.1103/PhysRevD.89.115014}{Phys. Rev. D {\bf 89}
  (2014) 115014} [\href{http://arxiv.org/abs/1308.0592}{{\ttfamily
  1308.0592}}].

\bibitem{Kopp:2014tsa}
J.~Kopp, L.~Michaels, and J.~Smirnov, {\em{Loopy Constraints on Leptophilic
  Dark Matter and Internal Bremsstrahlung},}
  \href{http://dx.doi.org/10.1088/1475-7516/2014/04/022}{JCAP {\bf 04} (2014)
  022} [\href{http://arxiv.org/abs/1401.6457}{{\ttfamily 1401.6457}}].

\bibitem{Garny:2015wea}
M.~Garny, A.~Ibarra, and S.~Vogl, {\em{Signatures of Majorana dark matter with
  t-channel mediators},}
  \href{http://dx.doi.org/10.1142/S0218271815300190}{Int. J. Mod. Phys. D {\bf
  24} (2015) 1530019} [\href{http://arxiv.org/abs/1503.01500}{{\ttfamily
  1503.01500}}].

\bibitem{Arina:2020udz}
C.~Arina, B.~Fuks, and L.~Mantani, {\em{A universal framework for t-channel
  dark matter models},}
  \href{http://dx.doi.org/10.1140/epjc/s10052-020-7933-7}{Eur. Phys. J. C {\bf
  80} (2020) 409} [\href{http://arxiv.org/abs/2001.05024}{{\ttfamily
  2001.05024}}].

\bibitem{CMS:2017abv}
{\bfseries CMS} Collaboration, A.~M. Sirunyan {\em et~al.}, {\em{Search for
  supersymmetry in multijet events with missing transverse momentum in
  proton-proton collisions at 13 TeV},}
  \href{http://dx.doi.org/10.1103/PhysRevD.96.032003}{Phys. Rev. D {\bf 96}
  (2017) 032003} [\href{http://arxiv.org/abs/1704.07781}{{\ttfamily
  1704.07781}}].

\bibitem{CMS:2017mbm}
{\bfseries CMS} Collaboration, A.~M. Sirunyan {\em et~al.}, {\em{Search for
  direct production of supersymmetric partners of the top quark in the all-jets
  final state in proton-proton collisions at $ \sqrt{s}=13 $ TeV},}
  \href{http://dx.doi.org/10.1007/JHEP10(2017)005}{JHEP {\bf 10} (2017) 005}
  [\href{http://arxiv.org/abs/1707.03316}{{\ttfamily 1707.03316}}].

\bibitem{ATLAS:2017eoo}
{\bfseries ATLAS} Collaboration, M.~Aaboud {\em et~al.}, {\em{Search for
  top-squark pair production in final states with one lepton, jets, and missing
  transverse momentum using 36 fb$^{-1}$ of $\sqrt{s}=13 $ TeV pp collision
  data with the ATLAS detector},}
  \href{http://dx.doi.org/10.1007/JHEP06(2018)108}{JHEP {\bf 06} (2018) 108}
  [\href{http://arxiv.org/abs/1711.11520}{{\ttfamily 1711.11520}}].

\bibitem{ATLAS:2021twp}
{\bfseries ATLAS} Collaboration, G.~Aad {\em et~al.}, {\em{Search for squarks
  and gluinos in final states with one isolated lepton, jets, and missing
  transverse momentum at $\sqrt{s}=13$~ with the ATLAS detector},}
  \href{http://dx.doi.org/10.1140/epjc/s10052-021-09748-8}{Eur. Phys. J. C {\bf
  81} (2021) 600} [\href{http://arxiv.org/abs/2101.01629}{{\ttfamily
  2101.01629}}].

\bibitem{CMS:2020fia}
{\bfseries CMS} Collaboration, A.~M. Sirunyan {\em et~al.}, {\em{Search for
  supersymmetry in proton-proton collisions at $\sqrt{s} =$ 13 TeV in events
  with high-momentum Z bosons and missing transverse momentum},}
  \href{http://dx.doi.org/10.1007/JHEP09(2020)149}{JHEP {\bf 09} (2020) 149}
  [\href{http://arxiv.org/abs/2008.04422}{{\ttfamily 2008.04422}}].

\bibitem{CMS:2020pyk}
{\bfseries CMS} Collaboration, A.~M. Sirunyan {\em et~al.}, {\em{Search for top
  squark pair production using dilepton final states in ${\text {p}}{\text
  {p}}$ collision data collected at $\sqrt{s}=13\,\text {TeV} $},}
  \href{http://dx.doi.org/10.1140/epjc/s10052-020-08701-5}{Eur. Phys. J. C {\bf
  81} (2021) 3} [\href{http://arxiv.org/abs/2008.05936}{{\ttfamily
  2008.05936}}].

\bibitem{ATLAS:2019gti}
{\bfseries ATLAS} Collaboration, G.~Aad {\em et~al.}, {\em{Search for direct
  stau production in events with two hadronic $\tau$-leptons in $\sqrt{s} = 13$
  TeV $pp$ collisions with the ATLAS detector},}
  \href{http://dx.doi.org/10.1103/PhysRevD.101.032009}{Phys. Rev. D {\bf 101}
  (2020) 032009} [\href{http://arxiv.org/abs/1911.06660}{{\ttfamily
  1911.06660}}].

\bibitem{ATLAS:2019lff}
{\bfseries ATLAS} Collaboration, G.~Aad {\em et~al.}, {\em{Search for
  electroweak production of charginos and sleptons decaying into final states
  with two leptons and missing transverse momentum in $\sqrt{s}=13$ TeV $pp$
  collisions using the ATLAS detector},}
  \href{http://dx.doi.org/10.1140/epjc/s10052-019-7594-6}{Eur. Phys. J. C {\bf
  80} (2020) 123} [\href{http://arxiv.org/abs/1908.08215}{{\ttfamily
  1908.08215}}].

\bibitem{Fermi-LAT:2013thd}
{\bfseries Fermi-LAT} Collaboration, M.~Ackermann {\em et~al.}, {\em{Search for
  Gamma-ray Spectral Lines with the Fermi Large Area Telescope and Dark Matter
  Implications},} \href{http://dx.doi.org/10.1103/PhysRevD.88.082002}{Phys.
  Rev. D {\bf 88} (2013) 082002}
  [\href{http://arxiv.org/abs/1305.5597}{{\ttfamily 1305.5597}}].

\bibitem{HESS:2013rld}
{\bfseries H.E.S.S.} Collaboration, A.~Abramowski {\em et~al.}, {\em{Search for
  Photon-Linelike Signatures from Dark Matter Annihilations with H.E.S.S.},}
  \href{http://dx.doi.org/10.1103/PhysRevLett.110.041301}{Phys. Rev. Lett. {\bf
  110} (2013) 041301} [\href{http://arxiv.org/abs/1301.1173}{{\ttfamily
  1301.1173}}].

\bibitem{HESS:2018cbt}
{\bfseries HESS} Collaboration, H.~Abdallah {\em et~al.}, {\em{Search for
  $\gamma$-Ray Line Signals from Dark Matter Annihilations in the Inner
  Galactic Halo from 10 Years of Observations with H.E.S.S.},}
  \href{http://dx.doi.org/10.1103/PhysRevLett.120.201101}{Phys. Rev. Lett. {\bf
  120} (2018) 201101} [\href{http://arxiv.org/abs/1805.05741}{{\ttfamily
  1805.05741}}].

\bibitem{Matsubara:1955ws}
T.~Matsubara, {\em{A New approach to quantum statistical mechanics},}
  \href{http://dx.doi.org/10.1143/PTP.14.351}{Prog. Theor. Phys. {\bf 14}
  (1955) 351}.

\bibitem{Ginsparg:1980ef}
P.~H. Ginsparg, {\em{First Order and Second Order Phase Transitions in Gauge
  Theories at Finite Temperature},}
  \href{http://dx.doi.org/10.1016/0550-3213(80)90418-6}{Nucl. Phys. {\bf B170}
  (1980) 388}.

\bibitem{Appelquist:1981vg}
T.~Appelquist and R.~D. Pisarski, {\em{High-Temperature Yang-Mills Theories and
  Three-Dimensional Quantum Chromodynamics},}
  \href{http://dx.doi.org/10.1103/PhysRevD.23.2305}{Phys.\ Rev. {\bf D23}
  (1981) 2305}.

\bibitem{Farakos:1994kx}
K.~Farakos, K.~Kajantie, K.~Rummukainen, and M.~E. Shaposhnikov, {\em{3-D
  physics and the electroweak phase transition: Perturbation theory},}
  \href{http://dx.doi.org/10.1016/0550-3213(94)90173-2}{Nucl. Phys. {\bf B425}
  (1994) 67} [\href{http://arxiv.org/abs/hep-ph/9404201}{{\ttfamily
  hep-ph/9404201}}].

\bibitem{Braaten:1995cm}
E.~Braaten and A.~Nieto, {\em{Effective field theory approach to high
  temperature thermodynamics},}
  \href{http://dx.doi.org/10.1103/PhysRevD.51.6990}{Phys.\ Rev. {\bf D51}
  (1995) 6990} [\href{http://arxiv.org/abs/hep-ph/9501375}{{\ttfamily
  hep-ph/9501375}}].

\bibitem{Braaten:1995jr}
E.~Braaten and A.~Nieto, {\em{Free energy of QCD at high temperature},}
  \href{http://dx.doi.org/10.1103/PhysRevD.53.3421}{Phys.\ Rev. D {\bf 53}
  (1996) 3421} [\href{http://arxiv.org/abs/hep-ph/9510408}{{\ttfamily
  hep-ph/9510408}}].

\bibitem{Kajantie:1995dw}
K.~Kajantie, M.~Laine, K.~Rummukainen, and M.~E. Shaposhnikov, {\em{Generic
  rules for high temperature dimensional reduction and their application to the
  standard model},} \href{http://dx.doi.org/10.1016/0550-3213(95)00549-8}{Nucl.
  Phys. B {\bf 458} (1996) 90}
  [\href{http://arxiv.org/abs/hep-ph/9508379}{{\ttfamily hep-ph/9508379}}].

\bibitem{Ekstedt:2022bff}
A.~Ekstedt, P.~Schicho, and T.~V.~I. Tenkanen, {\em{{\tt DRalgo}: a package for
  effective field theory approach for thermal phase transitions},}
  [\href{http://arxiv.org/abs/2205.08815}{{\ttfamily 2205.08815}}].

\bibitem{Niemi:2021qvp}
L.~Niemi, P.~Schicho, and T.~V.~I. Tenkanen, {\em{Singlet-assisted electroweak
  phase transition at two loops},}
  \href{http://dx.doi.org/10.1103/PhysRevD.103.115035}{Phys. Rev. D {\bf 103}
  (2021) 115035} [\href{http://arxiv.org/abs/2103.07467}{{\ttfamily
  2103.07467}}].

\bibitem{Schicho:2021gca}
P.~M. Schicho, T.~V.~I. Tenkanen, and J.~\"Osterman, {\em{Robust approach to
  thermal resummation: Standard Model meets a singlet},}
  \href{http://dx.doi.org/10.1007/JHEP06(2021)130}{JHEP {\bf 06} (2021) 130}
  [\href{http://arxiv.org/abs/2102.11145}{{\ttfamily 2102.11145}}].

\bibitem{Laine:2019uua}
M.~Laine, P.~Schicho, and Y.~Schr\"oder, {\em{A QCD Debye mass in a broad
  temperature range},}
  \href{http://dx.doi.org/10.1103/PhysRevD.101.023532}{Phys. Rev. D {\bf 101}
  (2020) 023532} [\href{http://arxiv.org/abs/1911.09123}{{\ttfamily
  1911.09123}}].

\bibitem{Laine:2000kv}
M.~Laine and M.~Losada, {\em{Two loop dimensional reduction and effective
  potential without temperature expansions},}
  \href{http://dx.doi.org/10.1016/S0550-3213(00)00298-4}{Nucl. Phys. B {\bf
  582} (2000) 277} [\href{http://arxiv.org/abs/hep-ph/0003111}{{\ttfamily
  hep-ph/0003111}}].

\bibitem{Brauner:2016fla}
T.~Brauner, T.~V.~I. Tenkanen, A.~Tranberg, A.~Vuorinen, and D.~J. Weir,
  {\em{Dimensional reduction of the Standard Model coupled to a new singlet
  scalar field},} \href{http://dx.doi.org/10.1007/JHEP03(2017)007}{JHEP {\bf
  03} (2017) 007} [\href{http://arxiv.org/abs/1609.06230}{{\ttfamily
  1609.06230}}].

\bibitem{Laine:2017hdk}
M.~Laine, M.~Meyer, and G.~Nardini, {\em{Thermal phase transition with full
  2-loop effective potential},}
  \href{http://dx.doi.org/10.1016/j.nuclphysb.2017.04.023}{Nucl. Phys. B {\bf
  920} (2017) 565} [\href{http://arxiv.org/abs/1702.07479}{{\ttfamily
  1702.07479}}].

\bibitem{Niemi:2018asa}
L.~Niemi, H.~H. Patel, M.~J. Ramsey-Musolf, T.~V.~I. Tenkanen, and D.~J. Weir,
  {\em{Electroweak phase transition in the real triplet extension of the SM:
  Dimensional reduction},}
  \href{http://dx.doi.org/10.1103/PhysRevD.100.035002}{Phys. Rev. D {\bf 100}
  (2019) 035002} [\href{http://arxiv.org/abs/1802.10500}{{\ttfamily
  1802.10500}}].

\bibitem{Schicho:2022wty}
P.~Schicho, T.~V.~I. Tenkanen, and G.~White, {\em{Combining thermal resummation
  and gauge invariance for electroweak phase transition},}
  [\href{http://arxiv.org/abs/2203.04284}{{\ttfamily 2203.04284}}].

\bibitem{Arnold:1992rz}
P.~B. Arnold and O.~Espinosa, {\em{The Effective potential and first order
  phase transitions: Beyond leading-order},}
  \href{http://dx.doi.org/10.1103/PhysRevD.47.3546}{Phys. Rev. D {\bf 47}
  (1993) 3546} [\href{http://arxiv.org/abs/hep-ph/9212235}{{\ttfamily
  hep-ph/9212235}}].

\bibitem{Basler:2021kgq}
P.~Basler, M.~M\"uhlleitner, and J.~M\"uller, {\em{Electroweak Baryogenesis in
  the CP-Violating Two-Higgs Doublet Model},}
  [\href{http://arxiv.org/abs/2108.03580}{{\ttfamily 2108.03580}}].

\bibitem{Kajantie:1996mn}
K.~Kajantie, M.~Laine, K.~Rummukainen, and M.~E. Shaposhnikov, {\em{Is there a~
  hot electroweak phase transition at $m_H \gtrsim m_W$?},}
  \href{http://dx.doi.org/10.1103/PhysRevLett.77.2887}{Phys. Rev. Lett. {\bf
  77} (1996) 2887} [\href{http://arxiv.org/abs/hep-ph/9605288}{{\ttfamily
  hep-ph/9605288}}].

\bibitem{Kajantie:1995kf}
K.~Kajantie, M.~Laine, K.~Rummukainen, and M.~E. Shaposhnikov, {\em{The
  Electroweak phase transition: A Nonperturbative analysis},}
  \href{http://dx.doi.org/10.1016/0550-3213(96)00052-1}{Nucl. Phys. B {\bf 466}
  (1996) 189} [\href{http://arxiv.org/abs/hep-lat/9510020}{{\ttfamily
  hep-lat/9510020}}].

\bibitem{Kajantie:1996qd}
K.~Kajantie, M.~Laine, K.~Rummukainen, and M.~E. Shaposhnikov, {\em{A
  Nonperturbative analysis of the finite T phase transition in
  SU(2)$\times$U(1) electroweak theory},}
  \href{http://dx.doi.org/10.1016/S0550-3213(97)00164-8}{Nucl. Phys. B {\bf
  493} (1997) 413} [\href{http://arxiv.org/abs/hep-lat/9612006}{{\ttfamily
  hep-lat/9612006}}].

\bibitem{Csikor:1998eu}
F.~Csikor, Z.~Fodor, and J.~Heitger, {\em{Endpoint of the hot electroweak phase
  transition},} \href{http://dx.doi.org/10.1103/PhysRevLett.82.21}{Phys. Rev.
  Lett. {\bf 82} (1999) 21}
  [\href{http://arxiv.org/abs/hep-ph/9809291}{{\ttfamily hep-ph/9809291}}].

\bibitem{Aoki:1999fi}
Y.~Aoki, F.~Csikor, Z.~Fodor, and A.~Ukawa, {\em{The Endpoint of the first
  order phase transition of the SU(2) gauge Higgs model on a four-dimensional
  isotropic lattice},}
  \href{http://dx.doi.org/10.1103/PhysRevD.60.013001}{Phys. Rev. D {\bf 60}
  (1999) 013001} [\href{http://arxiv.org/abs/hep-lat/9901021}{{\ttfamily
  hep-lat/9901021}}].

\bibitem{Niemi:2020hto}
L.~Niemi, M.~J. Ramsey-Musolf, T.~V.~I. Tenkanen, and D.~J. Weir,
  {\em{Thermodynamics of a Two-Step Electroweak Phase Transition},}
  \href{http://dx.doi.org/10.1103/PhysRevLett.126.171802}{Phys. Rev. Lett. {\bf
  126} (2021) 171802} [\href{http://arxiv.org/abs/2005.11332}{{\ttfamily
  2005.11332}}].

\bibitem{Farakos:1994xh}
K.~Farakos, K.~Kajantie, K.~Rummukainen, and M.~E. Shaposhnikov, {\em{3-d
  physics and the electroweak phase transition: A Framework for lattice Monte
  Carlo analysis},} \href{http://dx.doi.org/10.1016/0550-3213(95)80129-4}{Nucl.
  Phys. B {\bf 442} (1995) 317}
  [\href{http://arxiv.org/abs/hep-lat/9412091}{{\ttfamily hep-lat/9412091}}].

\bibitem{Gould:2019qek}
O.~Gould, J.~Kozaczuk, L.~Niemi, M.~J. Ramsey-Musolf, T.~V.~I. Tenkanen, and
  D.~J. Weir, {\em{Nonperturbative analysis of the gravitational waves from a
  first-order electroweak phase transition},}
  \href{http://dx.doi.org/10.1103/PhysRevD.100.115024}{Phys. Rev. D {\bf 100}
  (2019) 115024} [\href{http://arxiv.org/abs/1903.11604}{{\ttfamily
  1903.11604}}].

\bibitem{Andersen:2017ika}
J.~O. Andersen, T.~Gorda, A.~Helset, {\em et~al.}, {\em{Nonperturbative
  Analysis of the Electroweak Phase Transition in the Two Higgs Doublet
  Model},} \href{http://dx.doi.org/10.1103/PhysRevLett.121.191802}{Phys. Rev.
  Lett. {\bf 121} (2018) 191802}
  [\href{http://arxiv.org/abs/1711.09849}{{\ttfamily 1711.09849}}].

\bibitem{Gorda:2018hvi}
T.~Gorda, A.~Helset, L.~Niemi, T.~V.~I. Tenkanen, and D.~J. Weir,
  {\em{Three-dimensional effective theories for the two Higgs doublet model at
  high temperature},} \href{http://dx.doi.org/10.1007/JHEP02(2019)081}{JHEP
  {\bf 02} (2019) 081} [\href{http://arxiv.org/abs/1802.05056}{{\ttfamily
  1802.05056}}].

\bibitem{Cline:1999wi}
J.~M. Cline, G.~D. Moore, and G.~Servant, {\em{Was the electroweak phase
  transition preceded by a color broken phase?},}
  \href{http://dx.doi.org/10.1103/PhysRevD.60.105035}{Phys. Rev. D {\bf 60}
  (1999) 105035} [\href{http://arxiv.org/abs/hep-ph/9902220}{{\ttfamily
  hep-ph/9902220}}].

\bibitem{Bodeker:2017cim}
D.~Bodeker and G.~D. Moore, {\em{Electroweak Bubble Wall Speed Limit},}
  \href{http://dx.doi.org/10.1088/1475-7516/2017/05/025}{JCAP {\bf 05} (2017)
  025} [\href{http://arxiv.org/abs/1703.08215}{{\ttfamily 1703.08215}}].

\bibitem{Kajantie:1998zn}
K.~Kajantie, M.~Laine, T.~Neuhaus, J.~Peisa, A.~Rajantie, and K.~Rummukainen,
  {\em{Vortex tension as an order parameter in three-dimensional U(1) + Higgs
  theory},} \href{http://dx.doi.org/10.1016/S0550-3213(99)00033-4}{Nucl. Phys.
  B {\bf 546} (1999) 351}
  [\href{http://arxiv.org/abs/hep-ph/9809334}{{\ttfamily hep-ph/9809334}}].

\bibitem{Gondolo:1990dk}
P.~Gondolo and G.~Gelmini, {\em{Cosmic abundances of stable particles: Improved
  analysis},} \href{http://dx.doi.org/10.1016/0550-3213(91)90438-4}{Nucl. Phys.
  {\bf B360} (1991) 145}.

\bibitem{Griest:1990kh}
K.~Griest and D.~Seckel, {\em{Three exceptions in the calculation of relic
  abundances},} \href{http://dx.doi.org/10.1103/PhysRevD.43.3191}{Phys. Rev.
  {\bf D43} (1991) 3191}.

\bibitem{McDonald:2001vt}
J.~McDonald, {\em{Thermally generated gauge singlet scalars as selfinteracting
  dark matter},} \href{http://dx.doi.org/10.1103/PhysRevLett.88.091304}{Phys.
  Rev. Lett. {\bf 88} (2002) 091304}
  [\href{http://arxiv.org/abs/hep-ph/0106249}{{\ttfamily hep-ph/0106249}}].

\bibitem{Hall:2009bx}
L.~J. Hall, K.~Jedamzik, J.~March-Russell, and S.~M. West, {\em{Freeze-In
  Production of FIMP Dark Matter},}
  \href{http://dx.doi.org/10.1007/JHEP03(2010)080}{JHEP {\bf 03} (2010) 080}
  [\href{http://arxiv.org/abs/0911.1120}{{\ttfamily 0911.1120}}].

\bibitem{Garny:2017rxs}
M.~Garny, J.~Heisig, B.~L\"ulf, and S.~Vogl, {\em{Coannihilation without
  chemical equilibrium},}
  \href{http://dx.doi.org/10.1103/PhysRevD.96.103521}{Phys. Rev. D {\bf 96}
  (2017) 103521} [\href{http://arxiv.org/abs/1705.09292}{{\ttfamily
  1705.09292}}].

\bibitem{Junius:2019dci}
S.~Junius, L.~Lopez-Honorez, and A.~Mariotti, {\em{A feeble window on
  leptophilic dark matter},}
  \href{http://dx.doi.org/10.1007/JHEP07(2019)136}{JHEP {\bf 07} (2019) 136}
  [\href{http://arxiv.org/abs/1904.07513}{{\ttfamily 1904.07513}}].

\bibitem{Enqvist:1992va}
K.~Enqvist, K.~Kainulainen, and I.~Vilja, {\em{Phase transitions in the singlet
  majoron model},} \href{http://dx.doi.org/10.1016/0550-3213(93)90369-Z}{Nucl.
  Phys. B {\bf 403} (1993) 749}.

\bibitem{Enqvist:2014zqa}
K.~Enqvist, S.~Nurmi, T.~Tenkanen, and K.~Tuominen, {\em{Standard Model with a
  real singlet scalar and inflation},}
  \href{http://dx.doi.org/10.1088/1475-7516/2014/08/035}{JCAP {\bf 08} (2014)
  035} [\href{http://arxiv.org/abs/1407.0659}{{\ttfamily 1407.0659}}].

\bibitem{Tenkanen:2016idg}
T.~Tenkanen, K.~Tuominen, and V.~Vaskonen, {\em{A Strong Electroweak Phase
  Transition from the Inflaton Field},}
  \href{http://dx.doi.org/10.1088/1475-7516/2016/09/037}{JCAP {\bf 09} (2016)
  037} [\href{http://arxiv.org/abs/1606.06063}{{\ttfamily 1606.06063}}].

\bibitem{Bernal:2017kxu}
N.~Bernal, M.~Heikinheimo, T.~Tenkanen, K.~Tuominen, and V.~Vaskonen, {\em{The
  Dawn of FIMP Dark Matter: A Review of Models and Constraints},}
  \href{http://dx.doi.org/10.1142/S0217751X1730023X}{Int. J. Mod. Phys. {\bf
  A32} (2017) 1730023} [\href{http://arxiv.org/abs/1706.07442}{{\ttfamily
  1706.07442}}].

\bibitem{Garny:2018ali}
M.~Garny and J.~Heisig, {\em{Interplay of super-WIMP and freeze-in production
  of dark matter},} \href{http://dx.doi.org/10.1103/PhysRevD.98.095031}{Phys.
  Rev. {\bf D98} (2018) 095031}
  [\href{http://arxiv.org/abs/1809.10135}{{\ttfamily 1809.10135}}].

\bibitem{Biondini:2020ric}
S.~Biondini and J.~Ghiglieri, {\em{Freeze-in produced dark matter in the
  ultra-relativistic regime},}
  \href{http://dx.doi.org/10.1088/1475-7516/2021/03/075}{JCAP {\bf 03} (2021)
  075} [\href{http://arxiv.org/abs/2012.09083}{{\ttfamily 2012.09083}}].

\bibitem{Edsjo:1997bg}
J.~Edsjo and P.~Gondolo, {\em{Neutralino relic density including
  coannihilations},} \href{http://dx.doi.org/10.1103/PhysRevD.56.1879}{Phys.
  Rev. D {\bf 56} (1997) 1879}
  [\href{http://arxiv.org/abs/hep-ph/9704361}{{\ttfamily hep-ph/9704361}}].

\bibitem{Ellis:2014ipa}
J.~Ellis, K.~A. Olive, and J.~Zheng, {\em{The Extent of the Stop Coannihilation
  Strip},} \href{http://dx.doi.org/10.1140/epjc/s10052-014-2947-7}{Eur. Phys.
  J. C {\bf 74} (2014) 2947} [\href{http://arxiv.org/abs/1404.5571}{{\ttfamily
  1404.5571}}].

\bibitem{Biondini:2019int}
S.~Biondini and S.~Vogl, {\em{Scalar dark matter coannihilating with a coloured
  fermion},} \href{http://dx.doi.org/10.1007/JHEP11(2019)147}{JHEP {\bf 11}
  (2019) 147} [\href{http://arxiv.org/abs/1907.05766}{{\ttfamily 1907.05766}}].

\bibitem{DeSimone:2016fbz}
A.~De~Simone and T.~Jacques, {\em{Simplified models vs. effective field theory
  approaches in dark matter searches},}
  \href{http://dx.doi.org/10.1140/epjc/s10052-016-4208-4}{Eur. Phys. J. C {\bf
  76} (2016) 367} [\href{http://arxiv.org/abs/1603.08002}{{\ttfamily
  1603.08002}}].

\bibitem{Bollig:2021psb}
J.~Bollig and S.~Vogl, {\em{Impact of bound states on non-thermal dark matter
  production},} [\href{http://arxiv.org/abs/2112.01491}{{\ttfamily
  2112.01491}}].

\bibitem{L3:2003fyi}
{\bfseries L3} Collaboration, P.~Achard {\em et~al.}, {\em{Search for scalar
  leptons and scalar quarks at LEP},}
  \href{http://dx.doi.org/10.1016/j.physletb.2003.10.010}{Phys. Lett. B {\bf
  580} (2004) 37} [\href{http://arxiv.org/abs/hep-ex/0310007}{{\ttfamily
  hep-ex/0310007}}].

\bibitem{ALEPH:2003acj}
{\bfseries ALEPH} Collaboration, A.~Heister {\em et~al.}, {\em{Absolute mass
  lower limit for the lightest neutralino of the MSSM from $e^+ e^-$ data at
  $\sqrt{s}$ up to 209-GeV},}
  \href{http://dx.doi.org/10.1016/j.physletb.2003.12.066}{Phys. Lett. B {\bf
  583} (2004) 247}.

\bibitem{ALEPH:2001oot}
{\bfseries ALEPH} Collaboration, A.~Heister {\em et~al.}, {\em{Search for
  scalar leptons in e+ e- collisions at center-of-mass energies up to
  209-GeV},} \href{http://dx.doi.org/10.1016/S0370-2693(01)01494-0}{Phys. Lett.
  B {\bf 526} (2002) 206}
  [\href{http://arxiv.org/abs/hep-ex/0112011}{{\ttfamily hep-ex/0112011}}].

\bibitem{Biondini:2017ufr}
S.~Biondini and M.~Laine, {\em{Re-derived overclosure bound for the inert
  doublet model},} \href{http://dx.doi.org/10.1007/JHEP08(2017)047}{JHEP {\bf
  08} (2017) 047} [\href{http://arxiv.org/abs/1706.01894}{{\ttfamily
  1706.01894}}].

\bibitem{Biondini:2018pwp}
S.~Biondini and M.~Laine, {\em{Thermal dark matter co-annihilating with a
  strongly interacting scalar},}
  \href{http://dx.doi.org/10.1007/JHEP04(2018)072}{JHEP {\bf 04} (2018) 072}
  [\href{http://arxiv.org/abs/1801.05821}{{\ttfamily 1801.05821}}].

\bibitem{salpeter}
E.~E. {Salpeter}, {\em{Electrons Screening and Thermonuclear Reactions},}
  \href{http://dx.doi.org/10.1071/PH540373}{Australian Journal of Physics {\bf
  7} (1954) 373}.

\bibitem{Laine:2015kra}
M.~Laine and M.~Meyer, {\em{Standard Model thermodynamics across the
  electroweak crossover},}
  \href{http://dx.doi.org/10.1088/1475-7516/2015/07/035}{JCAP {\bf 07} (2015)
  035} [\href{http://arxiv.org/abs/1503.04935}{{\ttfamily 1503.04935}}].

\bibitem{Feng:2010zp}
J.~L. Feng, M.~Kaplinghat, and H.-B. Yu, {\em{Sommerfeld Enhancements for
  Thermal Relic Dark Matter},}
  \href{http://dx.doi.org/10.1103/PhysRevD.82.083525}{Phys. Rev. D {\bf 82}
  (2010) 083525} [\href{http://arxiv.org/abs/1005.4678}{{\ttfamily
  1005.4678}}].

\bibitem{Iengo:2009ni}
R.~Iengo, {\em{Sommerfeld enhancement: General results from field theory
  diagrams},} \href{http://dx.doi.org/10.1088/1126-6708/2009/05/024}{JHEP {\bf
  05} (2009) 024} [\href{http://arxiv.org/abs/0902.0688}{{\ttfamily
  0902.0688}}].

\bibitem{Detmold:2014qqa}
W.~Detmold, M.~McCullough, and A.~Pochinsky, {\em{Dark Nuclei I: Cosmology and
  Indirect Detection},}
  \href{http://dx.doi.org/10.1103/PhysRevD.90.115013}{Phys. Rev. D {\bf 90}
  (2014) 115013} [\href{http://arxiv.org/abs/1406.2276}{{\ttfamily
  1406.2276}}].

\bibitem{Ghiglieri:2016xye}
J.~Ghiglieri and M.~Laine, {\em{Neutrino dynamics below the electroweak
  crossover},} \href{http://dx.doi.org/10.1088/1475-7516/2016/07/015}{JCAP {\bf
  07} (2016) 015} [\href{http://arxiv.org/abs/1605.07720}{{\ttfamily
  1605.07720}}].

\bibitem{Kim:2016kxt}
S.~Kim and M.~Laine, {\em{On thermal corrections to near-threshold
  annihilation},} \href{http://dx.doi.org/10.1088/1475-7516/2017/01/013}{JCAP
  {\bf 01} (2017) 013} [\href{http://arxiv.org/abs/1609.00474}{{\ttfamily
  1609.00474}}].

\bibitem{Rogers:1970xx}
{Rogers, F. J. and Graboske, H. C. and Harwood, D. J.}, {\em{Bound Eigenstates
  of the Static Screened Coulomb Potential},}
  \href{http://dx.doi.org/10.1103/PhysRevA.1.1577}{Phys. Rev. A {\bf 1} (1970)
  1577}.

\bibitem{Kharzeev:1994pz}
D.~Kharzeev and H.~Satz, {\em{Quarkonium interactions in hadronic matter},}
  \href{http://dx.doi.org/10.1016/0370-2693(94)90604-1}{Phys. Lett. B {\bf 334}
  (1994) 155} [\href{http://arxiv.org/abs/hep-ph/9405414}{{\ttfamily
  hep-ph/9405414}}].

\bibitem{Grandchamp:2001pf}
L.~Grandchamp and R.~Rapp, {\em{Thermal versus direct $J/\Psi$ production in
  ultrarelativistic heavy ion collisions},}
  \href{http://dx.doi.org/10.1016/S0370-2693(01)01311-9}{Phys. Lett. B {\bf
  523} (2001) 60} [\href{http://arxiv.org/abs/hep-ph/0103124}{{\ttfamily
  hep-ph/0103124}}].

\bibitem{Petraki:2015hla}
K.~Petraki, M.~Postma, and M.~Wiechers, {\em{Dark-matter bound states from
  Feynman diagrams},} \href{http://dx.doi.org/10.1007/JHEP06(2015)128}{JHEP
  {\bf 06} (2015) 128} [\href{http://arxiv.org/abs/1505.00109}{{\ttfamily
  1505.00109}}].

\bibitem{Binder:2020efn}
T.~Binder, B.~Blobel, J.~Harz, and K.~Mukaida, {\em{Dark matter bound-state
  formation at higher order: a non-equilibrium quantum field theory approach},}
  \href{http://dx.doi.org/10.1007/JHEP09(2020)086}{JHEP {\bf 09} (2020) 086}
  [\href{http://arxiv.org/abs/2002.07145}{{\ttfamily 2002.07145}}].

\bibitem{Biondini:2021ycj}
S.~Biondini and V.~Shtabovenko, {\em{Bound-state formation, dissociation and
  decays of darkonium with potential non-relativistic Yukawa theory for scalar
  and pseudoscalar mediators},}
  \href{http://dx.doi.org/10.1007/JHEP03(2022)172}{JHEP {\bf 03} (2022) 172}
  [\href{http://arxiv.org/abs/2112.10145}{{\ttfamily 2112.10145}}].

\bibitem{LL}
V.~B. Berestetskii, E.~M. Lifshitz, and V.~B. Pitaevskii, {\em {Relativistic
  quantum theory, 1971}}.

\bibitem{Planck:2018nkj}
{\bfseries Planck} Collaboration, N.~Aghanim {\em et~al.}, {\em{Planck 2018
  results. I. Overview and the cosmological legacy of Planck},}
  \href{http://dx.doi.org/10.1051/0004-6361/201833880}{Astron. Astrophys. {\bf
  641} (2020) A1} [\href{http://arxiv.org/abs/1807.06205}{{\ttfamily
  1807.06205}}].

\bibitem{Biondini:2018ovz}
S.~Biondini and S.~Vogl, {\em{Coloured coannihilations: Dark matter
  phenomenology meets non-relativistic EFTs},}
  \href{http://dx.doi.org/10.1007/JHEP02(2019)016}{JHEP {\bf 02} (2019) 016}
  [\href{http://arxiv.org/abs/1811.02581}{{\ttfamily 1811.02581}}].

\bibitem{CMS:2013czn}
{\bfseries CMS} Collaboration, S.~Chatrchyan {\em et~al.}, {\em{Searches for
  Long-Lived Charged Particles in $pp$ Collisions at $\sqrt{s}$=7 and 8 TeV},}
  \href{http://dx.doi.org/10.1007/JHEP07(2013)122}{JHEP {\bf 07} (2013) 122}
  [\href{http://arxiv.org/abs/1305.0491}{{\ttfamily 1305.0491}}].

\bibitem{ATLAS:2017tny}
{\bfseries ATLAS} Collaboration, M.~Aaboud {\em et~al.}, {\em{Search for
  long-lived, massive particles in events with displaced vertices and missing
  transverse momentum in $\sqrt{s}$ = 13 TeV $pp$ collisions with the ATLAS
  detector},} \href{http://dx.doi.org/10.1103/PhysRevD.97.052012}{Phys. Rev. D
  {\bf 97} (2018) 052012} [\href{http://arxiv.org/abs/1710.04901}{{\ttfamily
  1710.04901}}].

\bibitem{Hessler:2014ssa}
A.~G. Hessler, A.~Ibarra, E.~Molinaro, and S.~Vogl, {\em{Impact of the Higgs
  boson on the production of exotic particles at the LHC},}
  \href{http://dx.doi.org/10.1103/PhysRevD.91.115004}{Phys. Rev. D {\bf 91}
  (2015) 115004} [\href{http://arxiv.org/abs/1408.0983}{{\ttfamily
  1408.0983}}].

\bibitem{Baker:2016xzo}
M.~J. Baker and J.~Kopp, {\em{Dark Matter Decay between Phase Transitions at
  the Weak Scale},}
  \href{http://dx.doi.org/10.1103/PhysRevLett.119.061801}{Phys. Rev. Lett. {\bf
  119} (2017) 061801} [\href{http://arxiv.org/abs/1608.07578}{{\ttfamily
  1608.07578}}].

\bibitem{Baker:2017zwx}
M.~J. Baker, M.~Breitbach, J.~Kopp, and L.~Mittnacht, {\em{Dynamic Freeze-In:
  Impact of Thermal Masses and Cosmological Phase Transitions on Dark Matter
  Production},} \href{http://dx.doi.org/10.1007/JHEP03(2018)114}{JHEP {\bf 03}
  (2018) 114} [\href{http://arxiv.org/abs/1712.03962}{{\ttfamily 1712.03962}}].

\bibitem{Dvorkin:2019zdi}
C.~Dvorkin, T.~Lin, and K.~Schutz, {\em{Making dark matter out of light:
  freeze-in from plasma effects},}
  \href{http://dx.doi.org/10.1103/PhysRevD.99.115009}{Phys. Rev. D {\bf 99}
  (2019) 115009} [\href{http://arxiv.org/abs/1902.08623}{{\ttfamily
  1902.08623}}].

\bibitem{Darme:2019wpd}
L.~Darm\'e, A.~Hryczuk, D.~Karamitros, and L.~Roszkowski, {\em{Forbidden
  frozen-in dark matter},}
  \href{http://dx.doi.org/10.1007/JHEP11(2019)159}{JHEP {\bf 11} (2019) 159}
  [\href{http://arxiv.org/abs/1908.05685}{{\ttfamily 1908.05685}}].

\bibitem{Landau:1953gr}
L.~Landau and I.~Pomeranchuk, {\em{Electron cascade process at very
  high-energies},} Dokl.Akad.Nauk Ser.Fiz. {\bf 92} (1953) 735.

\bibitem{Landau:1953um}
L.~Landau and I.~Pomeranchuk, {\em{Limits of applicability of the theory of
  bremsstrahlung electrons and pair production at high-energies},}
  Dokl.Akad.Nauk Ser.Fiz. {\bf 92} (1953) 535.

\bibitem{Migdal:1956tc}
A.~B. Migdal, {\em{Bremsstrahlung and pair production in condensed media at
  high-energies},} \href{http://dx.doi.org/10.1103/PhysRev.103.1811}{Phys.Rev.
  {\bf 103} (1956) 1811}.

\bibitem{Anisimov:2010gy}
A.~Anisimov, D.~Besak, and D.~B{\"o}deker, {\em{Thermal production of
  relativistic Majorana neutrinos: Strong enhancement by multiple soft
  scattering},} \href{http://dx.doi.org/10.1088/1475-7516/2011/03/042}{JCAP
  {\bf 03} (2011) 042} [\href{http://arxiv.org/abs/1012.3784}{{\ttfamily
  1012.3784}}].

\bibitem{Besak:2012qm}
D.~Besak and D.~B{\"o}deker, {\em{Thermal production of ultrarelativistic
  right-handed neutrinos: Complete leading-order results},}
  \href{http://dx.doi.org/10.1088/1475-7516/2012/03/029}{JCAP {\bf 03} (2012)
  029} [\href{http://arxiv.org/abs/1202.1288}{{\ttfamily 1202.1288}}].

\bibitem{Ghisoiu:2014mha}
I.~Ghisoiu and M.~Laine, {\em{Interpolation of hard and soft dilepton rates},}
  \href{http://dx.doi.org/10.1007/JHEP10(2014)083}{JHEP {\bf 10} (2014) 083}
  [\href{http://arxiv.org/abs/1407.7955}{{\ttfamily 1407.7955}}].

\bibitem{Bringmann:2021sth}
T.~Bringmann, S.~Heeba, F.~Kahlhoefer, and K.~Vangsnes, {\em{Freezing-in a hot
  bath: resonances, medium effects and phase transitions},}
  \href{http://dx.doi.org/10.1007/JHEP02(2022)110}{JHEP {\bf 02} (2022) 110}
  [\href{http://arxiv.org/abs/2111.14871}{{\ttfamily 2111.14871}}].

\bibitem{Feng:2003xh}
J.~L. Feng, A.~Rajaraman, and F.~Takayama, {\em{Superweakly interacting massive
  particles},} \href{http://dx.doi.org/10.1103/PhysRevLett.91.011302}{Phys.
  Rev. Lett. {\bf 91} (2003) 011302}
  [\href{http://arxiv.org/abs/hep-ph/0302215}{{\ttfamily hep-ph/0302215}}].

\bibitem{Feng:2003uy}
J.~L. Feng, A.~Rajaraman, and F.~Takayama, {\em{SuperWIMP dark matter signals
  from the early universe},}
  \href{http://dx.doi.org/10.1103/PhysRevD.68.063504}{Phys. Rev. D {\bf 68}
  (2003) 063504} [\href{http://arxiv.org/abs/hep-ph/0306024}{{\ttfamily
  hep-ph/0306024}}].

\bibitem{Gould:2021ccf}
O.~Gould and J.~Hirvonen, {\em{Effective field theory approach to thermal
  bubble nucleation},}
  \href{http://dx.doi.org/10.1103/PhysRevD.104.096015}{Phys. Rev. D {\bf 104}
  (2021) 096015} [\href{http://arxiv.org/abs/2108.04377}{{\ttfamily
  2108.04377}}].

\bibitem{Ekstedt:2021kyx}
A.~Ekstedt, {\em{Higher-order corrections to the bubble-nucleation rate at
  finite temperature},}
  \href{http://dx.doi.org/10.1140/epjc/s10052-022-10130-5}{Eur. Phys. J. C {\bf
  82} (2022) 173} [\href{http://arxiv.org/abs/2104.11804}{{\ttfamily
  2104.11804}}].

\bibitem{Ekstedt:2022tqk}
A.~Ekstedt, {\em{Bubble nucleation to all orders},}
  \href{http://dx.doi.org/10.1007/JHEP08(2022)115}{JHEP {\bf 08} (2022) 115}
  [\href{http://arxiv.org/abs/2201.07331}{{\ttfamily 2201.07331}}].

\bibitem{Friedrich:2022cak}
L.~S. Friedrich, M.~J. Ramsey-Musolf, T.~V.~I. Tenkanen, and V.~Q. Tran,
  {\em{Addressing the Gravitational Wave - Collider Inverse Problem},}
  [\href{http://arxiv.org/abs/2203.05889}{{\ttfamily 2203.05889}}].

\bibitem{Abdallah:2015ter}
J.~Abdallah {\em et~al.}, {\em{Simplified Models for Dark Matter Searches at
  the LHC},} \href{http://dx.doi.org/10.1016/j.dark.2015.08.001}{Phys. Dark
  Univ. {\bf 9-10} (2015) 8} [\href{http://arxiv.org/abs/1506.03116}{{\ttfamily
  1506.03116}}].

\bibitem{Arcadi:2019lka}
G.~Arcadi, A.~Djouadi, and M.~Raidal, {\em{Dark Matter through the Higgs
  portal},} \href{http://dx.doi.org/10.1016/j.physrep.2019.11.003}{Phys. Rept.
  {\bf 842} (2020) 1} [\href{http://arxiv.org/abs/1903.03616}{{\ttfamily
  1903.03616}}].

\bibitem{Luo:2002ti}
M.-x. Luo, H.-w. Wang, and Y.~Xiao, {\em{Two loop renormalization group
  equations in general gauge field theories},}
  \href{http://dx.doi.org/10.1103/PhysRevD.67.065019}{Phys. Rev. D {\bf 67}
  (2003) 065019} [\href{http://arxiv.org/abs/hep-ph/0211440}{{\ttfamily
  hep-ph/0211440}}].

\bibitem{Workman:2022ynf}
{\bfseries Particle Data Group} Collaboration, R.~L. Workman, {\em{Review of
  Particle Physics},} PTEP {\bf 2022} (2022) 083C01.

\bibitem{Kainulainen:2019kyp}
K.~Kainulainen, V.~Keus, L.~Niemi, K.~Rummukainen, T.~V.~I. Tenkanen, and
  V.~Vaskonen, {\em{On the validity of perturbative studies of the electroweak
  phase transition in the Two Higgs Doublet model},}
  \href{http://dx.doi.org/10.1007/JHEP06(2019)075}{JHEP {\bf 06} (2019) 075}
  [\href{http://arxiv.org/abs/1904.01329}{{\ttfamily 1904.01329}}].

\bibitem{Ivanov:2018jmz}
I.~P. Ivanov, M.~K\"opke, and M.~M\"uhlleitner, {\em{Algorithmic
  Boundedness-From-Below Conditions for Generic Scalar Potentials},}
  \href{http://dx.doi.org/10.1140/epjc/s10052-018-5893-y}{Eur. Phys. J. C {\bf
  78} (2018) 413} [\href{http://arxiv.org/abs/1802.07976}{{\ttfamily
  1802.07976}}].

\bibitem{Cline:2013gha}
J.~M. Cline, K.~Kainulainen, P.~Scott, and C.~Weniger, {\em{Update on scalar
  singlet dark matter},}
  \href{http://dx.doi.org/10.1103/PhysRevD.88.055025}{Phys. Rev. D {\bf 88}
  (2013) 055025} [\href{http://arxiv.org/abs/1306.4710}{{\ttfamily
  1306.4710}}].

\bibitem{Ellis:2015vaa}
J.~Ellis, F.~Luo, and K.~A. Olive, {\em{Gluino Coannihilation Revisited},}
  \href{http://dx.doi.org/10.1007/JHEP09(2015)127}{JHEP {\bf 09} (2015) 127}
  [\href{http://arxiv.org/abs/1503.07142}{{\ttfamily 1503.07142}}].

\bibitem{Binder:2021vfo}
T.~Binder, A.~Filimonova, K.~Petraki, and G.~White, {\em{Saha equilibrium for
  metastable bound states and dark matter freeze-out},}
  \href{http://dx.doi.org/10.1016/j.physletb.2022.137323}{Phys. Lett. B {\bf
  833} (2022) 137323} [\href{http://arxiv.org/abs/2112.00042}{{\ttfamily
  2112.00042}}].

\bibitem{Garny:2021qsr}
M.~Garny and J.~Heisig, {\em{Bound-state effects on dark matter coannihilation:
  Pushing the boundaries of conversion-driven freeze-out},}
  \href{http://dx.doi.org/10.1103/PhysRevD.105.055004}{Phys. Rev. D {\bf 105}
  (2022) 055004} [\href{http://arxiv.org/abs/2112.01499}{{\ttfamily
  2112.01499}}].

\end{thebibliography}

}
\end{document}